\documentclass[aps,prx,twocolumn,notitlepage,showkeys,showpacs,superscriptaddress,longbibliography]{revtex4-1}
\usepackage{epsf}
\usepackage{bm}
\usepackage{times}
\usepackage{amsfonts}
\usepackage{amssymb}
\usepackage{amsmath,mathtools}
\usepackage{array}
\usepackage{enumerate,dsfont,here}
\usepackage{dcolumn,multirow}
\usepackage{tikz-cd}
\usepackage{bbding}
\usepackage[colorlinks=true,citecolor=blue,linkcolor=blue]{hyperref} 
\usepackage[normalem]{ulem}

\newcommand{\bk}{\bm{k}}

\newcommand{\br}{\bm{r}}

\newcommand{\bb}{\bm{b}}

\newcommand{\mZ}{\mathbb{Z}}
\newcommand{\calT}{\mathcal{T}}
\newcommand{\calC}{\mathcal{C}}
\newcommand{\calP}{\mathcal{P}}
\newcommand{\calJ}{\mathcal{J}}
\newcommand{\calG}{\mathcal{G}}
\newcommand{\calM}{\mathcal{M}}
\newcommand{\calGk}{\mathcal{G}_{\bk}}
\newcommand{\calA}{\mathcal{A}}

\newcommand{\tiG}{\widetilde{\mathcal{G}}}
\newcommand{\frakp}{\mathfrak{p}}
\newcommand{\frakn}{\mathfrak{n}}
\newcommand{\frakcn}{\mathfrak{N}}
\newcommand{\frakb}{\mathfrak{b}}

\newcommand{\add}[1]{{\color{black} #1}}

\begin{document}

\title{Symmetry-based approach to nodal structures: Unification of compatibility relations and gapless point classifications}
\author{Seishiro Ono}
\email{s-ono@g.ecc.u-tokyo.ac.jp}
\affiliation{Department of Applied Physics, University of Tokyo, Tokyo 113-8656, Japan}

\author{Ken Shiozaki}
\email{ken.shiozaki@yukawa.kyoto-u.ac.jp}
\affiliation{Yukawa Institute for Theoretical Physics, Kyoto University, Kyoto 606-8502, Japan}

\begin{abstract}
	Determination of the symmetry property of superconducting gaps has been a central issue in studies to understand the mechanisms of unconventional superconductivity. Although it is often difficult to completely achieve the aforementioned goal, the existence of superconducting nodes, one of the few important experimental signatures of unconventional superconductivity, plays a vital role in exploring the possibility of unconventional superconductivity. The interplay between superconducting nodes and topology has been actively investigated, and intensive research in the past decade has revealed various intriguing nodes out of the scope of the pioneering work to classify superconducting order parameters based on the point groups. However, a systematic and unified description of superconducting nodes for arbitrary symmetry settings is still elusive. In this paper, we develop a systematic framework to comprehensively classify superconducting nodes pinned to any line in momentum space. While most previous studies have been based on the homotopy theory, our theory is on the basis of the symmetry-based analysis of band topology, which enables systematic diagnoses of nodes in all nonmagnetic and magnetic space groups. Furthermore, our framework can readily provide a highly effective scheme to detect nodes in a given \add{superconductor by using density functional theory and assuming symmetry properties of Cooper pairs (called pairing symmetries), which can reduce candidates of pairing symmetries}. We substantiate the power of our method through the time-reversal broken and noncentrosymmetric superconductor CaPtAs. Our work establishes a unified theory for understanding superconducting nodes and facilitates determining superconducting gaps in materials combined with experimental observations.
\end{abstract}

\maketitle

\section{Introduction}
\label{sec1}
While it is often difficult to determine the symmetry property of Cooper pairs (called pairing symmetry in this work)~\cite{Ishida:1998aa,Luke:1998aa,PhysRevLett.110.077003,PhysRevB.90.220502,PhysRevX.7.011032,PhysRevB.96.180507,doi:10.7566/JPSJ.87.093703,Pustogow:2019aa,PhysRevB.100.094530,doi:10.7566/JPSJ.89.034712,Kivelson:2020aa,chronister2020evidence,Ran684,PhysRevLett.123.217001,PhysRevLett.123.217002,PhysRevB.100.220504,Jiao:2020aa,PhysRevResearch.2.032014,bae2020anomalous,hayes2020weyl,ishizuka2020periodic,PhysRevB.86.100507,PhysRevB.87.180503,PhysRevB.89.020509,PhysRevB.89.140504,doi:10.7566/JPSJ.84.054705}, superconducting nodes---geometry of gapless regions in the Bogoliubov quasiparticle spectrum---are key ingredients to identify pairing symmetries. 
For example, power-law behaviors of the specific heat and the magnetic penetration depth are signatures of nodal superconductivity.
Therefore, predictions of superconducting nodes by theoretical studies are helpful to clarify the possible properties of unconventional superconductivity.

Inspired by a series of the discovery of heavy-fermion superconductors such as CeCu$_2$Si$_2$~\cite{PhysRevLett.43.1892} and UPt$_3$~\cite{PhysRevLett.52.679}, superconducting order parameters are classified by irreducible representations of point groups~\cite{Volovik1984, PhysRevB.30.4000,10.1143/PTP.74.221,10.1143/PTP.75.442,Sigrist-Ueda}. 
Since the order parameters are described by basis functions of the irreducible representations in these theories, the intersection between Fermi surfaces and regions where the basis functions vanish is understood as superconducting nodes. Indeed, such analyses succeed in explaining nodes of certain superconductors like cuprate superconductors~\cite{RevModPhys.72.969}. 
%\add{These classifications succeeded in explaining nodes of particular superconductors like cuprate superconductors}~\cite{RevModPhys.72.969}. 
However, recent intensive studies have revealed that such analyses do not consider multiband (orbital) effects and the presence of nonsymmorphic symmetries. As a result, novel symmetry-protected nodes~\cite{PhysRevLett.116.177001,PhysRevLett.118.127001,PhysRevB.96.094526,PhysRevB.96.214514,Kimeaao4513,PhysRevLett.120.057002,PhysRevX.8.011029} have been missed in these theories. 
For example, although Ref.~\onlinecite{PhysRevB.32.2935} argued that symmetry-protected line nodes could not exist in odd-parity superconductors, several works provide counterexamples in the presence of nonsymmorphic symmetries~\cite{PhysRevB.52.15093,PhysRevB.80.100506,PhysRevLett.117.217002,PhysRevB.94.174502,PhysRevB.95.024508}. UPt$_3$ is a prototypical example of materials that exhibits such symmetry-protected line nodes~\cite{PhysRevB.52.15093,PhysRevB.80.100506,PhysRevLett.117.217002,PhysRevB.94.174502,PhysRevB.95.024508}. Another example is surface nodes called Bogoliubov Fermi surfaces. When the time-reversal symmetry (TRS) is broken, the Bogoliubov Fermi surfaces can be realized by a pseudo magnetic field arising from interband Cooper pairs~\cite{PhysRevLett.118.127001,PhysRevB.98.224509}.

Recently, three approaches to overcoming the insufficiency of the previous studies have been proposed.
The first approach is based on the group-theoretical analysis of representations of the Cooper pair wave functions~\cite{PhysRevB.80.100506,PhysRevLett.118.207001,doi:10.7566/JPSJ.86.023703,PhysRevB.95.024508,Sumita-Yanase}. 
In the presence of the inversion symmetry, the theory tells us pairing symmetries that force gap functions to vanish on the mirror plane~\cite{PhysRevB.80.100506,Sumita-Yanase}. Thus, when Fermi surfaces are located on the mirror planes, line nodes exist in the mirror plane because of such pairing symmetries. 
The second approach is based on homotopy theory~\cite{Teo-Kane2010,PhysRevB.83.064505,PhysRevB.83.224511,doi:10.1143/JPSJ.81.011013, Matsuura_2013,PhysRevLett.110.240404,PhysRevB.90.024516,PhysRevB.90.205136,PhysRevLett.114.096804,PhysRevB.92.214514,PhysRevB.94.134512,AZ-node,Kobayashi-Sumita-Yanase-Sato,Sumita-Nomoto-Shiozaki-Yanase,kim2020linking}. In the presence of the inversion and internal symmetries, we define zero-, one-, and two-dimensional topological charges that protect nodes at generic points. Then, depending on the dimensions of the defined topological charges, the shapes of protected nodes, such as line and surface nodes, are determined.
The last approach is the $k\cdot p$ model analysis, which discusses the number of symmetry-allowed mass terms and dispersion in $k\cdot p$ models~\cite{PhysRevLett.108.140405,Yang:2014aa,PhysRevB.90.115111,Chen:2015aa,PhysRevX.5.011029,PhysRevLett.115.036807,PhysRevLett.116.186402}.
Despite the significant progress reported in these works, existing theories cover only simple symmetry settings such as generic points or the mirror planes. In other words, high-symmetry settings such as the rotation and the screw axes in the glide planes, which commonly happen in realistic materials, are out of their scope. Therefore, a comprehensive theory to classify and predict superconducting nodes for arbitrary symmetry classes has long been awaited.
\add{To achieve this goal, we need to answer the following two questions: 
	\begin{enumerate}
		\setlength{\itemsep}{-2pt}
		\item[(I)] Is there a way to comprehensively classify nodes pinned to high-symmetric momenta (often called symmetry-enforced nodes)?
		\item[(II)] Can we classify topologically protected nodes not pinned to particular momenta, which can freely move in planes or the entire Brillouin zone?
	\end{enumerate}
}

\add{In this work, we propose a novel approach to symmetry-enforced nodes on arbitrary lines in momentum space, which will answer the question (I).}
Our method is based on two techniques to clarify the shapes of nodes pinned to the lines.
First, we employ the symmetry-based analysis of band topology~\cite{Po2017,TQC,Watanabeeaat8685,PhysRevX.8.031069,PhysRevX.8.031070,QuantitativeMappings,Ono-Watanabe2018,SI_Adrian,TQC_review,MTQC,Ono-Yanase-Watanabe2019,Skurativska2019,SI_Shiozaki,Ono-Po-Watanabe2020,SI_Luka,Ono-Po-Shiozaki2020,huang2020faithful}. Symmetry representations of wave functions play a pivotal role in the theory. In particular, there exist necessary conditions of symmetry representations to be gapped phases, referred to as compatibility relations~\cite{PhysRevB.59.5998,refId0,MICHEL2001377,PhysRevX.7.041069,Po2017,TQC}. Conversely, if some compatibility relations are violated, the system should be gapless. Suppose that we find a gapless point on a line, which originates from 
a violated compatibility relation. When compatibility relations between the line and its neighborhood exist, we find that the region of violation of the compatibility relation is line or surface; that is, line or surface nodes must exist. Although compatibility relations are powerful tools for understanding nodes, they alone cannot provide complete information about the geometry of nodes. More precisely, when there are no compatibility relations between the line and its neighborhood, we cannot judge whether the gapless point on the line is a genuine point node.

Then, the classification of point nodes on the lines can compensate for the incompleteness of compatibility relations. The results are mainly classified into three types: (i) genuine point nodes, (ii) loop or surface nodes shrinking to a point, and (iii) no point nodes and such shrunk loop or surface nodes. If the classification result on the line is type (ii) or (iii), the gapless point on the line is considered a part of line or surface nodes.

There are two distinctions from existing works in this work. One is that our symmetry-based approach can be applied to any symmetry settings, for example, in the absence of the inversion symmetry and the presence of several nonsymmorphic symmetries. 
In fact, we apply the framework to all \add{nonmagnetic} and magnetic space groups, considering all the possible pairing symmetries that belong to one-dimensional single-valued representations of the point groups. The classification tables we obtained are tabulated in Supplementary Materials. 
Furthermore, the symmetry-based approach has a chance to be more refined to answer question (II), which will also be discussed in the present paper.

The other one is that our framework leads to an efficient algorithm to detect and diagnose nodes in realistic materials, \add{requiring only pairing symmetry and information of irreducible representations of Bloch wave functions at high-symmetry momenta}. Our results therefore will help reduce the candidates of pairing symmetries in realistic superconductors by comparing our results with experimental results on the existence or absence of nodes.
%As a demonstration, we apply our algorithm to CaPtAs, in which the broken TRS is observed~\cite{PhysRevLett.124.207001}. We show that this material is expected to have small Bogoliubov Fermi surfaces.

The remaining part of this paper is organized as follows. 
\add{In Sec.~\ref{sec2}, we provide an overview of our study, which enables readers who do not interested in all details to understand our ideas and results.}
In Sec.~\ref{sec3}, we introduce several ingredients used to formulate our theory. 
We devote Sec.~\ref{sec4} to establish the classification of point nodes on the lines in the presence of point group symmetries. 
In Sec.~\ref{sec5}, we integrate the point-node classifications into the symmetry-based analysis to classify nodal structures pinned to the lines. 
In Sec.~\ref{sec6}, we discuss how to apply our theory to detection of nodes in realistic superconductors. As a demonstration, we apply our algorithm to CaPtAs, in which the broken TRS is observed~\cite{PhysRevLett.124.207001}. We show that this material is expected to have small Bogoliubov Fermi surfaces.
%We comment on nodes at generic momenta and the relationship between such nodes and symmetry indicators in Sec.~\ref{sec7}. 
\add{In Sec.~\ref{sec7}, we comment on nodes at generic momenta and the relationship between such nodes and symmetry-based analysis, which will be an answer to question (II) toward a complete classification of topologically stable nodes.}
We conclude the paper with outlooks for the future works in Sec.~\ref{sec8}. 
Several details are included in appendices to avoid digressing from the main subjects.

\section{Overview of this study}
\label{sec2}
Our major goal is to establish a systematic framework to classify various nodes pinned to lines in momentum space. To achieve this, we will integrate compatibility relations and point-node classifications. In this section, we provide an overview of our strategy and results. 
\add{Throughout the present paper, \textit{gapless point} means a point of momentum space where the bulk gap in the Bogoliubov quasiparticle spectrum is closing. It does not imply that the gapless point is always a genuine point node. As shown in the following discussions, a gapless point on a line connecting two momenta is sometimes part of a line or surface node.} 

\textit{Emergent Altland-Zirnbauer classes and zero-dimensional topological invariants}.---In principle, a complete diagnosis of nodal structures requires computations of all topological charges to protect nodes. In this work, we adopt an alternative way:~we characterize any Bogoliubov quasiparticle spectrum by zero-dimensional topological invariants at various momenta. To accomplish this, we first identify emergent Altland-Zirnbauer (EAZ) classes at a point in momentum space. \add{Here, \textit{emergent} means that such a symmetry class is not a global internal symmetry class but a local one for an irreducible representation at a point in momentum space.} Once the EAZ classes are determined for each irreducible representations at various momenta, we define zero-dimensional topological invariants in the topological periodic table~\cite{PhysRevB.78.195125,Kitaev_bott,Ryu_2010} (see also Table~\ref{tab:EAZ}).

Let us illustrate the notion of EAZ classes through spinful space group $P2/m$ with $B_g$ pairing. In this symmetry setting, the system possesses  TRS $\calT$, the particle-hole symmetry (PHS) $\calC$, the two-fold rotation $C_{2}^{y}$ along the $y$-axis satisfying the anticommutation relations $\{\calC, C_{2}^{y}\}=0$~\cite{Note3}, and the inversion $I$ holding the commutation relation $[\calC, I]=0$. 
Let $H_{\bk}$ and $\psi_{m\bk}$ be the Hamiltonian and its eigenvectors, and $H_{\bk}\psi_{m\bk} = E_{m\bk}\psi_{m\bk}$, where the Bogoliubov quasiparticle spectrum $E_{m\bk}$ is labeled by the band index $m$ and the momentum $\bk$. Since the combined symmetries $I\calC$ and $I\calT$ do not change $\bk$, $(I\calC)\psi_{m\bk}$ and $(I\calT)\psi_{m\bk}$ are also eigenvectors of $H_{\bk}$ with the energies $-E_{m\bk}$ and $E_{m\bk}$, respectively. 

\begin{figure}[t]
	\begin{center}
		\includegraphics[width=0.99\columnwidth]{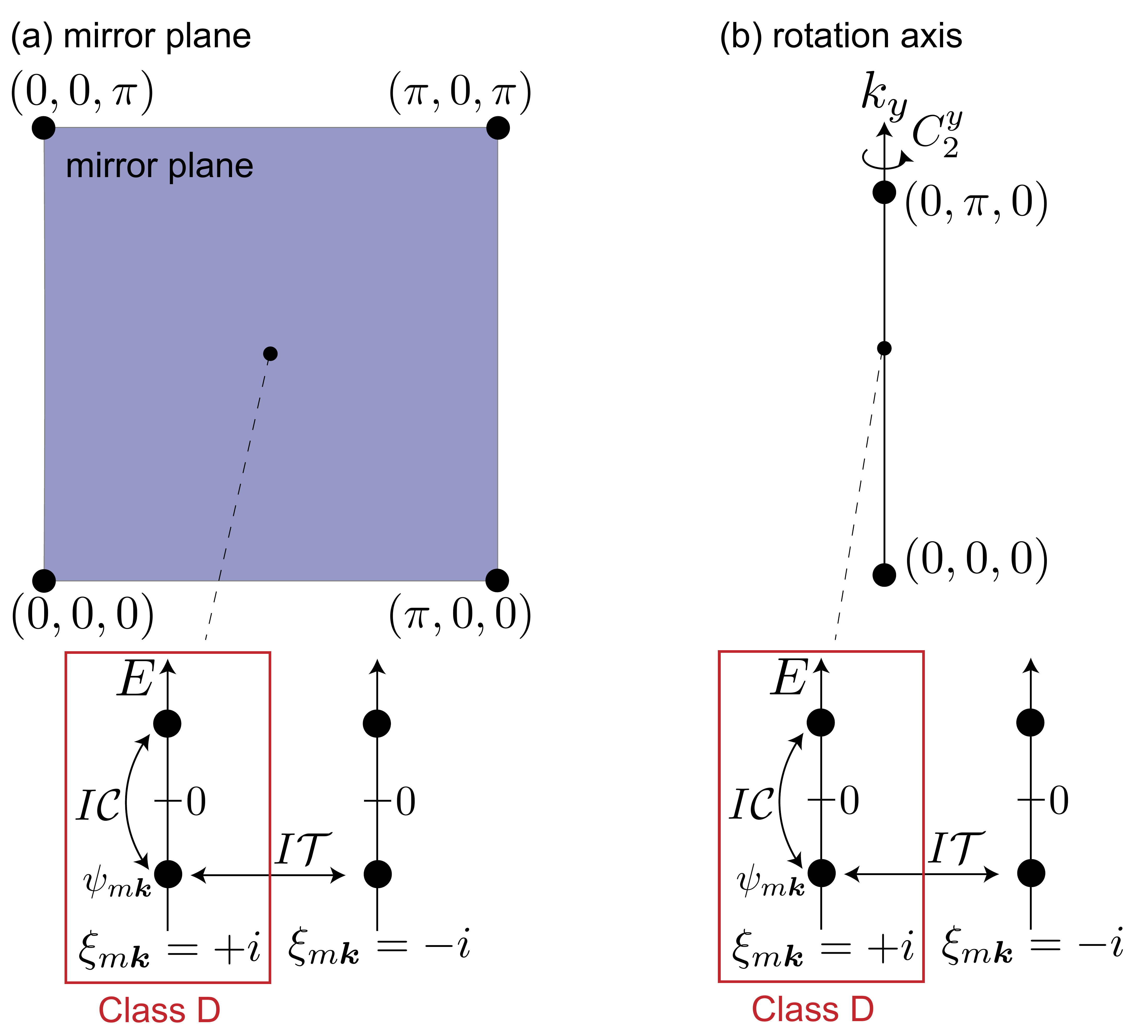}
		\caption{\label{fig:overview}Illustration of the action of symmetries discussed in Sec.~\ref{sec2}. There are two irreducible representations in the mirror plane ($k_z$-$k_x$ plane) [(a)] and the rotation axis ($k_y$-axis) [(b)]. They are invariant under $I\calC$ but exchanged by $I\calT$. As a result, the EAZ classes for the irreducible representations are class D, and thus two $\mZ_2$ topological invariants (Pfaffian invariants) are defined at every point in the mirror plane and the rotation axis (except for high-symmetry points).}
	\end{center}
\end{figure}

We begin by focusing on a generic momentum $\bk$ in the two-dimensional plane invariant under the mirror symmetry $M_{y} = IC_{2}^y$. In this plane, the eigenvectors $\psi_{m\bk}$ of $H_{\bk}$ are also those of $M_y$ with mirror eigenvalues $\xi_{m\bk} = \pm i$. Then $(I\calC)\psi_{m\bk}$ and $(I\calT)\psi_{m\bk}$ have the mirror eigenvalues $\xi_{m\bk}$ and $-\xi_{m\bk}$, respectively. This implies that the combined symmetry $I\calC$ does not change the mirror sector but $I\calT$ changes, which results in class D as the EAZ symmetry class of each mirror sector at the point $\bk$ (see Fig.~\ref{fig:overview} (a)). As is the case of the mirror plane, completely the same discussion can be applied to any point (except for higher-symmetry momenta) in the rotation symmetric line. Then, we find that the EAZ symmetry class of each rotation-eigenvalue sector at the point is class D (see Fig.~\ref{fig:overview} (b)). For EAZ class D, the Pfaffian invariants $p_{\bk}^{\pm i}$ are defined. 

\begin{figure}[t]
	\begin{center}
		\includegraphics[width=0.99\columnwidth]{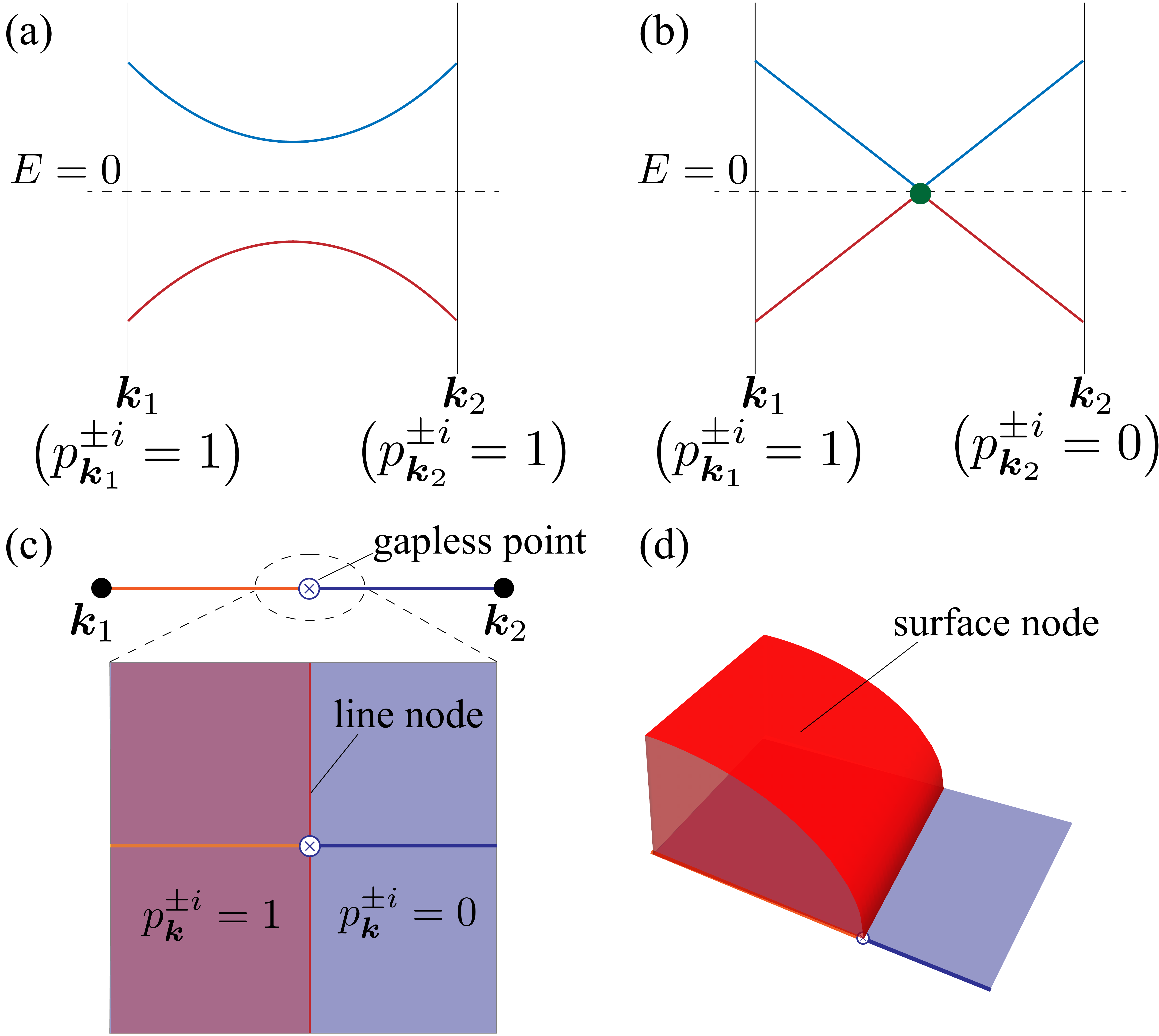}
		\caption{\label{fig:overview_CR}Illustration of diagnosis based on compatibility relations. 
			(a, b) Bogoliubov quasiparticle spectrum along the line connecting two momenta $\bk_1$ and $\bk_2$. The spectrum satisfies the compatibility relations on the line in (a), while does not in (b).
 			(c) Two divisions of the line connecting $\bk_1$ and $\bk_2$ and a nodal line in a plane containing the line. Here, the red shaded region and others have different values of the topological invariants, whose boundary results in a nodal line. (d) Surface node. When there are compatibility relations between the plane and its three-dimensional neighborhood, the regions in (c) are extended out of the plane, and the boundary surface is the surface node.
		}
	\end{center}
\end{figure}

\textit{Diagnosis of nodal structures based on compatibility relations}.---As seen in the preceding discussions, we show that zero-dimensional topological invariants are defined at each momentum. Then, the question is whether these zero-dimensional topological invariants are fully independent or not. In general, for the gapped region in momentum space, these zero-dimensional topological invariants are subject to symmetry constraints. Topological invariants do not change when the system in the same topological phase during the continuous deformation (see Fig.~\ref{fig:overview_CR}(a)). Thus, when we consider momentum as parameters of the deformation, the zero-dimensional topological invariants must be the same for the gapped region. In this work, we refer to such constraints on zero-dimensional topological invariants as \textit{compatibility relations}. Conversely, if the zero-dimensional topological invariants are changed between two points, the Bogoliubov quasiparticle spectrum must have gapless points on this line (see Fig.~\ref{fig:overview_CR}(b)).

The existence of a gapless point pinned to the line immediately implies that there are two regions in which the zero-dimensional topological invariants are different from each other (see the upper panel of Fig.~\ref{fig:overview_CR}(c)). Next, we discuss the diagnosis of the shape of nodes when we find a gapless point originating from the change of zero-dimensional topological invariants on a line. 
Suppose that there exist compatibility relations between the two subdivisions and their neighborhoods. Furthermore, since the gradient of dispersion does not usually diverge, it is natural to think that neighborhoods of the regions on the line are gapped. However, due to the compatibility relations, the two neighborhoods also have different topological invariants. Therefore, the boundary of these neighborhoods leads to a line node (see the lower panel of Fig.~\ref{fig:overview_CR}(c)). When the system is three-dimensional, the same discussion can be further applied to the line node and its three-dimensional neighborhood (see Fig.~\ref{fig:overview_CR}(d)). 

Again, we discuss the case for space group $P2/m$ with $B_g$ pairing. Let us start with the mirror plane. We pick two momenta $\bk_1$ and $\bk_2$, which are not the high-symmetry points. We also suppose that the different Pfaffian invariants are assigned, say, $p_{\bk_1}^{\pm i} = 1$ and $p_{\bk_2}^{\pm i} = 0$. Then, a gapless point must be on the line between $\bk_1$ and $\bk_2$, as discussed above. In the mirror plane, there exists a compatibility relation such that $p_{\bk}^{\pm i}$ must be the same for the gapped regions. As a result, we find that the situation is actually the same as Fig.~\ref{fig:overview_CR}(c) and that the gapless point is part of the line node. 
On the other hand, the situation for rotation axes is different from that for the mirror plane. There are no compatibility relations between a point in the rotation axis and generic momenta. 
In such a case, one might think that the point node is the only case. However, we cannot conclude that the gapless point is a genuine point node.
\add{The possibility of a line node protected by one-dimensional topological invariants, such as the Berry phase and the winding number, still remains since the absence of compatibility relations just guarantees that there are no line and surface nodes protected by zero-dimensional topological invariants.}
In summary, compatibility relations can tell us part of nodal structures but not completely determine them.

\textit{Gapless point classifications on lines}.---In such a case, we need another tool to distinguish two possibilities of a genuine point node or a line node. This is achieved by the classifications of two-dimensional massive Dirac Hamiltonians near gapless points on the line
\begin{align}
	\label{eq:overview_cc-ham}
	H_{(k_1, k_2)} &= k_1 \gamma_1 + k_2\gamma_2 + \delta k_3 \gamma_0,
\end{align}
where $k_1$ and $k_2$ are momenta in the directions perpendicular to the the line, and $\delta k_3$ is a displacement from the gapless point in the direction of the line. Gamma matrices $\gamma_{0}, \gamma_1$, and $\gamma_2$ anticommute with each others.
\begin{figure}[t]
	\begin{center}
		\includegraphics[width=0.9\columnwidth]{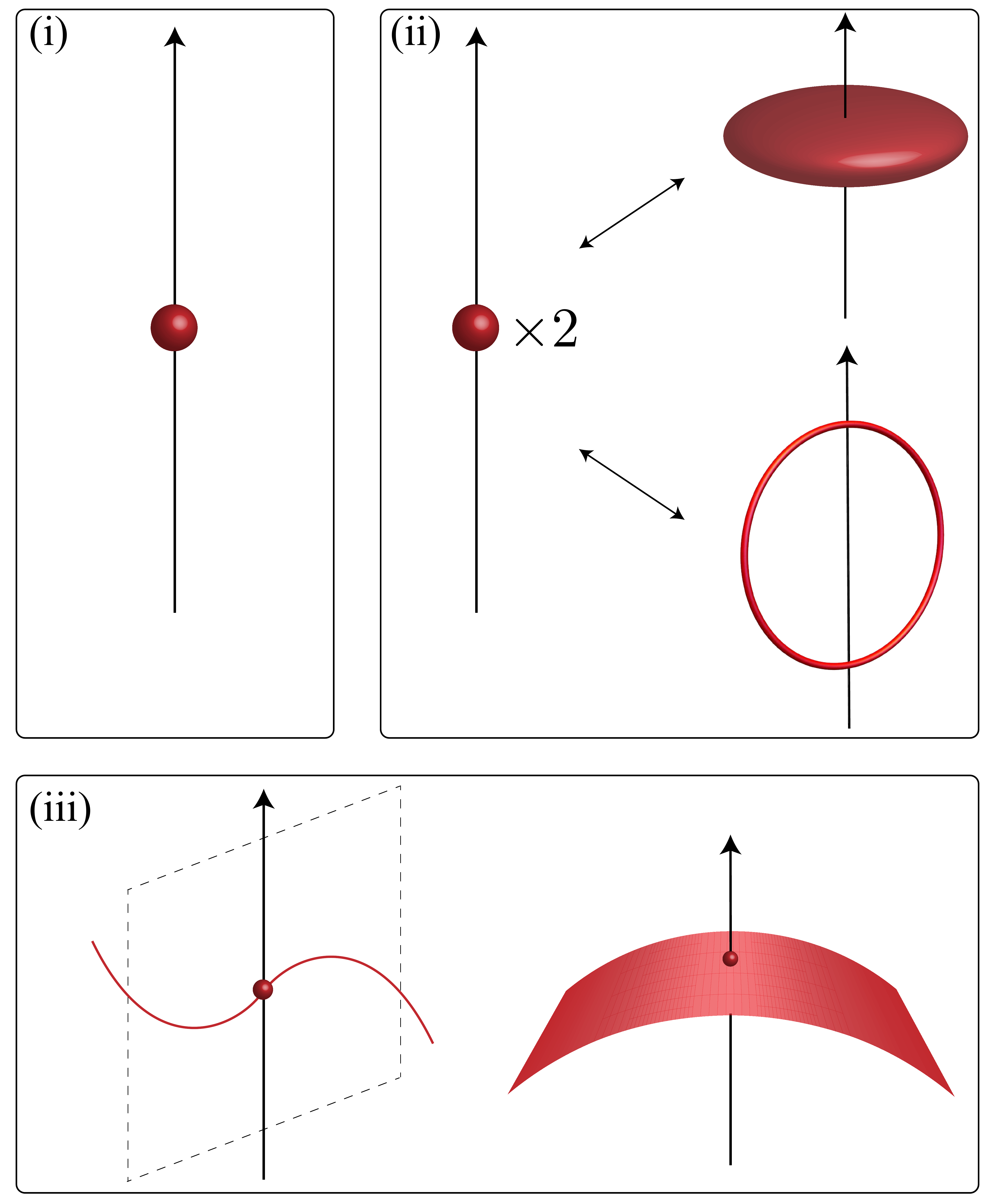}
		\caption{\label{fig:CC_result}Illustration of results of point-node classifications on a line:
			(i) The gapless point on the line is a genuine point node. (ii) A point node formed by multiple gapless points can be realized, but such a point node is actually a shrunk loop or surface node. Since there are no reasons why two gapless points are at the same position, it is natural to consider that a shrunk loop or surface node exists in such a case. (iii) The gapless point must be part of a line or surface node.
		}
	\end{center}
\end{figure}
After classifying the Dirac Hamiltonians, we find three types of gapless points: (i) a genuine point node [Fig.~\ref{fig:CC_result} (i)], (ii) a shrunk loop or surface node [Fig.~\ref{fig:CC_result} (ii)], and (iii) part of line or surface nodes [Fig.~\ref{fig:CC_result} (iii)]. It should be noted that the shrinking for case (ii) is not forced by symmetries. In other words, case (ii) indicates that such loop and surface nodes can shrink to a point just by deformations. In this work, we consider that such shrinkable nodes are realized as loop or surface nodes.

 Indeed, the classification result for the rotation axis in space group $P2/m$ with $B_g$ pairing is the case (iii), as shown in Sec.~\ref{sec4:p2m}. Thus, the gapless point on the rotation axis is not a genuine point node.

\begin{figure*}[t]
	\begin{center}
		\includegraphics[width=1.8\columnwidth]{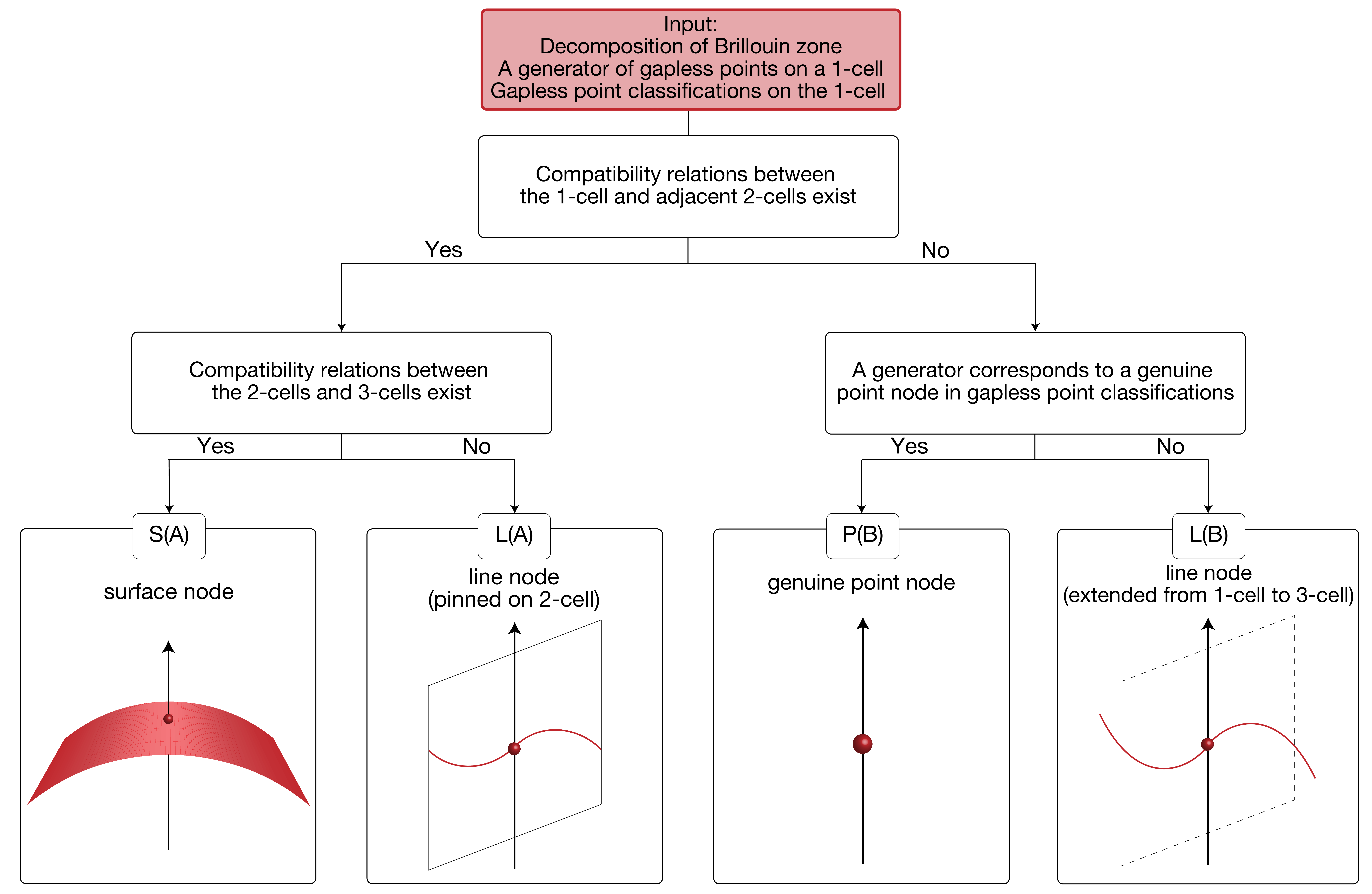}
		\caption{\label{fig:flow}A flowchart of our classification scheme. We focus on a generator of gapless points on a line (called 1-cell). We separately perform gapless point classifications on the 1-cell. Then, we ask if compatibility relations between the line and adjacent polygons exist or not. If yes, we examine whether compatibility relations are between the polygons and their three-dimensional neighborhoods. If they exist, the gapless point is part of a surface node, denoted by S(A). Otherwise, the gapless point is part of a line node pinned on the polygons, denoted by L(A).
		Next, we consider the case where the compatibility relations between the line and adjacent polygons do not exist. In such a case, we ask if the generator coincides with a genuine point node from the results of gapless point classifications on 1-cells. 
		When the gapless point belongs to case (i) of the point-node classification, it is a genuine point node, denoted by P(B).
		If the gapless point is not consistent with the existence of a genuine point node, i.e., the point-node classification result is the case (ii) or (iii), we conclude that the gapless point is part of a line node, denoted by L(B).
		}
	\end{center}
\end{figure*}

\textit{Unification of compatibility relations and point-node classifications}.---Unifying compatibility relations and point-node classifications, we finally arrive at our classification scheme for the gapless points on lines, which is summarized in Fig.~\ref{fig:flow}. 
As a preparation for the classifications, we decompose momentum space into points, lines, polygons, and polyhedrons (called $0$-cells, $1$-cells, $2$-cells, and $3$-cells, respectively in this work) [cf. Fig.~\ref{fig:p4mm}]. Suppose that we have a generator of gapless points on the line. 
\add{Here, \textit{generator} means a gapless point induced by a change of zero-dimensional topological invariants for irreducible representations. In other words, the generator has a minimum number of gapless states at a point in the line, which cannot be split due to symmetry constraints.} 
We first check whether compatibility relations between the line and adjacent polygons exist or not. 
Let us begin by discussing the case where they exist. Then, we further examine if compatibility relations are between the polygons and adjacent polyhedrons. If they exist, the gapless point is part of a surface node [S(A) in Fig.~\ref{fig:flow}]. Otherwise, the gapless point is part of a line node pinned on the polygons [L(A) in Fig.~\ref{fig:flow}].
On the other hand, when the compatibility relations between the line and adjacent polygons do not exist, we ask if the gapless point on the line is a genuine point node from the results of point-node classifications. 
When the gapless point belongs to case (i) of the point-node classification, it is a genuine point node [P(B) in Fig.~\ref{fig:flow}].
If the gapless point is not consistent with the existence of a genuine point node, i.e., the point-node classification result is the case (ii) or (iii), we conclude that the gapless point is part of a line node [L(B) in Fig.~\ref{fig:flow}]. Note that, since stable surface nodes require zero-dimensional topological charges, which are actually equivalent to zero-dimensional topological invariants, they are always diagnosable by compatibility relations.

In this work, we classify the nodes pinned to the lines in all nonmagnetic and magnetic space groups with concrete decomposition of momentum space. All results are summarized as tables in Supplemental Materials, which contain the information about positions and shapes of nodes. 

\begin{table}[t]
	\begin{center}
		\caption{\label{tab:overview_P2m}Part of classification table for space group $P2/m$ with $B_g$ pairing. 
			The first and second columns represent the boundary points of the line where a gapless point exists. 
			Labels of irreducible representations (irrep) are shown in the third column, which follows the notation in Ref.~\onlinecite{Bilbao}.
			The fourth column is the classification $\mZ$ or $\mZ_2$, and the fifth column means the type of nodes. Here P, L, and S denote point, line, and surface nodes, respectively. In addition, while (A) means that the shape of the node is determined only by compatibility relations, (B) indicates that gapless point classifications are necessary.
		}
		\begin{tabular}{c|c|c|c|c}
			\hline
			HSP1 & HSP2 & irrep & classification & type of node \\
			\hline\hline 
			$\left(0,0,0\right)$&$\left(\frac{1}{2},0,0\right)$&$\bar{F}_3$&$\mathbb{Z}_2$&\text{L(A)}\\
			$\left(0,0,0\right)$&$\left(0,\frac{1}{2},0\right)$&$\bar{\Lambda}_3$&$\mathbb{Z}_2$&\text{L(B)}\\
			\hline
		\end{tabular}
	\end{center}
\end{table}
\textit{Applications to materials}.---Our classification leads to an efficient way to diagnose nodal structures in realistic superconductors. 
There are two things that they have to do.
One is to perform density-functional theory (DFT) calculations and compute irreducible representations in the normal phase at high-symmetry points, which leads to zero-dimensional topological invariants at high-symmetry points in the weak-pairing assumptions~\cite{SI_Luka,Ono-Po-Shiozaki2020} (see Sec.~\ref{sec6} for more details). 
The other is to check if the obtained zero-dimensional topological invariants satisfy compatibility relations or not. 
Examining compatibility relations between zero-dimensional topological invariants at two high-symmetry points, we can detect the positions of gapless points on the line between two high-symmetry points.
Furthermore, referring to the classification tables, we can also understand the shape of nodes.

For example, let us suppose that we have a superconductor crystallized in space group $P2/m$ with $B_g$ pairing. 
In this space group, there are eight high-symmetry points, at which four irreducible representations are defined and labeled by $1,2,3,$ and $4$.
Then, their EAZ classes are class D, and four Pfaffian invariants are defined at these points.
We further suppose that the Pfaffian invariants for the irreducible representations $1$ and $2$ at $\Gamma$ are nontrivial and that others are trivial. After examining if compatibility relations are satisfied, we find various violated ones. Here, let us focus on the violated compatibility relations on $(0,0,0)$-$(1/2,0,0)$ and $(0,0,0)$-$(0,1/2,0)$. Then, referring to the Table~\ref{tab:overview_P2m}, we immediately see that the gapless points on these lines are part of line nodes.

It is worth noting that our framework have been implemented in an automatic program. In Ref.~\onlinecite{Tang-Ono-Wan-Watanabe2021}, the authors have developed a subroutine, which enable us to perform the diagnosis of nodal structures just by uploading particularly formatted results of DFT calculations.

\section{Formalism}
\label{sec3}
In Sec.~\ref{sec2}, we have provided an overview of our ideas to classify nodes on symmetric lines. In this section, we explain several ingredients of implementations of the systematic classifications, which will be discussed in Secs.~\ref{sec4} and \ref{sec5}. 

\subsection{BdG Hamiltonian and Symmetry representations}
In this work, we always consider superconductors which can be described by the Bogoliubov–de Gennes (BdG) Hamiltonian
\begin{align}
	\label{eq:BdG}
	H_{\bk}&=\begin{pmatrix}
		h_{\bk}& \Delta_{\bk} \\
		\Delta_{\bk}^{\dagger} & -h_{-\bm{k}}^{*}
	\end{pmatrix},
\end{align}
where $h_{\bk}$ and $\Delta_{\bk}$ denote the normal-phase Hamiltonian and the superconducting gap function, respectively~\cite{Note1}.
\add{Here, we choose the gauge such that the BdG Hamiltonian is periodic in $\bk$, i.e., $H_{\bk+\bm{G}} = H_{\bk}$ for reciprocal lattice vectors $\bm{G}$.}

Suppose that the normal phase is invariant under a magnetic space group (MSG) $\calM = \calG + \calA$, where $\calG$ is a space group and $\calA$ is an antiunitary part of $\calM$. \add{Note that the notion of MSG contains all ordinary space groups without and with TRS. For instance, when every element in $\calA$ is the product of TRS and an element of $\calG$, the MSG is no more than a space group with TRS.}
A MSG $\calM$ always has a subgroup $T$ consisted of all lattice translations. An element $g\in \calM$ transform a point $\br$ in the real space to $g\br = p_g\br+\bm{t}_g$, where $p_g$ is an element of $\text{O}(3)$ and $\bm{t}_g$ represents a lattice translation or a fractional translation. Because of the existence of PHS $\calC$ in the BdG Hamiltonian, the full symmetry group $G$ is divided by the following four parts
\begin{align}
	G &= \calM + \calM \calC \nonumber\\
	\label{eq:MSG}
	&= \calG + \calA + \mathcal{P} + \mathcal{J}
\end{align}
where $\mathcal{P}=\calG\calC$ and $\mathcal{J}=\calA\calC$ are sets of particle-hole like and chiral like symmetries.

We recall symmetry representations of $G$ in momentum space. We introduce two maps $c, \phi: G \rightarrow \mZ_2=\{-1,1\}$. Here, $\phi_g = +1\ (-1)$ means $g$ is unitary (antiunitary), and $c_g=+1\ (-1)$ represents $g$ commutes (anticommutes) with the Hamiltonian $H_{\bk}$. Accordingly, an element $g\in \calM$ transforms a point $\bk$ in momentum space into $g\bk=\phi_gp_g\bk$. In addition, the representation $\rho_{\bk}(g)$ is expressed by
\begin{align}
	\rho_{\bk}(g) &= \begin{cases}
		U_{\bk}(g) \quad \text{for }\phi_g = +1,\\
		U_{\bk}(g)K \quad \text{for }\phi_g = -1,
	\end{cases}
\end{align}
and $\rho_{\bk}(g)$ satisfies
\begin{align}
	\label{eq:trans_ham}
	\rho_{\bk}(g)H_{\bk} &= \begin{cases}
		H_{g\bk}\rho_{\bk}(g)\quad \text{for }c_g = +1,\\
		-H_{g\bk}\rho_{\bk}(g)\quad \text{for }c_g = -1, \end{cases}
\end{align}
where $U_{\bk}(g)$ and $K$ are a unitary matrix and the conjugation operator, respectively. Note that $U_{\bk}(g)$ is a projective representation, i.e., the following relation holds
\begin{align}
	\rho_{g'\bk}(g) \rho_{\bk}(g') = z_{g,g'}\rho_{\bk}(gg'),
\end{align}
where $z_{g,g'}\in \text{U}(1)$ is a projective factor of $G$. For spinless systems, we can always choose $z_{g, g'} = +1$ for $g,g' \in \calG$ or $\calA$.

Let us consider a point $\bk$ in momentum space. For this point, we introduce a little group $\calG_{\bk}= \{h \in  \calG| h\bk = \bk +^\exists\bm{G}\}$, where $\bm{G}$ is a reciprocal lattice vector. For $h \in \calGk$, since elements in $\calGk$ are symmetries of $H_{\bk}$, 
we can simultaneously block-diagonalize $H_{\bk}$ and $U_{\bk}(h)$ such that
\begin{align}
	\label{eq:block-diag}
	U_{\bk}(h) &=\text{diag}\left[U_{\bk}^{\alpha_1}(h)\otimes\mathds{1}_{m_1}, \cdots, U_{\bk}^{\alpha_n}(h)\otimes\mathds{1}_{m_n}\right],\\
	\label{eq:block-diag2}
	H_{\bk}&=\text{diag}\left[\mathds{1}_{d^{\alpha_1}}\otimes H^{\alpha_1}_{\bk}, \cdots, \mathds{1}_{d^{\alpha_n}}\otimes H^{\alpha_n}_{\bk}\right],
\end{align}
where $U_{\bk}^{\alpha}(h)$ is an irreducible representations of $\calG_{\bk}$. Here, $d^{\alpha}$ and $m_{\alpha}$ are dimensions of $U_{\bk}^{\alpha}(h)$ and $H^{\alpha}_{\bk}$, respectively~\cite{SI_Luka,Ono-Po-Shiozaki2020}.

One often considers the finite group $G_{\bk}/T$, where $G_{\bk}$ is a subgroup of $G$ and is defined in the same way as $\calGk$. In the literature~\cite{Bradley}, $G_{\bk}/T$ is referred to as ``little co-group.'' Importantly, $G_{\bk}/T$ is isomorphic to a magnetic point group with PHS. We can always relate representations of $G_{\bk}$ to those of $G_{\bk}/T$, and we define the representation $\sigma_{\bk}(g)$ of $G_{\bk}/T$ by 
\begin{align}
	\sigma_{\bk}(g) &= \begin{cases}
		U_{\bk}(g)e^{-i \bk \cdot \bm{t}_g} \quad \text{for }\phi_g = +1,\\
		U_{\bk}(g)e^{-i \bk \cdot \bm{t}_g} K \quad \text{for }\phi_g = -1,
	\end{cases}
\end{align}
where $\bm{t}_g$ is a fractional translation or zeros.
Correspondingly, projective factors also change as
\begin{align}
	\sigma_{\bk}(g)\sigma_{\bk}(h) &= z_{g,h}^{\bk}\sigma_{\bk}(gh),
\end{align}
where $z_{g,h}^{\bk}=z_{g,h}e^{-i \bk (p_g \bm{t}_h- \phi_g \bm{t}_h)}$. Using these projective factors $z_{g,h}^{\bk}$, we can obtain irreducible representations $u_{\bk}^{\alpha}$ of $\calGk/T$, which is simply related to irreducible representations $U_{\bk}^{\alpha}$ of $\calGk$ by
\begin{align}
	U_{\bk}^{\alpha}(g) &= u_{\bk}^{\alpha}(g)e^{-i\bk\cdot \bm{t}_g}.
\end{align}

\subsection{Cell decomposition}
\label{sec3:cell}
\begin{figure*}[t]
	\begin{center}
		\includegraphics[width=1.6\columnwidth]{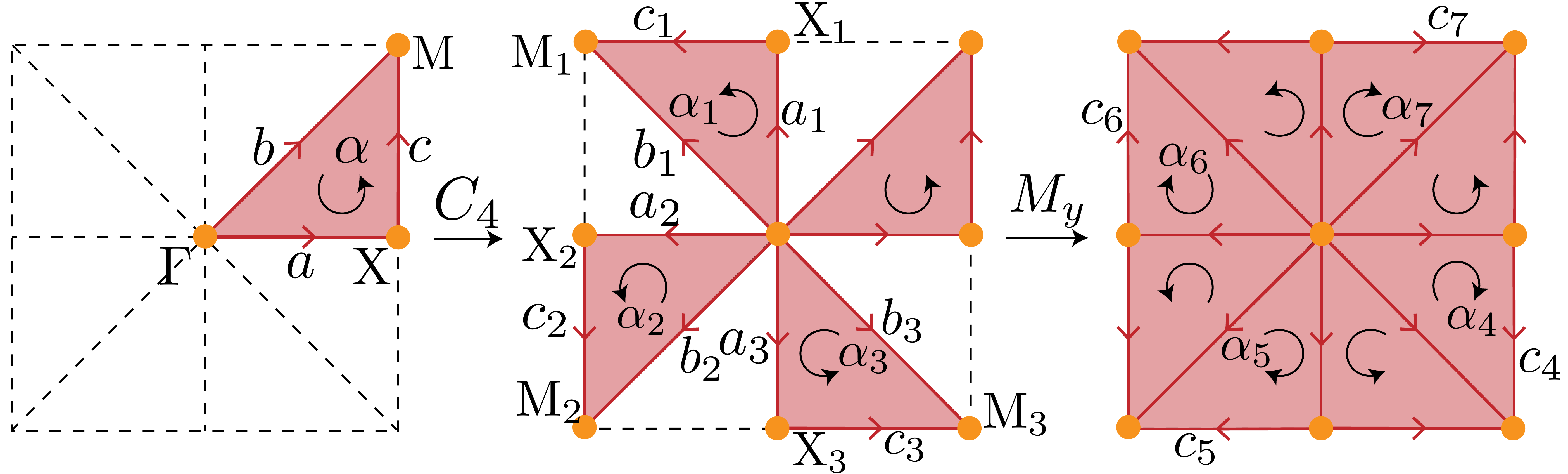}
		\caption{\label{fig:p4mm}Cell decomposition for $p4mm$. We first find a unit of BZ illustrated in the left panel. The red and black arrows signify orientations of 1-cells and 2-cells, respectively. Then, we rotate each p-cell in the unit by the four-fold rotation symmetry. Finally, mapping them by the mirror symmetry, we arrive at the cell decomposition shown in the right panel.}
	\end{center}
\end{figure*}
Here, we explain the \textit{cell decomposition} of the Brillouin zone (BZ)~\cite{Shiozaki2018}. In this work, we divide BZ into points, lines, polygons, and polyhedrons, which are called 0-cells, 1-cells, 2-cells, and 3-cells, respectively. Before moving on to the formal discussions, we begin by introducing an example.

Let us consider the wallpaper group $p4mm$ in two-dimension. Here we describe a way to find the cell decomposition shown in Fig.~\ref{fig:p4mm}. in which $0$-cells, $1$-cells, and $2$-cells are represented by orange circles, solid red lines, and pink polygons, respectively.
%A cell decomposition is given by Fig.~\ref{fig:p4mm}, where $0$-cells, $1$-cells, and $2$-cells are represented by orange circles, solid red lines, and pink polygons, respectively. Here we describe a way to find the cell decomposition. 
We first find an asymmetric unit of BZ, and then decompose the asymmetric unit into three $0$-cells (orange circles), three $1$-cells (solid red lines), and a $2$-cell (pink plane) in the left panel of Fig.~\ref{fig:p4mm}. Finally, we act symmetry operations on this asymmetric unit and obtain the cell decomposition of the entire BZ:
\begin{align}
 	\label{eq:C0_p4mm}
 	\calC_0 &= \{\Gamma, \text{X}, M, \text{X}_1, \text{M}_1, \text{X}_2, \text{M}_2, \text{X}_3, \text{M}_3 \}, \\
 	\label{eq:C1_p4mm}
 	\calC_1 &= \{a, b, c, a_1, b_1, c_1, a_2, b_2, c_2,a_3, b_3, c_3, c_4, c_5, c_6, c_7\}, \\
 	\label{eq:C2_p4mm}
 	\calC_2 &= \{\alpha, \alpha_1, \alpha_2, \alpha_3, \alpha_4, \alpha_5, \alpha_6, \alpha_7\},
\end{align}
where $\calC_p\ (p = 0, 1,2)$ represents the set of $p$-cells.
Note that, although various $p$-cells are equivalent or symmetry-related to other $p$-cells, we here assign different labels to them. For example, $\text{X}_2=(-\pi,0)$ is equivalent to $\text{X}=(\pi,0)$ and  $\text{X}_1=(0,\pi)$ is symmetry-related to $\text{X}$.

%Here, we explain the decomposition of the first Brillouin zone (BZ), which is called \textit{cell decomposition}~\cite{Shiozaki2018}. Let $T^d$ be the $d$-dimensional BZ. We introduce a series of subspace of $T^d$ such that
%\begin{align}
%	X_0 \subset X_1 \subset \cdots \subset X_d = T^d,
%\end{align}
%where $X_p$ is referred to as $p$-skelton. Each $X_p$ is invariant under $G$, i.e., $g\bk \in X_p$ for $\forall g \in G$ if and only if $\bk \in X_p$. We remark that the choice of $X_p$ is not unique. 

We proceed to explain a construction for arbitrary symmetry settings. 
%In the following, we will explain a way to get $X_p$ that enable us to perform procedures in Sec.~\ref{sec5} systematically. 
As is the case of the above example, we first find an asymmetric unit of BZ and divide the asymmetric unit into the set of $p$-cells $\{D^{p}_{i}\}_i$ for $p=0,1,\cdots, d$. Next, we copy the decomposition of the asymmetric unit throughout the entire BZ by using crystalline symmetries. In other words, we define the entire set of $p$-cells by
%Next, we extend such division to the entire BZ. To achieve this, we act symmetry operations in $G/T$ on each $p$-cell, and then we can define the entire set of $p$-cells by
\begin{align}
	\calC_p \equiv \bigcup_{i} \bigcup_{g \in G/T} D_{g(i)}^{p},
\end{align}
where $D_{g(i)}^{p} = gD^{p}_{i}$. Note that, in this construction, some $p$-cells are equivalent or symmetry-related to others up to reciprocal lattice vectors. However, we do not identify such $p$-cells with others in the procedures of cell decomposition, and we will take into account these identifications in the construction of $E_1$-pages in Sec.~\ref{sec3:E1}.

Each $p$-cell satisfies the following conditions: 
\begin{enumerate}
	\setlength{\itemsep}{-2pt}
	\item[(i)] The intersection of any two $p$-cells in $\calC_p$ is an empty set, i.e., $D^{p}_{i}\cap D^{p}_{j}=\emptyset\ (i\neq j)$.
	\item[(ii)] Any point in a $p$-cell $D_{i}^{p}$ is invariant under symmetries or transformed to points in different $p$-cells by symmetries, namely, $g\bk = \bk+^{\exists}\bm{G}$ or $g\bk \in D_{g(i)}^{p}$ if $\bk \in D_{i}^{p}$.
	\item[(iii)] The boundary $\partial D_{i}^{p}$ consists of $(p-1)$-cells for $p\geq 1$.
	\item[(iv)] Each $p$-cell ($p\geq 1$) is oriented in a symmetric manner.
	\item[(v)] Any two of the boundary $p$-cells of the $(p+1)$-cell are not equivalent and symmetry-related to each other.
\end{enumerate}
For our purpose to systematically diagnose nodes pinned to lines in BZ, the condition (v) is crucial. In Appendix~\ref{app:cell_3D}, we provide units of 3D BZ for each type of lattices. 
%Finally, the $p$-skelton $X_p$ is determined by
%\begin{align}
%	X_0 = \calC_0, X_p= \calC_p \cup X_{p-1}\ \ (p \geq 1).
%\end{align}

\subsection{Emergent Altland-Zirnbauer classes}
\label{sec3:EAZ}
\begin{table}
	\begin{center}
		\caption{\label{tab:EAZ}The classification of zero-dimensional topological phases for each EAZ symmetry class. Topological indices $p_{\bk}^{\alpha}$ and $N_{\bk}^{\alpha}$ in the table are defined by Eq.~\eqref{eq:Pf} and \eqref{eq:int}. Here, $\mathcal{W}[\alpha]$ represents the triple of the result of Wigner criteria  $(W^{\alpha}_{\bk}(\calT) , W^{\alpha}_{\bk}(\calP), W^{\alpha}_{\bk}(\calJ))$ defined by Eqs.~\eqref{eq:wigner_C}-\eqref{eq:wigner_G}.
		}
		\begin{tabular}{c|c|c|c}
			\hline
			EAZ & $\mathcal{W}_{\bk}[\alpha]$ & classification & index  \\
			\hline\hline
			A & $(0,0,0)$ & $\mZ$ & $N_{\bk}^{\alpha}$ \\
			AIII & $(0,0,1)$ & $0$ & None \\
			AI & $(1,0,0)$ & $\mZ$ & $N_{\bk}^{\alpha}$ \\
			BDI & $(1,1,1)$ & $\mZ_2$ & $p_{\bk}^{\alpha}$ \\
			D & $(0,1,0)$ & $\mZ_2$ & $p_{\bk}^{\alpha}$ \\
			DIII & $(-1,1,1)$ & $0$ & None \\
			AII & $(-1,0,0)$ & $2\mZ$ & $N_{\bk}^{\alpha}$ \\
			CII & $(-1,-1,1)$ & $0$ & None \\
			C & $(0,-1,0)$ & $0$ & None \\
			CI & $(1,-1,1)$ & $0$ & None \\
			\hline
		\end{tabular}
	\end{center}
\end{table}
\add{Symmetries of $\calA, \mathcal{P},$ and $\mathcal{J}$ in Eq.~\eqref{eq:MSG} sometimes keep a sector $H_{\bk}^{\alpha}$ in Eq.~\eqref{eq:block-diag2} unchanged, and other times transform it to another sector. The symmetries that leave $H_{\bk}^{\alpha}$ unchanged lead to an effective internal symmetry class for each irreducible representation on $p$-cell, which is referred to as emergent Altland-Zirnbauer symmetry class (EAZ class).}
In the following, we discuss how to know the effects of symmetries in $\calA, \mathcal{P},$ and $\mathcal{J}$.

In our construction of the cell decomposition, the little groups $\calGk$ at any point $\bk$ in a $p$-cell $D^p$ are in common, and therefore the common little group is denoted by $\calG_{D^p}$. In the same way as $\calG_{D^p}$, we define a subset $\mathcal{V}_{D^p}$ of $\mathcal{V}$ by $\mathcal{V}_{D^p} = \{v \in  \mathcal{V}| v\bk = \bk +^\exists\bm{G}\ \text{for}\ \forall \bk \in D^p\}$, where $\mathcal{V}=\calA, \mathcal{P}, \mathcal{J}$. Then, we identify actions of time-reversal like, particle-hole like, and chiral like symmetries on each $H_{\bk}^{\alpha}$ by the Wigner criteria~\cite{Bradley,Shiozaki2018}
\begin{align}
	\label{eq:wigner_C}
	W^{\alpha}_{D^p}(\mathcal{P}) &=\frac{1}{\vert \mathcal{P}_{\bk}/T \vert}\sum_{c \in \calP_{\bk}/T }z^{\bk}_{c, c}\chi_{\bk}^{\alpha}(c^2) \in \{0, \pm 1\},\\
	W^{\alpha}_{D^p}(\calA) &=\frac{1}{\vert \calA_{\bk}/T \vert}\sum_{a \in \calT_{\bk}/T }z^{\bk}_{a, a}\chi_{\bk}^{\alpha}(a^2) \in \{0, \pm 1\},\\
	\label{eq:wigner_G}
	W^{\alpha}_{D^p}(\mathcal{J}) &= \frac{1}{\vert \calG_{\bk}/T \vert} \sum_{g \in \calG_{\bk}/T } \frac{z^{\bk}_{\gamma, \gamma^{-1}g\gamma}}{z^{\bk}_{g, \gamma}} [\chi^{\alpha}_{\bk}(\gamma^{-1} g \gamma)]^{*}\chi^{\alpha}_{\bk}(g)\\
	&\in\{0,1\}\nonumber,
\end{align}
where $\chi_{\bk}^{\alpha}(g) =\mathrm{tr}[u_{\bk}^{\alpha}(g)]$ for $\bk \in D^p$ and $\gamma$ is a chiral like symmetry. Note that, in fact, it is enough for our purpose to consider a point $\bk$ in $D^p$. 
When $W_{D^p}^{\alpha}(\mathcal{V}) = 0$, additional symmetries in $\mathcal{V}_{D^p}$ transform $H^{\alpha}_{\bk}$ into another sector $H^{\beta}_{\bk}$. On the other hand, when $W_{D^p}^{\alpha}(\mathcal{V}) = \pm 1$, $H^{\alpha}_{\bk}$ is invariant under the additional symmetries. Then, the EAZ symmetry class for $H_{\bk}^{\alpha}$ is determined by
\begin{align}
	\label{eq:wigner}
	\mathcal{W}_{D^p}[\alpha]\equiv(W^{\alpha}_{D^p}(\calA) , W^{\alpha}_{D^p}(\mathcal{P}), W^{\alpha}_{D^p}(\mathcal{J})).
\end{align}
Depending on the EAZ symmetry classes, the following zero-dimensional topological invariants are assigned to each sector $H_{\bk}^{\alpha}$~\cite{SI_Luka, Ono-Po-Shiozaki2020} (see Table~\ref{tab:EAZ}):
\begin{align}
	\label{eq:Pf}
	p_{\bk}^{\alpha} &\equiv \frac{1}{i\pi}\mathrm{log}\frac{\mathrm{Pf}[U(H_{\bk}^{\alpha})]}{\mathrm{Pf}[U(H_{\bk}^{\alpha})^{\text{vac}}]}\mod 2,\\
	\label{eq:int}
	N_{\bk}^{\alpha} &\equiv n_{\bk}^{\alpha} - (n_{\bk}^{\alpha})^{\text{vac}}.
\end{align}
To define the above topological invariants, we introduce a reference Hamiltonian $H_{\bk}^{\text{vac}}$ in the same symmetry setting~\cite{Skurativska2019,SI_Shiozaki,Ono-Po-Watanabe2020,SI_Luka,Ono-Po-Shiozaki2020}. 
In Eqs.~\eqref{eq:Pf} and \eqref{eq:int}, $(H_{\bk}^{\alpha})^{\text{vac}}$ denotes the counterpart of $H_{\bk}^{\alpha}$ for $H_{\bk}^{\text{vac}}$, and $U$ is the particle-hole like symmetry for $H_{\bk}^{\alpha}$ satisfying $(UU^*)=+1$ and $U (H_{\bk}^{\alpha})^*=-H_{\bk}^{\alpha} U$. We also define $n_{\bk}^{\alpha}$ and $(n_{\bk}^{\alpha})^{\text{vac}}$ by the number of occupied states in $H_{\bk}^{\alpha}$ and $(H_{\bk}^{\alpha})^{\text{vac}}$. 
Practically, we can always choose an appropriate reference $H_{\bk}^{\text{vac}}$ using $H_{\bk}$. For example, since the vacuum is always topologically trivial, $H_{\bk}$ in the limit of infinite chemical potential is often used as $H_{\bk}^{\text{vac}}$~\cite{Skurativska2019,Ono-Po-Shiozaki2020}.
In fact, we will adopt this definition of a reference Hamiltonian in Sec.~\ref{sec6}. 

\subsection{$E_1$-pages}
\label{sec3:E1}
As seen in the preceding discussions, the Wigner criteria in Eqs.~\eqref{eq:wigner_C}-\eqref{eq:wigner_G} tell us EAZ classes for each irreducible representation. Then, let us define abelian groups $E_{1}^{p,0}$ in the following, which can be interpreted as the classification of $\{H_{\bk}^{\alpha}\}_{\alpha}$ at points $\bk$ inside $p$-cells.%for the AZ class $n$ (see Table~\ref{tab:AZ} for the meaning of $n$). 

The total set $\calC_p$ of $p$-cells consists of $N_p$ subsets (so-called ``star'' in the literature~\cite{Bradley}) defined by $S_{D^{p}_{i}} =\{D_{g(i)}^{p}=gD^{p}_{i}\vert g \in G\}$, where $N_p$ the number of subsets and $D^{p}_{i}$ is a representative $p$-cell of the subset $S_{D^{p}_{i}}$. The representatives form a set of independent $p$-cells
\begin{align}
	F^p \equiv \{D^{p}_{i}\}_{i=1}^{N_p}.%,
\end{align}
%which is a set of independent $p$-cells.

\add{In Ref.~\onlinecite{Shiozaki2018}, the abelian groups $E_{1}^{p,0}$ (called $E_1$-pages) are defined by the direct sum of twisted equivalent $K$-groups~\cite{Freed2013} on $p$-cells in $F^p $. It turns out that $E_{1}^{p,0}$ is the direct sum of the classification of zero-dimensional topological phases of $\{H_{\bk}^{\alpha}\}_{\alpha}$ (defined in Eq.~\eqref{eq:block-diag2}) at a point $\bk$ in each $D^{p}_{i} \in F^p$. Then, $E_{1}^{p,0}$ is completely determined by $\mathcal{W}_{D^{p}_{i}}[\alpha]$ for each irreducible representation and each $D^{p}_{i} \in F^p$. In other words, $E_{1}^{p,0}$ is defined by}
\begin{align}
	\label{eq:E1-def1}
	E_{1}^{p,0} &\equiv \bigoplus_{i\vert_{D^{p}_{i}\in F^p}}\left(\mZ_{2}^{\oplus_{\alpha}}\oplus \mZ^{\oplus_{\beta}} \right),
\end{align}
%where we perform the summation about labels of irreducible representations $\alpha$ and $\beta$ only when $\mathcal{W}_{D^{p}_{i}}[\alpha]\in\{(0,1,0), (1,1,1)\}$ and $\mathcal{W}_{D^{p}_{i}}[\beta]\in\{(0,0,0), (1,0,0), (-1,0,0)\}$ (see Eq.~\eqref{eq:wigner}). 
\add{where we perform the summation about labels of irreducible representations $\alpha$ and $\beta$ with the following conditions:
\begin{enumerate}
	\setlength{\itemsep}{-2pt}
	\item[(a)] $\mathcal{W}_{D^{p}_{i}}[\alpha]\in\{(0,1,0), (1,1,1)\}$ and $\mathcal{W}_{D^{p}_{i}}[\beta]\in\{(0,0,0), (1,0,0), (-1,0,0)\}$ (see Eq.~\eqref{eq:wigner});
	\item[(b)] When an irreducible representation on $D^{p}_{i}$ is related to other ones by antiunitary and chiral-like symmetries, only one of irreducible representation on $D^{p}_{i}$ is taken into account.
\end{enumerate}
}

\begin{figure*}[t]
	\begin{center}
		\includegraphics[width=1.9\columnwidth]{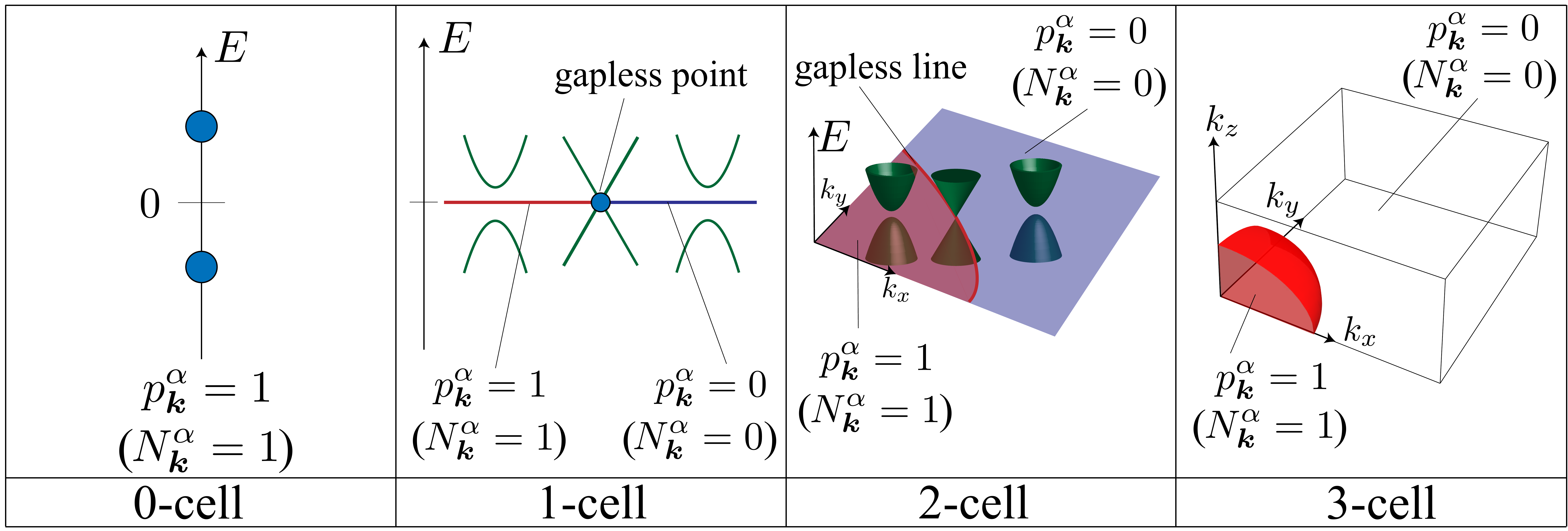}
		\caption{\label{fig:E1}Illustration of elements of $E_{1}^{p,0}$. For $p=0$, the elements are gapped states at $0$-cells. As for $p\ (p \geq 1)$, there are two $p$-dimensional regions in which the zero-dimensional topological invariants are different from each other. Since the zero-dimensional topological invariants must be the same for gapped regions, the boundary results in gapless states on $p$-cells.}
	\end{center}
\end{figure*}

As discussed in Ref.~\onlinecite{Shiozaki2018}, $E_{1}^{p,0}$ for $p\geq 1$ has several different interpretations. 
%$E_{1}^{0,0}$ is the set of zero-dimensional gapped Hamiltonians on 0-cells. 
\add{For $p\geq 1$, $E_{1}^{p,0}$ represents the set of gapless states with $(p-1)$-dimensional gapless regions in the Bogoliubov quasiparticle spectrum on $p$-cells.}
%\change{$E_{1}^{0,0}$ and $E_{1}^{1,0}$}{$E_{1}^{p,0}$} represent the set\change{s of zero-dimensional gapped Hamiltonians on 0-cells and gapless states on 1-cells, respectively}{ of gapless states with $p$-dimensional gapless regions in the Bogoliubov quasiparticle spectrum on $p$-cells}.  
\add{
%These interpretations originate from the bulk-boundary correspondence. $E_{1}^{p,0}$ can be also understood as the direct sum of twisted equivalent $K$-groups on $p$-dimensional spheres $S_{j}^{p}$, where $S_{j}^{p}$ is obtained from the $p$-cell $D^{p}_{j}$ by identifying its boundaries~\cite{Shiozaki2018, Shiozaki-Sato-Gomi2017}. For $p=0$, the twisted $K$-groups on $S_{j}^{p}$ represents 0D gapped topological states. As shown in Ref.~\cite{Shiozaki-Sato-Gomi2017}, due to the bulk-boundary correspondence, the twisted $K$-groups on $S_{j}^{p}$ for $p=1$ correspond to gapless states on the $1$-cells.
%More 
Intuitively, it can also be understood as changes of zero-dimensional topological invariants on $p$-cells. 
Let us focus on a $1$-cell. Then, we define the same zero-dimensional topological invariants for any point on the $1$-cell, as explained in Sec.~\ref{sec3:EAZ}. However, it is not necessary to have the same values of them at all points in the $1$-cell. When we consider momentum as parameters of the deformation, the system must have gapless points on the 1-cell if the zero-dimensional topological invariants at points on the line are different [See Fig.~\ref{fig:E1}]. 
Possible changes of zero-dimensional topological invariants on the 1-cell are equivalent to the classifications of zero-dimensional topological phases of $\{H_{\bk}^{\alpha}\}_{\alpha}$ at a point $\bk$ on the 1-cell, which is the first interpretation of $E_{1}^{1,0}$.
%Therefore, $E_{1}^{1,0}$ can also be understood as the set of gapless states on 1-cells. 
In the same way as $E_{1}^{1,0}$, $E_{1}^{2,0}$ and $E_{1}^{3,0}$ can be understood as the sets of gapless lines and surfaces on $2$- and $3$-cells, respectively [See Fig.~\ref{fig:E1}]. Note that gapless points and lines for $E_{1}^{1,0}$ and $E_{1}^{2,0}$ are not always the genuine point and line nodes. In other words, they are often part of higher-dimensional nodes.
}

Based on these interpretations, we can characterize any system by a list of band labels
\begin{align}
	\mathfrak{n}^{(p)}  &=(\frakp_{D^{p}_{1}}^{\alpha_1}, \frakp_{D^{p}_{1}}^{\alpha_2}, \cdots \frakcn_{D^{p}_{1}}^{\beta_1}, \cdots, \frakp_{D^{p}_{2}}^{\alpha'_1}, \cdots,\frakcn_{D^{p}_{2}}^{\beta'_1}, \cdots),
\end{align}
where $\frakp_{D_{i}^{p}}^{\alpha}$ and $\frakcn_{D_{i}^{p}}^{\beta}$ are $\mZ_2$-valued and $\mZ$-valued band labels, respectively.
\add{While band labels for $p=0$ are no more than the zero-dimensional topological invariants in Eqs.~\eqref{eq:Pf} and \eqref{eq:int}, those for $p$-cells $(p \geq 1)$ represent changes of the zero-dimensional topological invariants.}
Correspondingly, the abelian group $E_{1}^{p,0}$ is formulated by 
\begin{align}
	\label{eq:E1-topo}
	E^{p, 0}_{1} =\bigoplus_{i\vert_{D^{p}_{i}\in F^p}}\left(\bigoplus_{\alpha}\mZ_{2}[\frakb^{(p)}_{D^{p}_{i},\alpha}] \oplus \bigoplus_{\beta} \mZ[\frakb^{(p)}_{D^{p}_{i},\beta}]  \right),
\end{align}
where $\{\frakb^{(p)}_{D^{p}_{i},\alpha}\}$ denotes the set of \add{generators of $E^{p, 0}_{1}$} which can expand an arbitrary $\frakn^{(p)}$, and the summation about $\alpha$ and $\beta$ are the same in Eq.~\eqref{eq:E1-def1}. \add{In addition, $\mZ_{2}[\frakb^{(p)}_{D^{p}_{i},\alpha}]$ and $\mZ[\frakb^{(p)}_{D^{p}_{i},\beta}]$ represent abelian groups generated by $\frakb^{(p)}_{D^{p}_{i},\alpha}$ and $\frakb^{(p)}_{D^{p}_{i},\beta}$}

In this work, we construct the generator $\frakb^{(p)}_{D^{p}_{i},\alpha}$ as follows. Each $\frakb^{(p)}_{D^{p}_{i},\alpha}$ is generated by an irreducible representation $U_{D^{p}_{i}}^{\alpha}$ at a $p$-cell $D_{i}^p$ in $\calC_p$. As explained in Sec.~\ref{sec3:cell}, we include the equivalence or symmetry relations among $p$-cells in the basis. We consider a $p$-cell $D^{p}_{i}$, and suppose that we have a nontrivial band label $\frakp_{D^{p}_{i}}^{\alpha} = 1$ or $\frakcn_{D^{p}_{i}}^{\alpha} = 1$ for an irreducible representation $U_{D^{p}_{i}}^{\alpha}$. Then, band labels on equivalent or symmetry-related $p$-cells are determined by those on $D^{p}_{i}$. We first derive the relation between irreducible representations $U_{D^{p}_{g(i)}}^{\alpha'}$ and $U_{D^{p}_{i}}^{\alpha}$%. Irreducible representations at $g\bk$ are related to those at $\bk$ as
\begin{align}
	U_{D^{p}_{g(i)}}^{\alpha'}(h') &= \begin{cases}
		\frac{z_{h', g}}{z_{g, g^{-1} h' g}}U_{D^{p}_{i}}^{\alpha}(g^{-1}h'g)\quad\text{for }\phi_g = +1 \\
		\frac{z_{h', g}}{z_{g, g^{-1} h' g}}[U_{D^{p}_{i}}^{\alpha}(g^{-1}h'g)]^{*}\quad\text{for }\phi_g = -1
	\end{cases},
\end{align}
where $g\in G$ and $h' \in \calG_{D^{p}_{g(i)}}$. Since the spectrum of $H_{D^{p}_{g(i)}}^{\alpha'}$ is the same as that of $H_{D^{p}_{i}}^{\alpha}$, band labels at $D^{p}_{g(i)}$ then straightforwardly follow
\begin{align}
	\label{eq:pf_rel}
	\frakp_{D^{p}_{g(i)}}^{\alpha'} &= \frakp_{D^{p}_{i}}^{\alpha}, \\
	\frakcn_{D^{p}_{g(i)}}^{\alpha'}&= \begin{cases}
		\frakcn_{D^{p}_{i}}^{\alpha}\quad\text{for }c_g = +1 \\
		-\frakcn_{D^{p}_{i}}^{\alpha}\quad\text{for }c_g = -1
	\end{cases}.
\end{align}
As a result, we can obtain  the set of band labels such that only $\frakp_{D^{p}_{i}}^{\alpha}$ $(\frakcn_{D^{p}_{i}}^{\alpha})$ and associated band labels are $1$ ($1$ or $-1$ \add{for EAZ class A and AI; $2$ or $-2$ for EAZ class AII}). Indeed, this is exactly what we call $\frakb^{(p)}_{D^{p}_{i},\alpha}$.

\begin{figure}[t]
	\begin{center}
		\includegraphics[width=1.0\columnwidth]{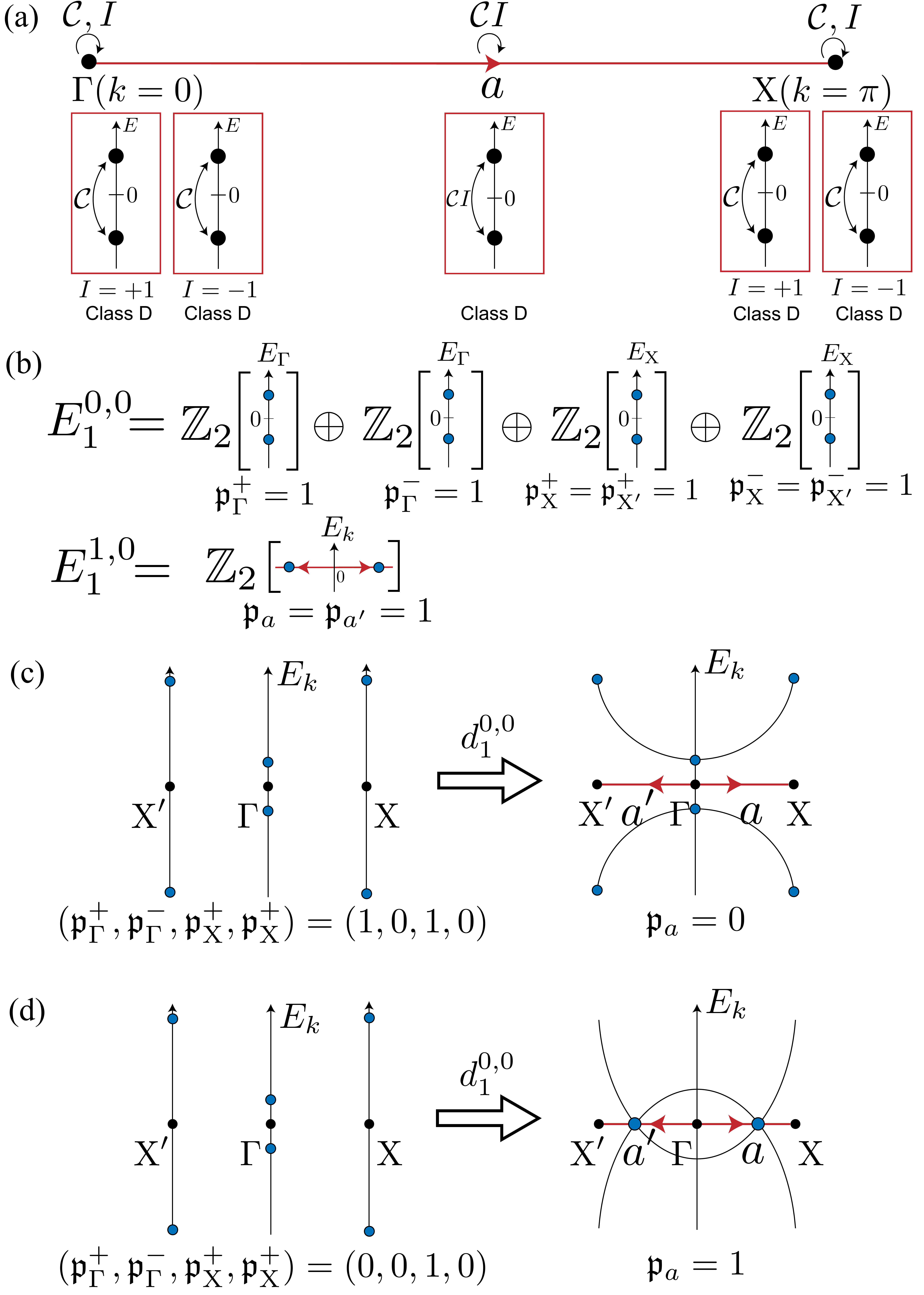}
		\caption{\label{fig:1Dex}Illustration of the 1D even-parity superconductors. (a) An asymmetric unit of BZ and EAZ classes for cells in $F^0=\{\Gamma,\text{X}\}$ and $F^1=\{a\}$. Here, the red arrows signify orientations of the 1-cell. (b) Illustrative description of $E_{1}^{0,0}$ and $E_{1}^{1,0}$. The entries in brackets represent the band structures of generators. (c,d) The physical process of $d_{1}^{0,0}$.
		For the system with $(\frakp_{\Gamma}^{+},\frakp_{\Gamma}^{-},\frakp_{\text{X}}^{+},\frakp_{\text{X}}^{-}) = (1,0,1,0)$, $d_{1}^{0,0}$ does not generate a gapless point on the 1-cells [(c)]. On the other hand, for the system with $(\frakp_{\Gamma}^{+},\frakp_{\Gamma}^{-},\frakp_{\text{X}}^{+},\frakp_{\text{X}}^{-}) = (0,0,1,0)$, $d_{1}^{0,0}$ there exist a gapless point on each 1-cell [(d)]. In the figure, we omit $\frakp_{\text{X}'}^{+},\frakp_{\text{X}'}^{-},$ and $\frakp_{a'}$ since $\frakp_{\text{X}'}^{\pm}=\frakp_{\text{X}}^{\pm}$ and $\frakp_{a'}=\frakp_a$
	}
	\end{center}
\end{figure}

To make our understanding clearer, let us discuss a simple example: a one-dimensional even-parity superconductor in class D. We first decompose an asymmetric unit into two 0-cells $\Gamma$, X and a 1-cell $a$ as illustrated in Fig.~\ref{fig:1Dex} (a). By acting the inversion symmetry $I$ on the unit, we find the cell decomposition:
\begin{align}
	\calC_0 &\equiv \{\Gamma, \text{X}, \text{X}'=I\text{X}\},\\
	\calC_1 &\equiv \{a, a'=Ia\},
\end{align}
where $F^{0}=\{\Gamma,\text{X}\}$ and $F^1=\{a\}$.
We then obtain the classifications of each irreducible representation at $0$-cells and $1$-cells.
Figure~\ref{fig:1Dex} (a) illustrates the action of the particle-hole like symmetries on each sector of Hamiltonians at each cell in $F^p$, and we find that the EAZ classes for each inversion eigenvalue at $\Gamma$ and X are class D and the EAZ class at $a$ is also class D. Therefore, $E_{1}^{0,0}=(\mZ_2)^4$ and $E_{1}^{1,0}=\mZ_2$.

Next, we formulate $E_1$-pages in the form of Eq.~\eqref{eq:E1-topo}.
We define the Pfaffian invariants $\frakp_{D^{0}\in \calC_0}^{\alpha=\pm}$~\cite{Ryu_2010} for each inversion eigenvalue at the 0-cells, and they form the set of band labels $(\frakp_{\Gamma}^{+},\frakp_{\Gamma}^{-},\frakp_{\text{X}}^{+},\frakp_{\text{X}}^{-},\frakp_{\text{X}'}^{+},\frakp_{\text{X}'}^{-})$.
On the other hand, since the 1-cells are invariant under the combination of PHS $\calC$ and the inversion symmetry $I$ with $(\calC I)^2=+1$, the Pfaffian invariant can also be defined on the $1$-cells $a$ and $a'$. Correspondingly, the set of band labels for the 1-cells is $(\frakp_{a},\frakp_{a'})$. We then construct the basis vectors of $E_{1}^{0,0}$ and $E_{1}^{1,0}$. From Eq.~\eqref{eq:pf_rel}, we find $\frakp_{\text{X}'}^{\pm} = \frakp_{\text{X}}^{\pm}$ and $\frakp_{a'} = \frakp_{a}$. Therefore, we obtain
\begin{align}
	\label{eq:basis1}
	\frakb_{\Gamma,+}^{(0)} &= (1,0,0,0,0,0);\\
	\frakb_{\Gamma,-}^{(0)} &= (0,1,0,0,0,0);\\
	\frakb_{\text{X},+}^{(0)} &= (0,0,1,0,1,0);\\
	\frakb_{\text{X},-}^{(0)} &= (0,0,0,1,0,1);\\
	\label{eq:basis5}
	\frakb_{a}^{(1)} &= (1,1),
\end{align}
and they generate $E_{1}^{0,0}$ and $E_{1}^{1,0}$ as 
\begin{align}
	E_{1}^{0,0}&=\mZ_2[\frakb_{\Gamma,+}^{(0)} ]\oplus\mZ_2[\frakb_{\Gamma,-}^{(0)} ]\oplus\mZ_2[\frakb_{\text{X},+}^{(0)} ]\oplus\mZ_2[\frakb_{\text{X},-}^{(0)} ],\\
	E_{1}^{1,0}&=\mZ_2[\frakb_{a}^{(1)}],
\end{align}
which are illustrated in Fig.~\ref{fig:1Dex} (b).

\subsection{Compatibility relations}
\label{sec3:CR}
In this subsection, we discuss constraints on the zero-dimensional topological invariants, which are called \textit{compatibility relations} developed in Refs.~\onlinecite{PhysRevX.7.041069,Po2017,TQC,SI_Luka,Ono-Po-Shiozaki2020}. 
Compatibility relations will be utilized in Sec.~\ref{sec5}.

Before moving on to the general discussion, we begin by showing compatibility relations in the 1D even-parity superconductors discussed in Sec.~\ref{sec3:E1}. As shown in Sec.~\ref{sec3:E1}, the Pfaffian invariants are defined for each inversion-eigenvalue sector at $\Gamma$ and X. Note that the sum of the Pfaffian invariants $p_{k}^{+}+p_{k}^{-}\ (k=\Gamma,\text{X})$ is also the Pfaffian invariant defined for total Hamiltonian, not each inversion-eigenvalue sector. Thus, when the system is fully gapped, $p_{k}^{+}+p_{k}^{-}\ (k=\Gamma,\text{X})$ should be the same value as the Pfaffian invariant at any point in 1-cell, i.e., 
\begin{align}
	p_{k \in a}&=p_{\Gamma}^{+}+p_{\Gamma}^{-} = p_{\text{X}}^{+}+p_{\text{X}}^{-}, \\
	p_{k \in a'}&=p_{\Gamma}^{+}+p_{\Gamma}^{-} = p_{\text{X}'}^{+}+p_{\text{X}'}^{-}.
\end{align}
This is what we refer to as compatibility relations. 

Compatibility relations also lead to the relations between band labels on $0$-cells and $1$-cells. 
Since the band label for $1$-cells can be understood as the change of the zero-dimensional topological invariants, the difference of Pfaffian invariants between $\Gamma$ and $\text{X}\ (\text{X}')$ results in $\frakp_{a}\ (\frakp_{a'})$, i.e., 
\begin{align}
	\label{eq:CR_1D}
	\begin{pmatrix}
		\frakp_{a}\\\frakp_{a'}
	\end{pmatrix}&=
	\begin{pmatrix}
		1 & 1 & -1 & -1 & 0 & 0 \\
		1 & 1 & 0 & 0 & -1 & -1 \\
	\end{pmatrix}
	\begin{pmatrix}
		\frakp_{\Gamma}^{+}\\\frakp_{\Gamma}^{-}\\\frakp_{\text{X}}^{+}\\\frakp_{\text{X}}^{-}\\\frakp_{\text{X}'}^{+} \\ \frakp_{\text{X}'}^{-}
	\end{pmatrix}.
\end{align}

Then, we generalize the above discussions. 
Let $D^{(p+1)}$ be a $(p+1)$-cell, and let $D^{p}$ be a boundary $p$-cell of $D^{(p+1)}$. Since $\calG_{D^{(p+1)}}$ is a subgroup of $\calG_{D^{p}}$ or the same as $\calG_{D^{p}}$ in our cell decomposition, an irreducible representation $U_{D^{p}}^{\alpha}$ of $\calG_{D^{p}}$ can always be constructed by irreducible representations on $D^{(p+1)}$ 
\begin{align}
	\label{eq:irrep_decomposition}
	U_{D^{p}}^{\alpha}(g) = \bigoplus_{\beta}c_{D^p,D^{p+1}}^{\alpha\beta}U^{\beta}_{D^{(p+1)}}(g),
\end{align}
where $c_{D^p,D^{p+1}}^{\alpha\beta}$ is a non-negative integer and obtained by the orthogonality of irreducible representations $\sum_{g\in\calG_{D^{(p+1)}}/T}\left(\chi_{D^{(p+1)}}^{\beta}(g)\right)^{*}\chi_{D^{p}}^{\alpha}(g)$. 
When we have the decomposition in Eq.~\eqref{eq:irrep_decomposition}, we know of the number of irreducible representations $U^{\beta}_{D^{p+1}}$ included in $U_{\bk}(g)$ from those at $D^p$ (denoted by $n_{D^{p}}^{\alpha}$). %, denoted by $n_{D^{p}}^{\alpha}$ and $n_{D^{p+1}}^{\beta}$, 
This relation is described by $n_{D^{(p+1)}}^{\beta} = \sum_{\alpha}n_{D^{p}}^{\alpha}c_{D^p,D^{p+1}}^{\alpha\beta}$~\cite{Bradley, Po2017}.

Accordingly, when the system is fully gapped, zero-dimensional topological invariants in Eqs.~\eqref{eq:Pf} and \eqref{eq:int} at $\bk' \in D^{(p+1)}$ are related to those at $\bk \in D^{p}$, which we refer to as compatibility relations. There exist the following four types of compatibility relations~\cite{Ono-Po-Shiozaki2020}:
\begin{align}
	\label{eq:CR-1}
	p_{\bk'}^{\beta} &= \sum_{\alpha}c_{D^p,D^{p+1}}^{\alpha\beta}p_{\bk}^{\alpha}+\sum_{\gamma}c_{D^p,D^{p+1}}^{\gamma\beta}N_{\bk}^{\gamma}, \mod 2\\
	\label{eq:CR-2}
	p_{\bk'}^{\beta} &= 0 \mod 2,\\
	\label{eq:CR-3}
	N_{\bk'}^{\beta} &= \sum_{\alpha}c_{D^p,D^{p+1}}^{\alpha\beta}N_{\bk}^{\alpha},\\
	\label{eq:CR-4}
	N_{\bk'}^{\beta} &= 0,
\end{align}

Using compatibility relations, we construct a map from $E_{1}^{p,0}$ to $E_{1}^{p+1,0}$.
Band labels at all boundary $p$-cells $D^{p}_{i}$ of $D^{p+1}$ contribute to those at $D^{p+1}$. Taking into account the orientations of cells, we have the following relations:
\begin{align}
	\label{eq:gCR-1}
	\frakp_{D^{p+1}}^{\beta} &= \sum_{i}\delta_{D_{i}^p,D^{p+1}}\left[\sum_{\alpha}c_{D^{p}_{i},D^{p+1}}^{\alpha\beta}\frakp_{\bk}^{\alpha}+\sum_{\gamma}c_{D^{p}_{i},D^{p+1}}^{\gamma\beta}\frakcn_{\bk}^{\gamma}\right], \\
	\label{eq:gCR-3}
	\frakcn_{D^{p+1}}^{\beta}&= \sum_{i}\sum_{\alpha}\delta_{D_{i}^p,D^{p+1}}c_{D^{p}_{i},D^{p+1}}^{\alpha\beta}\frakcn_{D^{p}_{i}}^{\alpha},
\end{align}
where $\delta_{D_{i}^p,D^{p+1}}=0$ when $D^{p+1}$ is not adjacent to $D^{p}_{i}$ and $\delta_{D_{i}^p,D^{p+1}}=1\ (-1)$ if $D^{p+1}$ is adjacent to $D^{p}_{i}$ and the orientation of $D^{p}_i$ agrees (disagrees) with that the orientation induced by $(p+1)$-cell $D^{p+1}$. Note that, while all coefficients are non-negative in Eqs.~\eqref{eq:CR-1}-\eqref{eq:CR-4}, some coefficients in Eqs.~\eqref{eq:gCR-1}-\eqref{eq:gCR-3} can be negative. 
By computing the above relations for all $(p+1)$-cells, one can construct a matrix in terms of band labels at $p$-cells. 

\begin{figure*}[t]
	\begin{center}
		\includegraphics[width=1.2\columnwidth]{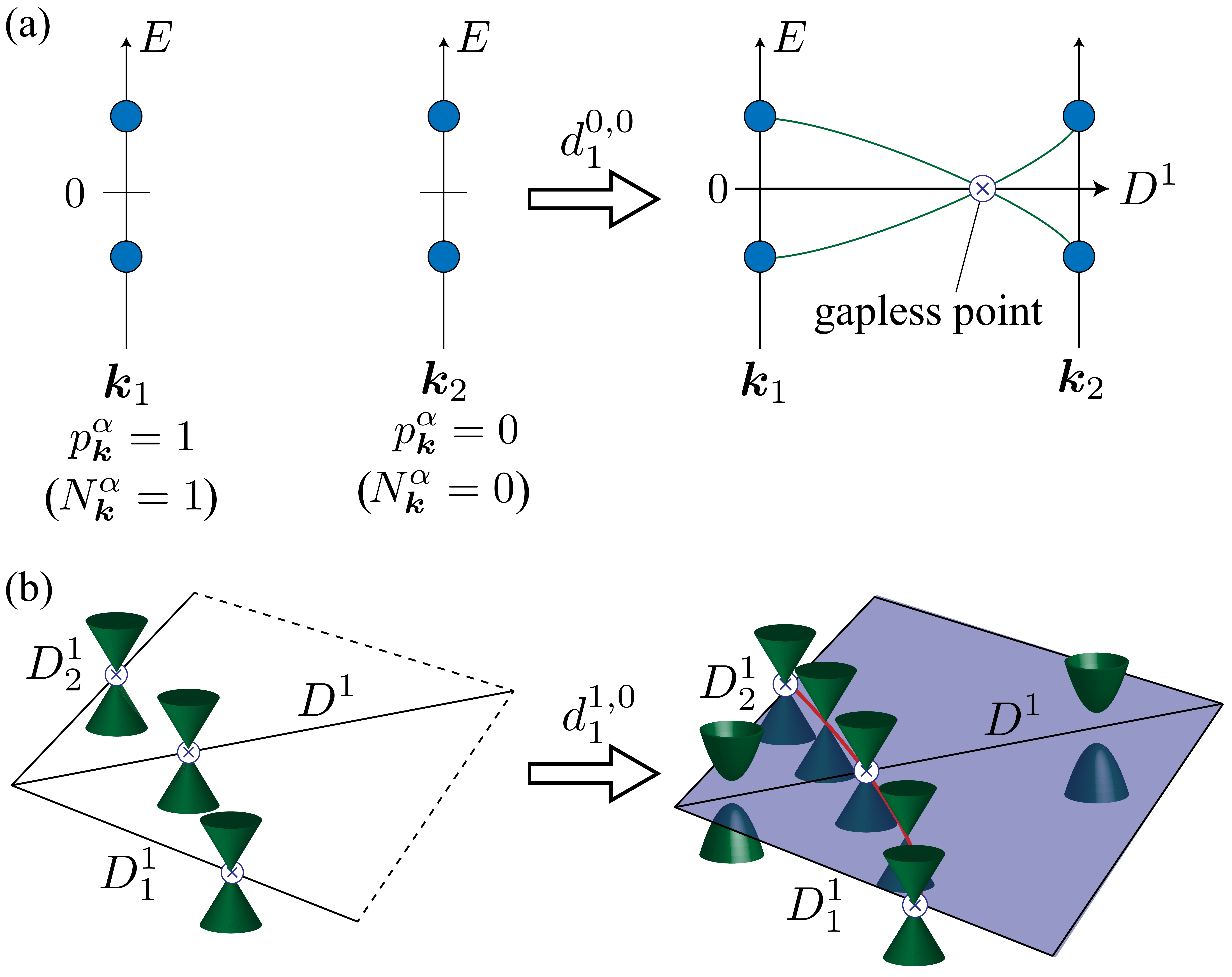}
		\caption{\label{fig:d1}Illustration of the the physical process of $d_{1}^{p,0}$. (a) For a given set of the zero-dimensional topological invariants at $0$-cells, $d_{1}^{0,0}$ determines whether gapless points should exist on the adjacent $1$-cells. In the figure, we focus on two $0$-cells (denoted by $\bk_1$ and $\bk_2$) and the 1-cell connecting $\bk_1$ to $\bk_2$. (b) For gapless points on $1$-cells, $d_{1}^{1,0}$ tells us whether the gapless points should be extended to the adjacent $2$-cells. In the figure, we discuss two $2$-cells adjacent to the 1-cell $D^1$ and illustrate the case where the gapless points are extended.
		}
	\end{center}
\end{figure*}

Rewriting the matrix constructed by Eqs.~\eqref{eq:gCR-1} and \eqref{eq:gCR-3} in terms of basis vectors of $E_{1}^{p,0}$ to $E_{1}^{p+1,0}$, we obtain a map from $E_{1}^{p,0}$ to $E_{1}^{p+1,0}$
\begin{align}
	d_{1}^{p,0}: E_{1}^{p,0} \rightarrow E_{1}^{p+1,0},
\end{align}
which is called \textit{first differential}~\cite{Shiozaki2018}. One can see that $d_{1}^{p,0}$ always satisfies $d_{1}^{p+1,0}\circ d_{1}^{p,0}=0$, that is, $d_{1}^{p+1,0}\left(d_{1}^{p,0}(\frakn^{(p)})\right) = \bm{0}$. 

\add{Physically, nontrivial $d_{1}^{p,0}$ connects states on $p$-cells to those on $(p+1)$-cells, as illustrated in Fig.~\ref{fig:d1}. Since $E_{1}^{p,0}$ has only local information about $p$-cells, the global structures are not known. Then, $d_{1}^{p,0}$ determines the relation between $p$ and $(p+1)$-cells.
For $p=0$, the nontrivial $d_{1}^{0,0}$ tells us whether gapped states at $0$-cells can be connected without closing the gap on $1$-cells. In other words, if $d_{1}^{0,0}(\frakn^{(0)})=\bm{0}$ holds, all zero-dimensional topological invariants at 0-cells satisfies all compatibility relations.
On the other hand, when $d_{1}^{0,0}(\frakn^{(0)})\neq\bm{0}$, some compatibility relations are violated, which implies that gapless points exist on the 1-cells. 
As for $p \geq 1$, nontrivial $d_{1}^{p,0}$ connects the gapless states on $p$-cells to those on $(p+1)$-cells. More concretely, nontrivial $d_{1}^{1,0}$ check if the gapless point on a $1$-cell, an element of $E_{1}^{1,0}$, is extended to the adjacent $2$-cells, which result in gapless lines on the $2$-cells, an element of $E_{1}^{2,0}$. In the same way, $d_{1}^{2,0}$ examines whether a gapless line on a $2$-cell, an element of $E_{1}^{2,0}$, is linked to gapless surfaces on the $3$-cells. 
}
In Secs.~\ref{sec5} A and B, we will explain more details on interpretation of $d_{1}^{p,0}$ and how to incorporate these first differentials into classifications of nodes.

Let us discuss $d_{1}^{0,0}$ for the above 1D example.  Using the bases in Eqs.~\eqref{eq:basis1}-\eqref{eq:basis5}, we rewrite the matrix in Eq.~\eqref{eq:CR_1D} by
\begin{align}
	\label{eq:d1_ex}
	M_{d_{1}^{0,0}}&=
	\begin{array}{c|cccc}
		& \frakb_{\Gamma,+}^{(0)} & \frakb_{\Gamma,-}^{(0)} & \frakb_{\text{X},+}^{(0)} & \frakb_{\text{X},-}^{(0)}\\
		\hline
		\frakb_{a}^{(1)} & 1 & 1 & -1 & -1 \\
	\end{array},
\end{align}
which is actually a matrix representation of $d_{1}^{0,0}$.
To see the physical meaning of $d_{1}^{0,0}$, let us discuss the band structures in Fig.~\ref{fig:1Dex} (c) and (d). We start with the band structure that corresponds to $\frakn^{(0)} = \frakb_{\Gamma,+}^{(0)} + \frakb_{\text{X},+}^{(0)}$ in Eqs.~\eqref{eq:basis1}-\eqref{eq:basis5}, which implies that $d_{1}^{0,0}(\frakn^{(0)})=0$. Thus, there are no gapless points on the 1-cells [Fig.~\ref{fig:1Dex} (c)].
On the other hand, let us suppose that a band-inversion at $\Gamma$ occurs and results in $\frakn'^{(0)} =  \frakb_{\text{X},+}^{(0)}$. From Eq.~\eqref{eq:d1_ex}, we find $d_{1}^{0,0}(\frakn'^{(0)}) = \frakb_{a}^{(1)}$. As shown in Fig.~\ref{fig:1Dex} (d), gapless points must exist on $1$-cells. This is what we have mentioned above.

\section{Classification of gapless points on 1-cell}
\label{sec4}
In this section, we discuss the method to classify locally stable point nodes on $1$-cells. The Hamiltonian near a gapless point on a $1$-cell is described by 
\begin{align}
	\label{eq:cc-ham}
	H_{(k_1, k_2)} &= k_1 \gamma_1 + k_2\gamma_2 + \delta k_3 \gamma_0,
\end{align}
where $k_1$ and $k_2$ are momenta in the directions perpendicular to the $1$-cell, and $\delta k_3$ is a displacement from the gapless point in the direction of $D^1$. Gamma matrices $\gamma_{0}, \gamma_1$, and $\gamma_2$ anticommute with each others. Then, the classification of the gapless points on 1-cells of 3D systems is equivalent to that of the above Dirac Hamiltonian.

Ref.~\onlinecite{Cornfeld-Chapman} has shown that one can redefine any point group symmetries as onsite symmetries with classifications of \add{massive Dirac Hamiltonians} unchanged, which we will refer to as Cornfeld-Chapman's method. Refs.~\onlinecite{Cornfeld-Chapman, Shiozaki-CC} also have classified 3D \add{massive Dirac Hamiltonians} in the presence of nonmagnetic and magnetic point group symmetries by using the method. 

\add{In the following, applying Cornfeld-Chapman's method~\cite{Cornfeld-Chapman} to classifications of 2D massive Dirac Hamiltonian on 1-cells, 
we will reveal that the results are classified into three cases: (i) The gapless point on the 1-cell is a genuine point node. (ii) The gapless point on the 1-cell is a shrunk loop or surface node. (iii) There are no stable point nodes and such shrunk nodes on the 1-cell.
}
This will be integrated into compatibility relations discussed in Sec.~\ref{sec5}.

\subsection{Cornfeld-Chapman's method for 2D systems}
\label{sec4:CC}
Suppose that there exists a gapless point on a $1$-cell (denoted by $D^1$).
Let us discuss the massive Dirac Hamiltonian in Eq.~\eqref{eq:cc-ham} near $D^1$.
To apply the Cornfeld-Chapman's method to the massive Dirac Hamiltonian, we consider the little co-group in the following discussion, and then Hamiltonian is symmetric under $G_{D^1}/T$, i.e., $H_{(k_1, k_2)}$ satisfies
\begin{align}
	\sigma_{\bk}(g)H_{(k_1, k_2)}&= \begin{cases}
		H_{r_g(k_1, k_2)}\sigma_{\bk}(g) \quad \text{for }c_g = +1,\\
		-H_{r_g(k_1, k_2)}\sigma_{\bk}(g) \quad \text{for }c_g = -1, \end{cases}
\end{align}
where $r_g$ is an element of $\text{O}(2)$. Generally, $r_g$ can be written by
\begin{align}
	r_g &= \begin{cases}
		\begin{pmatrix}
			\cos \theta_g & -\sin\theta_g\\
			\sin\theta_g & \cos \theta_g \\
		\end{pmatrix}
		\quad \text{for } \det r_g =+1,\\
		\begin{pmatrix}
			-\cos \theta_g & -\sin\theta_g\\
			-\sin\theta_g & \cos \theta_g \\
		\end{pmatrix}\quad \text{for }\det r_g = -1.
	\end{cases}
\end{align}
For simplicity, we thereafter use $s_g = \det r_g$. In the following, we will make all elements of $G_{D^1}/T$ onsite. 

First, we introduce onsite symmetries and define their representations by 
\begin{align}
	\label{eq:onsite-symm}
	\widetilde{\sigma}(g) \equiv \gamma_1^{\frac{1-s_g}{2}}e^{\frac{\theta_g}{2}\gamma_1\gamma_2}\sigma_{\bk}(g) \quad\text{for }\forall g\in G_{D^1}/T.
\end{align}
By performing explicit calculations, one can verify
\begin{align}
	\widetilde{\sigma}(g)H_{(k_1, k_2)} &= s_g c_g H_{(k_1, k_2)} \widetilde{\sigma}(g),\\
	\widetilde{\sigma}(g)\widetilde{\sigma}(h) &= (s_gc_g)^{\frac{1-s_h}{2}}z'_{g,h}z^{\bk}_{g,h}\widetilde{\sigma}(gh),
\end{align}
where $z'_{g,h}$ is determined by 
\begin{align}
	\gamma_1^{\frac{1-s_h}{2}}e^{\frac{\theta_h}{2}\gamma_1\gamma_2}\gamma_1^{\frac{1-s_g}{2}}e^{\frac{\theta_g}{2}\gamma_1\gamma_2} &= z'_{g,h}\gamma_1^{\frac{1-s_{gh}}{2}}e^{\frac{\theta_{gh}}{2}\gamma_1\gamma_2}.
\end{align}
Note that, when $\sigma_{\bk}(g)$ with $s_g=-1$ commutes (anticommutes) with $H_{(k_1, k_2)}$, $\widetilde{\sigma}(g)$ anticommutes (commutes) with $H_{(k_1, k_2)}$. In other words, unitary (chiral like) symmetries for $s_g =-1$ become onsite chiral (unitary) symmetries. The same thing happens to antiunitary symmetries. 
As a result, we have another decomposition of symmetry group $G_{D^1}/T=\tiG + \widetilde{\calA} + \widetilde{\calP}  + \widetilde{\calJ}$, where each subset is defined by
\begin{align}
	\tiG &= \{g \in G_{D^1}/T | s_gc_g=1, \phi_g=1\}, \\
	\widetilde{\calA} &= \{g \in G_{D^1}/T | s_gc_g=1,\phi_g=-1\}, \\
	\widetilde{\calP} &= \{g \in G_{D^1}/T | s_gc_g=-1,\phi_g=-1\}, \\
	\widetilde{\calJ} &= \{g \in G_{D^1}/T | s_gc_g=-1, \phi_g=1\}
\end{align}

\add{It is well known that the 2D Dirac Hamiltonians in the presence of onsite symmetries are classified by the second homotopy group of the classifying space~\cite{Teo-Kane2010}. Then, our next task is to identify the classifying space. Similar to Eqs.~\eqref{eq:block-diag} and \eqref{eq:block-diag2}, we can block-diagonalize $\widetilde{\sigma}(g)\ (g\in \tiG )$ and $H_{(k_1, k_2)}$ such that
\begin{align}
	\label{eq:CCblock-diag}
	&\widetilde{\sigma}(g) =\text{diag}\left[\tilde{u}^{\widetilde{\alpha}_1}(g)\otimes\mathds{1}_{m_1}, \cdots, \tilde{u}^{\widetilde{\alpha}_n}(g)\otimes\mathds{1}_{m_n}\right],\\
	\label{eq:CCCblock-diag2}
	&H_{(k_1, k_2)}=\text{diag}\left[\mathds{1}_{d^{\widetilde{\alpha}_1}}\otimes h^{\widetilde{\alpha}_1}, \cdots, \mathds{1}_{d^{\widetilde{\alpha}_n}}\otimes h^{\widetilde{\alpha}_n}\right],
\end{align}
where $\tilde{u}^{\widetilde{\alpha}}(g)$ is an irreducible representations of $\widetilde{\calG}$. Here, $d^{\widetilde{\alpha}}$ and $m_{\widetilde{\alpha}}$ are dimensions of $\tilde{u}^{\widetilde{\alpha}}(g)$ and $h^{\widetilde{\alpha}}_{\bk}$, respectively.
	
For each sector $h^{\widetilde{\alpha}}$, we again use Wigner criteria by replacing $z_{g,h}^{\bk}$ in Eqs.~\eqref{eq:wigner_C}-\eqref{eq:wigner_G} with $(s_gc_g)^{\frac{1-s_h}{2}}z'_{g,h}z^{\bk}_{g,h}$, i.e.,
	\begin{align}
		\label{eq:CCwigner_C}
		&\widetilde{W}^{\widetilde{\alpha}}(\widetilde{\mathcal{P}}) =\frac{1}{\vert \widetilde{\calP} \vert}\sum_{c \in \widetilde{\calP}}(s_cc_c)^{\frac{1-s_c}{2}}z'_{c,c}z_{c,c}^{\bk}\widetilde{\chi}^{\widetilde{\alpha}}(c^2) \in \{0, \pm 1\},\\
		&\widetilde{W}^{\widetilde{\alpha}}(\widetilde{\mathcal{A}}) =\frac{1}{\vert \widetilde{\calA} \vert}\sum_{c \in \widetilde{\calA}}(s_ac_a)^{\frac{1-s_a}{2}}z'_{a,a}z_{a,a}^{\bk}\widetilde{\chi}^{\widetilde{\alpha}}(a^2) \in \{0, \pm 1\},\\
		%W^{\widetilde{\alpha}}_{D^p}(\calA) &=\frac{1}{\vert \calA_{\bk}/T \vert}\sum_{a \in \calT_{\bk}/T }z^{\bk}_{a, a}\widetilde{\chi}_{\bk}^{\widetilde{\alpha}}(a^2) \in \{0, \pm 1\},\\
		\label{eq:CCwigner_G}
		&\widetilde{W}^{\widetilde{\alpha}}(\widetilde{\mathcal{J}}))= \frac{1}{\vert \widetilde{\calG} \vert} \sum_{g \in \widetilde{\calG}} \frac{(s_\gamma c_\gamma)^{\frac{1-s_{\gamma^{-1}g\gamma}}{2}}z'_{\gamma,\gamma^{-1}g\gamma}z^{\bk}_{\gamma, \gamma^{-1}g\gamma}}{(s_gc_g)^{\frac{1-s_\gamma}{2}}z'_{g,\gamma}z^{\bk}_{g, \gamma}} \nonumber \\
		&\quad\quad\quad\quad\quad\quad\quad\quad \times[\widetilde{\chi}^{\widetilde{\alpha}}(\gamma^{-1} g \gamma)]^{*}\widetilde{\chi}^{\widetilde{\alpha}}(g)\in\{0,1\},
	\end{align}
where $\widetilde{\chi}^{\widetilde{\alpha}}(g) = \mathrm{tr}[\widetilde{u}^{\widetilde{\alpha}}(g)]$ and $\gamma$ is an element of $\widetilde{\calJ}$. 
Correspondences between results of Wigner criteria and classifying spaces C$_s$ and R$_s$ are summarized in Table~\ref{tab:CC-EAZ}. As a result, we classify the Dirac Hamiltonian in Eq.~\eqref{eq:cc-ham} by $\pi_{2}(\text{C}_s)$ or $\pi_{2}(\text{R}_s)$ for each irreducible representation $\tilde{u}^{\widetilde{\alpha}}$ ~\cite{Teo-Kane2010}. 
}

\begin{table}
	\begin{center}
		\caption{\label{tab:CC-EAZ}Classification of EAZ symmetry classes. The subscripts $\calT,\calC$, and $\Gamma$ signify that irreducible representations are related by the onsite antiunitary and the chiral symmetries.}
		\begin{tabular}{c|c|c|c}
			\hline
			EAZ & $\widetilde{\mathcal{W}}[\widetilde{\alpha}]$ & classifying space & $\pi_2$\\
			\hline\hline
			A, A$_\calT$, A$_\calC$, A$_\Gamma$, A$_{\calT,\calC}$ & $(0,0,0)$ & C$_0$ & $\mZ$ \\
			AIII, AIII$_\calT$ & $(0,0,1)$ & C$_1$ & $0$ \\
			\hline
			AI, AI$_\calC$ & $(1,0,0)$ & R$_0$ & $\mZ_2$  \\
			BDI & $(1,1,1)$& R$_1$  & $0$ \\
			D, D$_\calT$ & $(0,1,0)$ & R$_2$ & $2\mZ$ \\
			DIII & $(-1,1,1)$& R$_3$  & $0$\\
			AII, AII$_\calC$ & $(-1,0,0)$ & R$_4$ & $0$ \\
			CII & $(-1,-1,1)$& R$_5$  & $0$  \\
			C, C$_\calT$ & $(0,-1,0)$ & R$_6$ & $\mZ$   \\
			CI & $(1,-1,1)$& R$_7$  & $\mZ_2$\\
			\hline
		\end{tabular}
	\end{center}
\end{table}

\subsection{Character decomposition formulas}
\label{sec4:cc-formulas}
As explained in the previous subsection, we can classify the two-dimensional Dirac Hamiltonians in Eq.~\eqref{eq:cc-ham} on 1-cells. The next step is to map the generating two-dimensional Dirac Hamiltonians to elements of  $E_{1}^{1,0}$. This can be achieved by the orthogonality of irreducible representations. In this subsection, we will derive formulas to obtain elements of $E_{1}^{1,0}$ corresponding to generating Dirac Hamiltonians. The formulas are summarized in Table~\ref{tab:formulas}.

Let us suppose that we have one of generating Dirac Hamiltonians on a 1-cell and onsite symmetries in Eq.~\eqref{eq:onsite-symm}. Then, we can construct symmetries of $G_{D^1}/T$ by
\begin{align}
	\label{eq:rep2}
	\sigma_{\bk}(g)&=e^{-\frac{\theta_g}{2}\gamma_1\gamma_2}\gamma_1^{\frac{1-s_g}{2}}\widetilde{\sigma}(g).
\end{align}
What we have to do is to obtain irreducible representations contained in the above representation $\sigma_{\bk}(g)$ in Eq.~\eqref{eq:rep2}, which result in band labels on the 1-cell. Using the orthogonality of irreducible representations, we obtain $\frakcn_{D^1}^{\alpha}$ and $\frakp_{D^1}^{\alpha}$ by
\begin{align}
	\label{eq:formula_Z}
	\frakcn_{D^1}^{\beta}&=\frac{1}{\vert\calG_{\bk}/T\vert}\sum_{g\in \calG_{\bk}/T}\chi^{\beta}_{\bk}(g)\mathrm{tr}[\gamma_0 e^{-\frac{\theta_g}{2}\gamma_1\gamma_2}\gamma_1^{\frac{1-s_g}{2}}\widetilde{\sigma}(g)],\\
	\label{eq:formula_Z2}
	\frakp_{D^1}^{\beta}&=\frac{1/2}{\vert\calG_{\bk}/T\vert}\sum_{g\in \calG_{\bk}/T}\chi^{\beta}_{\bk}(g)\mathrm{tr}[e^{-\frac{\theta_g}{2}\gamma_1\gamma_2}\gamma_1^{\frac{1-s_g}{2}}\widetilde{\sigma}(g)]\mod 2,
\end{align}
where occupied and unoccupied bands contribute to band labels with different signs by $\gamma_0$ in Eq.~\eqref{eq:formula_Z}. After performing the same procedures for all irreducible representations of $\calG_{\bk}$, we get an element of $E_{1}^{1,0}$ corresponding to one of generating Dirac Hamiltonians.

For each of EAZ classes, in fact, we can derive the formulas by fixing the form of generating Hamiltonians and representations, which is summarized in Table~\ref{tab:formulas}. Here, we show the formulas for class A$_\calC$ as an example. % Let us begin with class A$_C$. 
The generating Hamiltonian and representations can be represented by
\begin{align}
	\label{eq:gene_ham_AC}
	H_{(k_1, k_2)} &= k_1 \tau_1 + k_2\tau_2 + \delta k_3 \tau_3,\\
	\widetilde{\sigma}(\calC) &= \begin{cases}
		i \tau_2 \sigma_1K\quad \text{for } [\widetilde{\sigma}(\calC)]^2 = -1,\\
		\tau_2 \sigma_2 K \quad \text{for } [\widetilde{\sigma}(\calC)]^2 = +1,
	\end{cases} \\
	\label{eq:gene_rep_AC}
	\widetilde{\sigma}(g) &= \tau_0 \begin{pmatrix}
		\widetilde{u}^{\widetilde{\alpha}}(g) & \\
		& \widetilde{u}^{\widetilde{\calC}\widetilde{\alpha}}(g)
	\end{pmatrix}\quad \text{for }\forall\widetilde{g} \in \tiG,
\end{align}
where $\widetilde{u}^{\widetilde{\calC}\widetilde{\alpha}}$ denotes the particle-hole-related irreducible representation of $\tilde{u}^{\widetilde{\alpha}}$. In addition, $\widetilde{\calC}$ is the generator of $\widetilde{\calP}$, and $\sigma_{\mu}$ and $\tau_\mu (\mu=0,1,2,3)$ are Pauli matrices representing different degrees of freedom.
By substituting Eqs.~\eqref{eq:gene_ham_AC} and \eqref{eq:gene_rep_AC} into Eqs.~\eqref{eq:formula_Z} and \eqref{eq:formula_Z2}, we get
\begin{align}
	\label{eq:AC-Z}
	\frakcn_{D^1}^{\beta}&=\frac{-2i}{\vert G_{D^1} \vert}\sum_{g \in \calG_{D^1}/T}\delta_{s_g,1}\sin\frac{\theta_g}{2}[\chi_{D^1}^{\beta}(g)]^*\nonumber\\
	&\quad\quad\quad\quad\quad\quad\quad\quad\times\left(\widetilde{\chi}^{\widetilde{\alpha}}(g)+\widetilde{\chi}^{\widetilde{\calC}\widetilde{\alpha}}(g)\right),\\
	\label{eq:AC-Z2}
	\frakp_{D^1}^{\beta}&=\frac{1}{\vert G_{D^1} \vert}\sum_{g \in \calG_{D^1}/T}\delta_{s_g,1}\cos\frac{\theta_g}{2}[\chi_{D^1}^{\beta}(g)]^*\nonumber\\
	&\quad\quad\quad\quad\quad\quad\quad\times\left(\widetilde{\chi}^{\widetilde{\alpha}}(g)+\widetilde{\chi}^{\widetilde{\calC}\widetilde{\alpha}}(g)\right)\ \mathrm{mod}\ 2,
\end{align}
where $\widetilde{\chi}^{\widetilde{\alpha}}(g) = \mathrm{tr}[\widetilde{u}^{\widetilde{\alpha}}(g)]$.

\add{Finally, we find that the results are classified into three cases:
	\begin{enumerate}
		\setlength{\itemsep}{-2pt}
		\item[(i)] One of the generating Dirac Hamiltonians is mapped to a generator of $E_{1}^{1,0}$. In this case, the gapless point on the 1-cell is a genuine point node.
		\item[(ii)] The obtained element of $E_{1}^{1,0}$ for generating Dirac Hamiltonians does not coincide with any generator of $E_{1}^{1,0}$. In other words, the obtained element is composed of multiple bases of $E_{1}^{1,0}$, which implies that the realized point nodes must degenerate. However, these gapless points do not need to be at the same momentum. In such a case, gapless points are actually parts of shrunk loop or surface nodes. 
		\item[(iii)] The classification of Dirac Hamiltonians is trivial, i.e., the second homotopy group discussed in Sec.~\ref{sec4:CC} is trivial, which implies that any point and shrinkable nodes do not exist. Thus, the gapless point is part of line or surface nodes. 
	\end{enumerate}
}
\textcolor{black}{One might sometimes notice that the degeneracy of a point node is different from the dimension of corresponding Dirac Hamiltonians for case (i). In such a case, trivial gapped states exist in the energy spectrum. The existence of Dirac Hamiltonians in Eq.~\eqref{eq:cc-ham} ensures that the point node is stable in the sense of K-theory, i.e., against adding trivial degrees of freedom. It is tempting to think that our results have missed stable nodes in the sense of fragile topological phases~\cite{PhysRevLett.121.126402}, i.e., line or surface nodes when any trivial degree of freedom is not added. 
	However, when we consider quadratic and cubic terms, we can explicitly construct minimal dimension Dirac Hamiltonians. This implies that such fragile nodes do not exist on 1-cells.
	See Appendix~\ref{app:remark} for details.
}

\begin{table*}[t]
	\begin{center}
		\caption{\label{tab:formulas}Formulas to obtain elements of $E_{1}^{1,0}$ corresponding to generating Dirac Hamiltonians. Here, $\chi^{\alpha}$ and $\widetilde{\chi}^{\widetilde{\alpha}}$ are characters of the little co-group $\calG_{D^1}/T$ and the onsite symmetry group $\widetilde{\calG}$, respectively. The first column represent EAZ classes of irreducible representations $\tilde{u}^{\widetilde{\alpha}}$. In addition, $\widetilde{\calT}\widetilde{\alpha}$, $\widetilde{\calC}\widetilde{\alpha}$, and $\widetilde{\Gamma}\widetilde{\alpha}$ are labels of the time-reversal, the particle-hole, and the chiral symmetry related irreducible representations. Derivations of these formulas are included in Appendix~\ref{app:formulas}.}
		\begin{tabular}{c|l|l}
			\hline
			EAZ &  \quad\quad\quad\quad\quad\quad \quad Formula for the map to $\mZ$ &   \quad \quad\quad\quad\quad\quad\quad Formula for the map to $\mZ_2$ \\
			\hline\hline
			A & $\frac{-2i}{\vert G_{D^1} \vert}\sum_{g \in G_{D^1}}\delta_{s_g,1}\sin\frac{\theta_g}{2}[\chi_{D^1}^{\beta}(g)]^*\widetilde{\chi}^{\widetilde{\alpha}}(g)$ & $\frac{1}{\vert G_{D^1} \vert}\sum_{g \in G_{D^1}}\delta_{s_g,1}\cos\frac{\theta_g}{2}[\chi_{D^1}^{\beta}(g)]^*\widetilde{\chi}^{\widetilde{\alpha}}(g)$\\
			\hline 
			A$_\calT$ & $\frac{-2i}{\vert G_{D^1} \vert}\sum_{g \in G_{D^1}}\delta_{s_g,1}\sin\frac{\theta_g}{2}[\chi_{D^1}^{\beta}(g)]^*\left(\widetilde{\chi}^{\widetilde{\alpha}}(g)-\widetilde{\chi}^{\widetilde{\calT}\widetilde{\alpha}}(g)\right)$  & $\frac{1}{\vert G_{D^1} \vert}\sum_{g \in G_{D^1}}\delta_{s_g,1}\cos\frac{\theta_g}{2}[\chi_{D^1}^{\beta}(g)]^*\left(\widetilde{\chi}^{\widetilde{\alpha}}(g)+\widetilde{\chi}^{\widetilde{\calT}\widetilde{\alpha}}(g)\right)$ \\
			\hline
			A$_\calC$ & $\frac{-2i}{\vert G_{D^1} \vert}\sum_{g \in G_{D^1}}\delta_{s_g,1}\sin\frac{\theta_g}{2}[\chi_{D^1}^{\beta}(g)]^*\left(\widetilde{\chi}^{\widetilde{\alpha}}(g)+\widetilde{\chi}^{\widetilde{\calC}\widetilde{\alpha}}(g)\right)$  & $\frac{1}{\vert G_{D^1} \vert}\sum_{g \in G_{D^1}}\delta_{s_g,1}\cos\frac{\theta_g}{2}[\chi_{D^1}^{\beta}(g)]^*\left(\widetilde{\chi}^{\widetilde{\alpha}}(g)+\widetilde{\chi}^{\widetilde{\calC}\widetilde{\alpha}}(g)\right)$ \\
			\hline
			A$_\Gamma$ & $\frac{-2i}{\vert G_{D^1} \vert}\sum_{g \in G_{D^1}}\delta_{s_g,1}\sin\frac{\theta_g}{2}[\chi_{D^1}^{\beta}(g)]^*\left(\widetilde{\chi}^{\widetilde{\alpha}}(g)-\widetilde{\chi}^{\widetilde{\Gamma}\widetilde{\alpha}}(g)\right)$  & $\frac{1}{\vert G_{D^1} \vert}\sum_{g \in G_{D^1}}\delta_{s_g,1}\cos\frac{\theta_g}{2}[\chi_{D^1}^{\beta}(g)]^*\left(\widetilde{\chi}^{\widetilde{\alpha}}(g)+\widetilde{\chi}^{\widetilde{\Gamma}\widetilde{\alpha}}(g)\right)$ \\
			\hline
			\multirow{2}{*}{A$_{\calT,\calC}$} & $\frac{-2i}{\vert G_{D^1} \vert}\sum_{g \in G_{D^1}}\delta_{s_g,1}\sin\frac{\theta_g}{2}[\chi_{D^1}^{\beta}(g)]^*$ & $\frac{1}{\vert G_{D^1} \vert}\sum_{g \in G_{D^1}}\delta_{s_g,1}\cos\frac{\theta_g}{2}[\chi_{D^1}^{\beta}(g)]^*$\\
			&\quad\quad\quad\quad\quad$\times\left(\widetilde{\chi}^{\widetilde{\alpha}}(g)-\widetilde{\chi}^{\widetilde{\calT}\widetilde{\alpha}}(g)+\widetilde{\chi}^{\widetilde{\calC}\widetilde{\alpha}}(g)-\widetilde{\chi}^{\widetilde{\Gamma}\widetilde{\alpha}}(g)\right)$ & \quad\quad\quad\quad\quad$\times\left(\widetilde{\chi}^{\widetilde{\alpha}}(g)+\widetilde{\chi}^{\widetilde{\calT}\widetilde{\alpha}}(g)+\widetilde{\chi}^{\widetilde{\calC}\widetilde{\alpha}}(g)+\widetilde{\chi}^{\widetilde{\Gamma}\widetilde{\alpha}}(g)\right)$ \\
			\hline
			C & $\frac{-2i}{\vert G_{D^1} \vert}\sum_{g \in G_{D^1}}\delta_{s_g,1}\sin\frac{\theta_g}{2}[\chi_{D^1}^{\beta}(g)]^*\widetilde{\chi}^{\widetilde{\alpha}}(g)$ & $\frac{1}{\vert G_{D^1} \vert}\sum_{g \in G_{D^1}}\delta_{s_g,1}\cos\frac{\theta_g}{2}[\chi_{D^1}^{\beta}(g)]^*\widetilde{\chi}^{\widetilde{\alpha}}(g)$\\
			\hline 
			C$_\calT$ & $\frac{-2i}{\vert G_{D^1} \vert}\sum_{g \in G_{D^1}}\delta_{s_g,1}\sin\frac{\theta_g}{2}[\chi_{D^1}^{\beta}(g)]^*\left(\widetilde{\chi}^{\widetilde{\alpha}}(g)-\widetilde{\chi}^{\widetilde{\calT}\widetilde{\alpha}}(g)\right)$  & $\frac{1}{\vert G_{D^1} \vert}\sum_{g \in G_{D^1}}\delta_{s_g,1}\cos\frac{\theta_g}{2}[\chi_{D^1}^{\beta}(g)]^*\left(\widetilde{\chi}^{\widetilde{\alpha}}(g)+\widetilde{\chi}^{\widetilde{\calT}\widetilde{\alpha}}(g)\right)$ \\
			\hline
			D & $\frac{-4i}{\vert G_{D^1} \vert}\sum_{g \in G_{D^1}}\delta_{s_g,1}\sin\frac{\theta_g}{2}[\chi_{D^1}^{\beta}(g)]^*\widetilde{\chi}^{\widetilde{\alpha}}(g)$ & $\frac{2}{\vert G_{D^1} \vert}\sum_{g \in G_{D^1}}\delta_{s_g,1}\cos\frac{\theta_g}{2}[\chi_{D^1}^{\beta}(g)]^*\widetilde{\chi}^{\widetilde{\alpha}}(g)$\\
			\hline 
			D$_\calT$ & $\frac{4i}{\vert G_{D^1} \vert}\sum_{g \in G_{D^1}}\delta_{s_g,1}\sin\frac{\theta_g}{2}[\chi_{D^1}^{\beta}(g)]^*\left(\widetilde{\chi}^{\widetilde{\alpha}}(g)-\widetilde{\chi}^{\widetilde{\calT}\widetilde{\alpha}}(g)\right)$  & $\frac{2}{\vert G_{D^1} \vert}\sum_{g \in G_{D^1}}\delta_{s_g,1}\cos\frac{\theta_g}{2}[\chi_{D^1}^{\beta}(g)]^*\left(\widetilde{\chi}^{\widetilde{\alpha}}(g)+\widetilde{\chi}^{\widetilde{\calT}\widetilde{\alpha}}(g)\right)$ \\
			\hline
			AI &  \quad\quad\quad\quad\quad\quad\quad\quad\quad\quad\quad$-$ & $\frac{2}{\vert G_{D^1} \vert}\sum_{g \in G_{D^1}}\delta_{s_g,1}\cos\frac{\theta_g}{2}[\chi_{D^1}^{\beta}(g)]^*\widetilde{\chi}^{\widetilde{\alpha}}(g)$\\
			\hline
			AI$_\calC$ &  \quad\quad\quad\quad\quad\quad\quad\quad\quad\quad\quad$-$ & $\frac{2}{\vert G_{D^1} \vert}\sum_{g \in G_{D^1}}\delta_{s_g,1}\cos\frac{\theta_g}{2}[\chi_{D^1}^{\beta}(g)]^*\left(\widetilde{\chi}^{\widetilde{\alpha}}(g)+\widetilde{\chi}^{\widetilde{\calC}\widetilde{\alpha}}(g)\right)$\\
			\hline
			CI & \quad\quad\quad\quad\quad\quad\quad\quad\quad\quad\quad$-$ & $\frac{2}{\vert G_{D^1} \vert}\sum_{g \in G_{D^1}}\delta_{s_g,1}\cos\frac{\theta_g}{2}[\chi_{D^1}^{\beta}(g)]^*\widetilde{\chi}^{\widetilde{\alpha}}(g)$\\
			\hline
		\end{tabular}
	\end{center}
\end{table*}

\subsection{Example}
It is instructive to discuss concrete symmetry settings.
Here we consider four examples in the presence of PHS $\calC$: MSGs $P2/m1'$ with $B_g$ pairing, $P21'$ with $B$ pairing, $P4$ with $^1E$ pairing, and $Pmc2_11'$ with $A_2$ pairing. 
After classifying the Dirac Hamiltonians in Eq.~\eqref{eq:cc-ham} as discussed in Sec.~\ref{sec4:CC}, we obtain elements of $E^{1,0}_{1}$ corresponding to generating Dirac Hamiltonians by using formulas in Sec.~\ref{sec4:cc-formulas}. 
The results in this subsection will be used in Sec.~\ref{sec5:example}, where the physical consequences will be also discussed.
\subsubsection{$P2/m1'$ with $B_g$ pairing}
\label{sec4:p2m}
\begin{figure}
	\begin{center}
		\includegraphics[width=0.6\columnwidth]{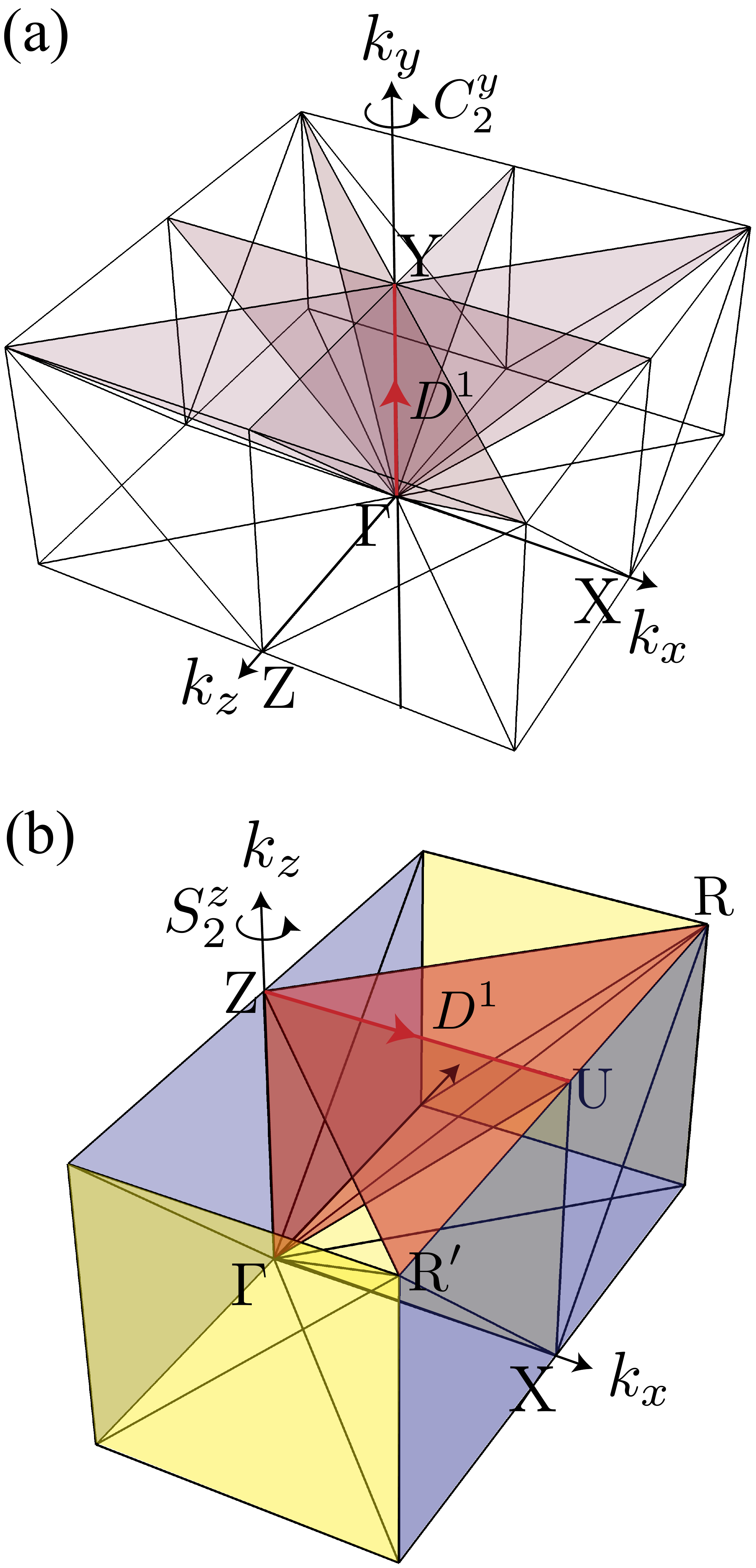}
		\caption{\label{fig:cell_p2m}Illustrations of cell decomposition for the half BZ in $P2/m1'$ (a) and the quarter BZ in $Pmc2_11'$ (b). Here we omit orientations except for the 1-cells denoted by $D^1$. In both (a) and (b), adjacent 2-cells to the 1-cell $D^1$ are colored by red. In (b), blue and yellow planes represent the mirror and glide planes of MSG $Pmc2_11'$, respectively.}
	\end{center}
\end{figure}

We fist discuss spinful MSG $P2/m1'$, and recall that this MSG has the two-fold rotation $C_{2}^{y}$ along the $y$-axis, the inversion $I$, and the TRS $\calT$. For $B_g$ pairing, $\{\sigma(\calC), \sigma(C_{2}^{y})\} = 0$ and $[\sigma(\calC), \sigma(I)]=0$ hold. 
Let us consider a two-fold rotation symmetric line as the 1-cell $D^1$ [see Fig.~\ref{fig:cell_p2m} (a)]. The little co-group is given by the following subsets:
\begin{align}
	\calG_{D^1}/T &=\{e, C_{2}^{y}\}, \\
	\calA_{D^1}/T &=\{I\calT,  (IC_{2}^{y})\calT\}, \\
	\calP_{D^1}/T &=\{I\calC, (IC_{2}^{y})\calC \}, \\
	\calJ_{D^1}/T &=\{\Gamma\equiv\calT\calC, C_{2}^{y}\Gamma\},
\end{align}
where $e$ denotes the identity element. 
To perform the procedures in Sec.~\ref{sec4:CC}, we define generators of the onsite symmetry group by 
\begin{align}
	\label{eq:tC2}
	\widetilde{\sigma}(C_{2}^{y}) &\equiv \gamma_1\gamma_2 \sigma(C_{2}^{y}),\\
	\label{eq:tIT}
	\widetilde{\sigma}(I\calT) &\equiv \sigma(I\calT),\\
	\label{eq:tIC}
	\widetilde{\sigma}(I\calC) &\equiv \sigma(I\calC),
\end{align}
One can verify that $s_g=+1$ for all elements in $\calG_{D^1}/T$, and then the onsite unitary symmetry group is $\tiG=\{e,C_{2}^{y} \}$. Since $[\widetilde{\sigma}(C_{2}^{y})]^2=-[\sigma(C_{2}^{y})]^2 =+1$, there are two one-dimensional irreducible representations $\widetilde{u}^{\widetilde{\alpha}}(C_{2}^{y}) = \alpha\ (\alpha=\pm 1)$. The representations in Eqs.~\eqref{eq:tC2}-\eqref{eq:tIC} possess the same commutation and anticommutation relations as $\sigma(C_{2}^{y})$, $\sigma(I\calT)$, and $\sigma(I\calC)$, i.e.,
\begin{align}
	\{\widetilde{\sigma}(C_{2}^{y}),\widetilde{\sigma}(I\calC)\} &= 0,\\
	[\widetilde{\sigma}(C_{2}^{y}), \widetilde{\sigma}(I\calT)] &= 0,\\
	[\widetilde{\sigma}(I\calT)]^2&=-1.
\end{align}
As a result, we find EAZ classes for $\alpha=\pm 1$ are class AII$_\calC$, whose classification is $\pi_2(\text{R}_4) = 0$. \add{This result is the case (iii), and therefore any point node is not stable on this line. We will see that the gapless point is part of line nodes in Sec.~\ref{sec5}.}

\subsubsection{$P21'$ with $B$ pairing}
\label{sec4:p2}
\begin{figure}[t]
	\begin{center}
		\includegraphics[width=0.9\columnwidth]{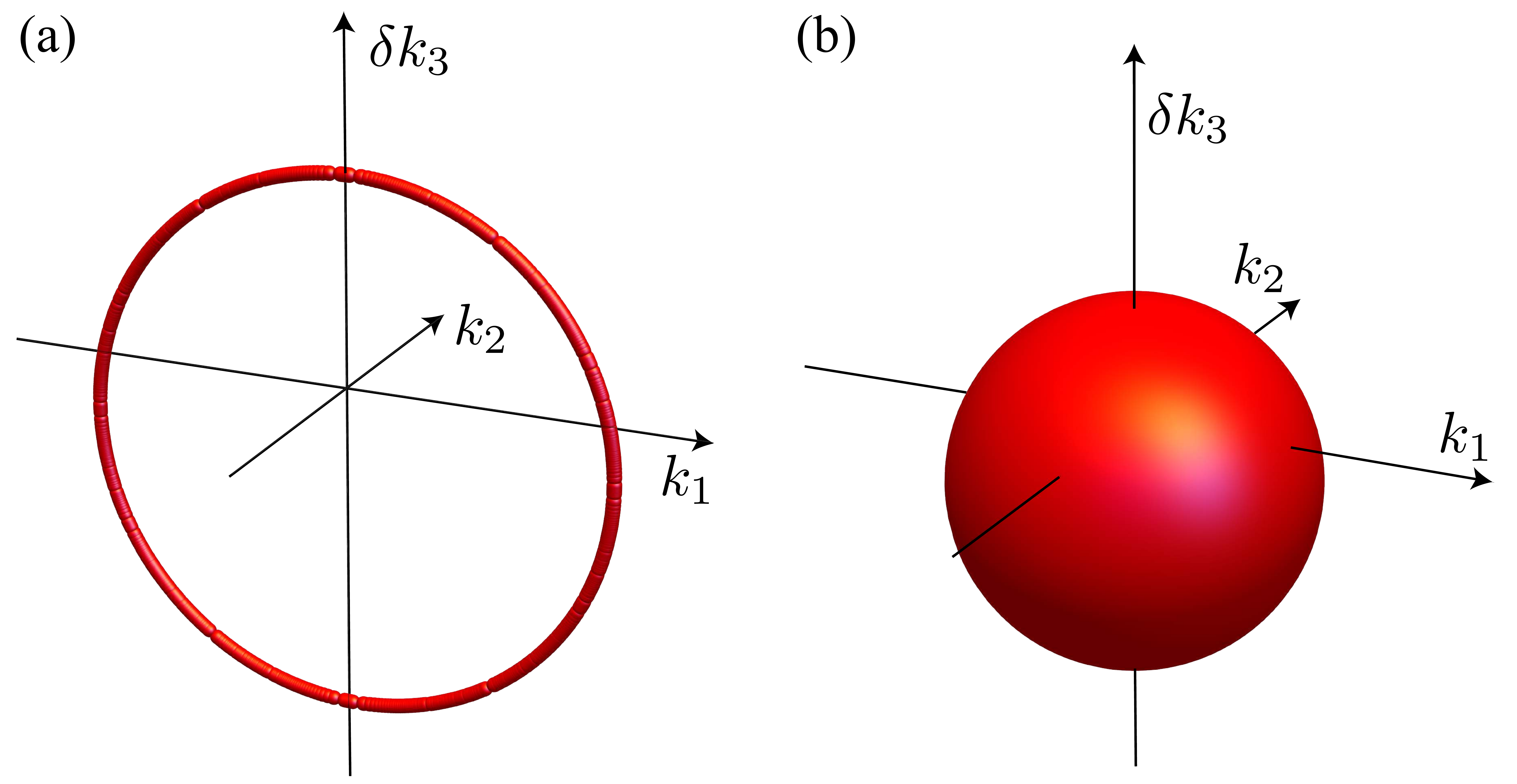}
		\caption{\label{fig:CC_example}Illustration of annihilation process of gapless points. Here white solid circles denote gapless points and $\pm$ represent the sign of the winding numbers.}
	\end{center}
\end{figure}

We next consider MSG $P21'$, which is generated by the two-fold rotation $C_{2}^{y}$ along the $y$-axis and the TRS $\calT$. For $B$ pairing, PHS anticommutes with the two-fold rotation, i.e., $\{\sigma(\calC), \sigma(C_{2}^{y})\} = 0$. 
Again, let us consider a two-fold rotation symmetric line as the 1-cell $D^1$ in Fig.~\ref{fig:cell_p2m} (a). Unlike the case of MSG $P2/m1'$, there exist only the following unitary and chiral parts in the little co-group
\begin{align}
	\calG_{D^1}/T &=\{e, C_{2}^{y}\}, \\
	\calJ_{D^1}/T &=\{\Gamma, C_{2}^{y}\Gamma\}.
\end{align}
To perform the procedures in Sec.~\ref{sec4:CC}, we define generators of onsite symmetries by Eq.~\eqref{eq:tC2} and $\widetilde{\sigma}(\Gamma) \equiv \sigma(\Gamma)$, and we find
\begin{align}
	\label{eq:tG}
	\{\widetilde{\sigma}(\Gamma), \widetilde{\sigma}(C_{2}^{y})\} = 0.
\end{align}
Since $[\widetilde{\sigma}(C_{2}^{y})]^2=+1$, we have two one-dimensional irreducible representations $\widetilde{U}^{\widetilde{\alpha}}(C_{2}^{y}) = \alpha\ (\alpha=\pm 1)$ whose EAZ classes are class A$_\Gamma$. Therefore, the Dirac Hamiltonians on the 1-cell are classified into $\pi_2(\text{C}_0)=\mZ$. The final step is to map the generating Dirac Hamiltonian of $\mZ$ to an elements of $E_{1}^{1,0}$. This can be accomplished by
\begin{align}
	\label{eq:A_G}
	\frakcn_{D^1}^{\beta}&=\frac{-2i}{\vert G_{D^1} \vert}\sum_{g \in G_{D^1}}\delta_{s_g,1}\sin\frac{\theta_g}{2}[\chi_{D^1}^{\beta}(g)]^*\nonumber\\
	&\quad\quad\quad\quad\quad\quad\quad\quad \times\left(\widetilde{\chi}^{\widetilde{\alpha}}(g)-\widetilde{\chi}^{\widetilde{\Gamma}\widetilde{\alpha}}(g)\right),
\end{align}
where $\widetilde{\chi}^{\widetilde{\Gamma}\widetilde{\alpha}}$ is the charcter of irreducible representation chiral-symmetry-related to $\widetilde{\chi}^{\widetilde{\alpha}}$.
By substituting irreducible representations in Table~\ref{tab:irreps_P2} into Eq.~\eqref{eq:A_G}, we obtain the band labels of the generating Dirac Hamiltonian
\begin{align}
	\label{eq:gene_P2}
	(\frakcn^{1},\frakcn^{2}) &= (2, -2),
\end{align} 
which corresponds to twice of a basis of $E_{1}^{1,0}$. 
\add{This result is the case (ii), which indicates that the gapless point is realized by a loop or surface node shrinking to the point. 
To see this, we consider a concrete Dirac Hamiltonian near the gapless point
\begin{align}
	\label{eq:gene_ham_AG}
	H_{(k_1, k_2)} &= k_1 \tau_1 + k_2 \tau_2\sigma_3 + \delta k_3\tau_3,\\
	\label{eq:gene_rep_AG}
	\widetilde{\sigma}(C_2) &= \sigma_3,\\
	\sigma(C_2) &= e^{-i\tfrac{\pi}{2}\tau_3\sigma_3}\widetilde{\sigma}(C_2) =-i \tau_3,\\
	\sigma(\Gamma) &=\widetilde{\sigma}(\Gamma) = \tau_2\sigma_2,
\end{align}
where we consider $\tilde{\alpha} =1$. Then, we add a symmetric perturbation 
\begin{align}
	M=m_0 \sigma_1 + m_1 \sigma_3 + m_2\tau_3 + m_3\tau_3\sigma_2
\end{align}
to the Dirac Hamiltonian in Eq.~\eqref{eq:gene_ham_AG}. As a result, we obtain a loop node shown in Fig.~\ref{fig:CC_example} (a). 
}

\begin{table}[t]
	\begin{center}
		\caption{\label{tab:irreps_P2}Irreducible representations of the onsite symmetry group $\widetilde{\calG}$ and $\calG_{D^1}/T$ for MSG $P2/m1'$ and $P21'$.}
		\begin{tabular}{c|c|c|c|c|c}
			\hline
			& EAZ of $P2/m1' (B_g)$ &	EAZ of $P21' (B)$ & irrep $\widetilde{\alpha}$ & $e$ &  $C_{2}^{y}$ \\
			\hline
			\multirow{2}{*}{$\tilde{\calG}$}& AII$_C$&  A$_\Gamma$ & $1$ & $1$ & $1$\\
			& AII$_C$ & A$_\Gamma$ & $2$ & $1$ & $-1$\\
			\hline\hline
			& EAZ of $P2/m1' (B_g)$ &	EAZ of $P21' (B)$  & irrep $\beta$ & $e$ &  $C_{2}^{y}$ \\
			\hline
			\multirow{2}{*}{$\calG_{D^1}/T$} &  D &	A & $1$ & $1$ & $i$ \\
			& D &	A & $2$ & $1$ & $-i$ \\
			\hline
		\end{tabular}
	\end{center}
\end{table}

\subsubsection{$P4$ with $^1E$ pairing}
\label{sec4:p4}
Next, we discuss the four-fold rotation symmetric line in spinful MSG $P4$, which is the same 1-cell $D^1$ in Fig.~\ref{fig:cell_p2m} (a) with the axes exchanged. Since this MSG does not have TRS, the little co-group $G_{D^1}/T$ has only a unitary part $\calG_{D^1}/T = \{e, C_{4}^z, (C_{4}^z)^2, (C_{4}^z)^3\}$. Then, the onsite symmetry group also has a unitary part generated by
\begin{align}
	\label{eq:tC4}
	\widetilde{\sigma}(C_{4}^{z}) &\equiv e^{\frac{\pi}{4}\gamma_1\gamma_2}\sigma(C_{4}^{z}),
\end{align}
where $[\widetilde{\sigma}(C_{4}^{z})]^4 = +1$. There are four irreducible representations of $\tiG$ in Table~\ref{tab:P4}, and therefore gapless points on the line are classified into $\mZ^4$. We can map the generating Dirac Hamiltonians to elements of $E_{1}^{1,0}$ by the following formula
\begin{align}
	\label{eq:A}
	\frakcn_{D^1}^{\beta}&=\frac{-2i}{\vert G_{D^1} \vert}\sum_{g \in G_{D^1}}\delta_{s_g,1}\sin\frac{\theta_g}{2}[\chi_{D^1}^{\beta}(g)]^*\widetilde{\chi}^{\widetilde{\alpha}}(g),
\end{align}
where $\beta$ represent to labels of irreducible representations of $\calG_{D^1}/T$ in Table~\ref{tab:P4}. 
As a result, we obtain the band labels
\begin{align}
	\label{eq:p4-point}
	(\frakcn^{1}_{D^1},\frakcn^{2}_{D^1}, \frakcn^{3}_{D^1},\frakcn^{4}_{D^1}) &= \begin{cases}
		(-1,0,0,1)\quad \text{for}\ \widetilde{\alpha} = 1\\
		(1,-1,0,0)\quad \text{for}\ \widetilde{\alpha} = 2\\
		(0,0,1,-1)\quad \text{for}\ \widetilde{\alpha} = 3\\
		(0,1,-1,0)\quad \text{for}\ \widetilde{\alpha} = 4,
	\end{cases}
\end{align}
which correspond to not any basis of $E_{1}^{1,0}$ but linear combinations of them. \add{The result is case (ii), i.e., the gapless point is actually a shrunk loop of surface node. To see this, let us discuss a concrete Dirac Hamiltonian near the gapless point
\begin{align}
	\label{eq:gene_ham_A}
	H_{(k_1, k_2)} &= k_1 \sigma_1 + k_2 \sigma_2 + \delta k_3\sigma_3,\\
	\label{eq:gene_rep_A}
	\widetilde{\sigma}(C_4) &= \sigma_0,\\
\sigma(C_4) &= e^{-\tfrac{\pi}{4}\sigma_1\sigma_2}\widetilde{\sigma}(C_4)  =\begin{pmatrix}
	e^{-i\tfrac{\pi}{4}} & \\
	& e^{i\tfrac{\pi}{4}}
\end{pmatrix},\
\end{align}
The Diac Hamiltonian and the symmetry representation correspond to the case of $\tilde{\alpha} =1$. We add a $C_4$-symmetric perturbation $M = \text{diag}(m_0, m_1)$ to the Dirac Hamiltonian. As shown in Fig.~\ref{fig:CC_example} (b), the Hamiltonian with the perturbation exhibits a surface node.
}

\begin{table}[t]
	\begin{center}
		\caption{\label{tab:P4}Irreducible representations of the onsite symmetry group $\widetilde{\calG}$ and $\calG_{D^1}/T$ for $P4$.}
		\begin{tabular}{c|c|c|cccc}
			\hline
			& EAZ & irrep $\widetilde{\alpha}$ & $e$ &  $C_{4}^{z}$ & $(C_{4}^{z})^2$ & $(C_{4}^{z})^3$ \\
			\hline
			\multirow{4}{*}{$\tilde{\calG}$} & A & $1$ & $1$ & $1$ & $1$ & $1$\\
			& A & $2$ & $1$ & $i$ & $-1$ & $-i$ \\
			& A & $3$ & $1$ & $-i$ & $-1$ & $i$ \\
			& A & $4$ & $1$ & $-1$ & $1$ & $-1$ \\
			\hline\hline
			& EAZ & irrep $\beta$ & $e$ &  $C_{4}^{z}$ & $(C_{4}^{z})^2$ & $(C_{4}^{z})^3$ \\
			\hline
			\multirow{4}{*}{$\calG_{D^1}/T$} & A & $1$ & $1$ & $e^{i \tfrac{\pi}{4}}$ & $i$ & $e^{i \tfrac{3\pi}{4}}$ \\
			& A & $2$ & $1$ & $e^{i \tfrac{3\pi}{4}}$ & $-i$ & $e^{i \tfrac{\pi}{4}}$ \\
			& A& $3$ & $1$ & $e^{-i \tfrac{3\pi}{4}}$ & $i$ & $e^{-i \tfrac{\pi}{4}}$ \\
			& A& $4$ & $1$ & $e^{-i \tfrac{\pi}{4}}$ & $-i$ & $e^{-i \tfrac{3\pi}{4}}$ \\
			\hline
		\end{tabular}
	\end{center}
\end{table}

\subsubsection{$Pmc2_11'$ with $A_2$ pairing}
\label{sec4:pmc2}
Last, we discuss nonsymmorphic and noncentrosymmetric MSG $Pmc2_11'$ with $A_2$ pairing. Here we consider the 1-cell on the boundary of BZ denoted by $D^1$ in Fig.~\ref{fig:cell_p2m} (b). 
The little co-group consists of the following four parts:
\begin{align}
	\label{eq:G/T_26}
	\calG_{D^1}/T &=\{e, M_{y}\}, \\
	\calA_{D^1}/T &=\{C_{2}^{z}\calT, M_{x}\calT\}, \\
	\calP_{D^1}/T &=\{C_{2}^{z}\calC, M_{x}\calC\}, \\
	\label{eq:Gamma/T_26}
	\calJ_{D^1}/T &=\{\Gamma, M_y\Gamma\}.
\end{align}
We define generators of the onsite symmetry group by
\begin{align}
	\label{eq:tMy}
	\widetilde{\sigma}(M_{y}) &\equiv \gamma_1\sigma(M_{y}),\\
	\widetilde{\sigma}(M_{x}\calT) &\equiv \gamma_1\gamma_2\sigma(M_{y}\calT),\\
	\widetilde{\sigma}(M_{x}\calC) &\equiv \gamma_1\gamma_2\sigma(M_{y}\calC).
\end{align}
Then, the onsite symmetry group is composed of the following symmetries:
\begin{align}
	\tiG &= \{e, M_y\Gamma\}, \\
	\widetilde{\calA} &= \{M_x\calT, M_y M_x\calC\},\\
	\widetilde{\calP} &= \{M_x \calC,M_y M_x\calT\},\\
	\widetilde{\mathcal{J}} &= \{\Gamma, M_y\}.
\end{align}
One can explicitly verify $[\widetilde{\sigma}(M_y\Gamma)]^2 = -1$ and $[\widetilde{\sigma}(M_y\Gamma), \widetilde{\sigma}(M_{x}\calT)] = [\widetilde{\sigma}(M_y\Gamma), \widetilde{\sigma}(M_{x}\calC)] = 0$. These relations imply that EAZ classes for irreducible representations $\widetilde{U}^{\widetilde{\alpha}}(M_y\Gamma) = i\alpha\ (\alpha=\pm 1)$ are class AIII$_\calT$, and therefore the classification is $\pi_2(\text{C}_1)=0$. \add{The result is the case (iii), and the gapless point on the 1-cell is part of line or surface nodes.}

\section{Unification of compatibility relations and point-node classifications}
\label{sec5}
In this section, we integrate classifications of gapless points discussed in Sec.~\ref{sec4} into compatibility relations in Sec.~\ref{sec3:CR}, which results in a unified way to diagnose the shapes of nodes. 
We first explain the general scheme to classify nodes on 1-cells, and then we apply the scheme to several symmetry settings: MSGs $P2/m1'$ with $B_g$ pairing, $P21'$ with $B$ pairing, $P4$ with $^1E$ pairing, and $Pmc2_11'$ with $A_2$ pairing.
\subsection{Revisiting compatibility relations and the first differential}
\label{sec5:d1}
\begin{figure}[t]
	\begin{center}
		\includegraphics[width=1\columnwidth]{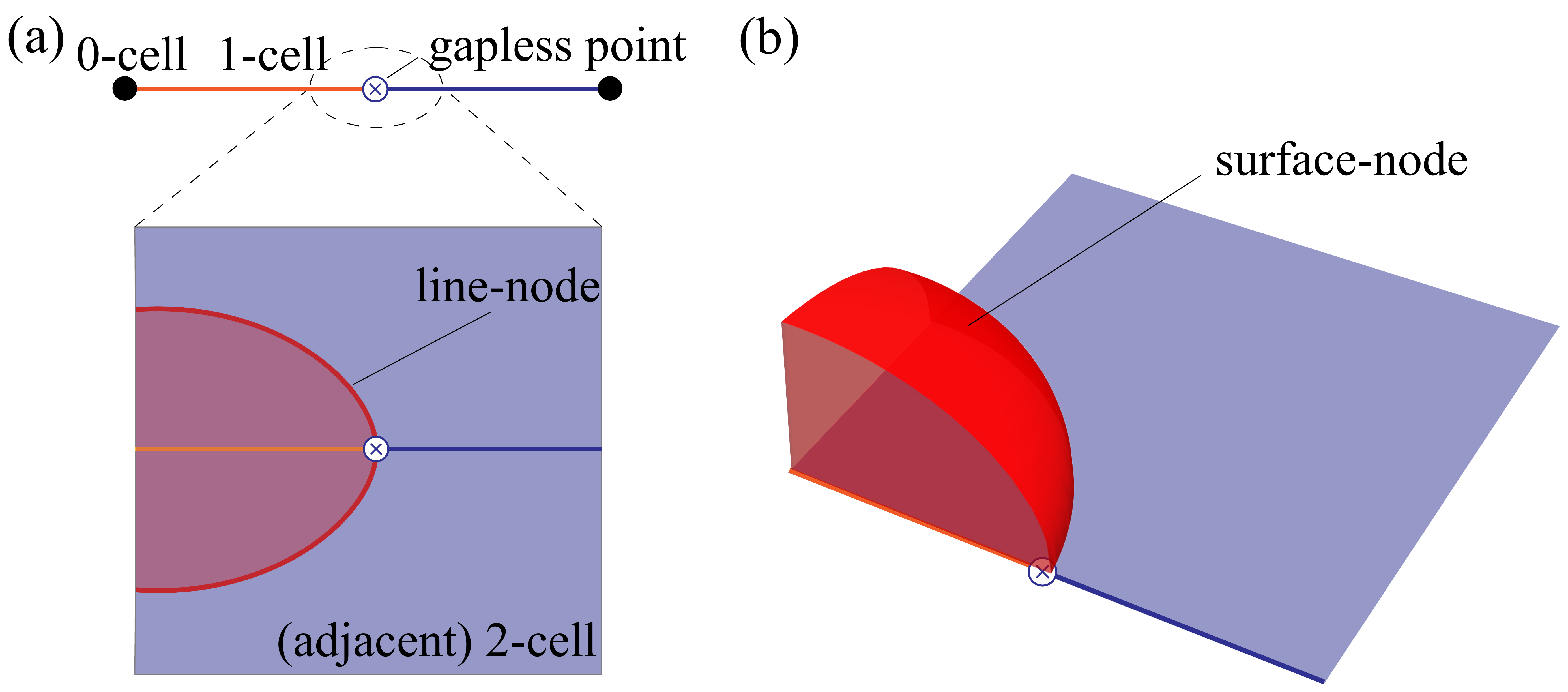}
		\caption{\label{fig:diff}Illustration of nodes near a gapless point at a 1-cell. (a) Two divisions of the 1-cell and nodal lines in adjacent 2-cells. Here, the red shaded region and others have different values of the topological invariants, whose boundary results in a nodal line. (b) Surface-node. When there are compatibility relations between the 2-cell and adjacent 3-cells, the regions in (a) are extended to 3-cells, and the boundary surface is the surface node.}
	\end{center}
\end{figure}
Before moving on to the scheme to classify nodes at 1-cells, let us revisit the first differentials $d_{1}^{p,0}$ for $p=1,2$. 
\add{Here, we discuss the reason why $d_{1}^{p,0}$ can be understood as the connection gapless states between $p$-cells and $(p+1)$-cells.}
	
Suppose that there exists a gapless point at a 1-cell, which involves the changes of zero-dimensional topological invariants at $\bk$-points in the 1-cell. In other words, there are two parts at the 1-cell which have different zero-dimensional topological invariants. It is not necessary that the gapless point at the 1-cell must be a genuine point node in BZ. In general, it might be part of line or surface nodes. We further assume that compatibility relations between points in the 1-cell and in adjacent 2-cells exist. 
\add{Although the zero-dimensional topological invariants for the above two parts in the $1$-cell are different, any points in the $1$-cell have common compatibility relations for points in the adjacent $2$-cells. Then, the compatibility relations and the different topological invariants of the two parts lead to two regions on the $2$-cell with different zero-dimensional topological invariants}
(see Fig.~\ref{fig:diff} (a)). As a result, the boundary line of these two regions results in the line node. In fact, $d_{1}^{1,0}$ informs us of the existence or absence of such line nodes.

Focusing on one of the adjacent 2-cells, we can apply the same discussion to this 2-cell. Namely, when compatibility relations between the 2-cell and adjacent 3-cells exist, $d_{1}^{2,0}$ examines whether the above two regions with different zero-dimensional topological invariants are extended to 3-cells (see Fig.~\ref{fig:diff} (b)). In the following, we formulate the above processes in a systematic manner based on $E_{1}^{p,0}$ and $d_{1}^{p,0}$.

\subsection{Classifications of nodes on 1-cell}
\label{sec5:general}
\begin{figure*}[t]
	\begin{center}
		\includegraphics[width=2\columnwidth]{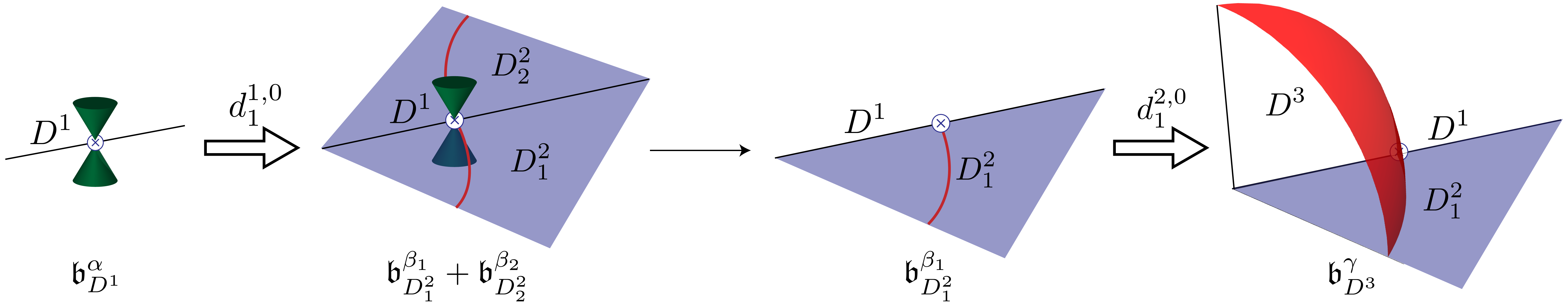}
		\caption{\label{fig:diagnosis}Illustration of the diagnostic scheme for case (A). We begin by acting $d_{1}^{1,0}$ on a generator of $E_{1}^{1,0}$ corresponding to a basis $\frakb_{D^1}^{\alpha}$. Then, we obtain the gapless lines on the $2$-cells adjacent to the 1-cell $D^1$, which is an element of $E_{1}^{2,0}$. Next, we focus on one of the adjacent $2$-cells. In the figure, we pick $D_1^2$ from the two $2$-cells. In other words, we consider only $\frakb_{D_1^2}^{\beta_1}$ of $\frakb_{D_1^2}^{\beta_1}+\frakb_{D_2^2}^{\beta_2}$ in the figure. Finally, we map $\frakb_{D_1^2}^{\beta_1}$ to an element $E_{1}^{3,0}$ by $d_{1}^{2,0}$ and examine whether the mapped element is trivial or not.}
	\end{center}
\end{figure*}
As discussed in the preceding section, compatibility relations tell us if the change of zero-dimensional topological invariants at a $p$-cell make domain walls of the changes at $(p+1)$-cells. This process is formulated in terms of $E_{1}^{p,0}$ and $d_{1}^{p,0}$. Recall that $E_{1}^{1,0}$ can be interpreted as the set of gapless states on 1-cells, and let us suppose that we have the set of band labels $\frakn^{(1)} =\frakb^{(1)}_{D^1,\alpha}$, where $\frakb^{(1)}_{D^1, \alpha}$ is a basis vector of $E_{1}^{1,0}$ generated by an irreducible representation $U_{D^1}^{\alpha}$ at a 1-cell $D^1$ (see Sec.~\ref{sec3:E1}). Applying the above strategy to the 1-cell, there are two cases: (A) $d_{1}^{1,0}(\frakn^{(1)})\neq \bm{0}$ and (B) $d_{1}^{1,0}(\frakn^{(1)})= \bm{0}$. 

We first consider case (A). Since $d_{1}^{1,0}(\frakn^{(1)})$ is an element of $E_{1}^{2,0}$, $d_{1}^{1,0}(\frakn^{(1)})$ can be expanded by the basis vectors of $E_{1}^{2,0}$ as
\begin{align}
	\label{eq:d1n}
	d_{1}^{1,0}(\frakn^{(1)}) &= \sum_{i}\left( \sum_{\alpha}r_{D^{2}_{i}, \alpha}^{(2)}\frakb_{D^{2}_{i},\alpha}^{(2)} + \sum_{\beta}m_{D^{2}_{i}, \beta}^{(2)}\frakb_{D^{2}_{i},\beta}^{(2)}\right),
\end{align}
where $r_{D^{2}_{i}, \alpha}^{(2)}\in \mZ_2$ and $m_{D^{2}_{i}, \beta}^{(2)} \in \mZ$. This equation tells us that the gapless point on the 1-cell is extended to adjacent 2-cells with the nontrivial coefficients in Eq.~\eqref{eq:d1n}, which results in the gapless lines on the $2$-cells. As a result, the gapless point on the 1-cell is part of line nodes or surface nodes. To distinguish between these two possibilities, we further examine whether $d_{1}^{2,0}$ is nontrivial. One might recall the relation $d_{1}^{2,0}\circ d_{1}^{1,0}=0$ and think that $d_{1}^{2,0}$ is useless for this purpose.
\add{However, when we focus on only one of the adjacent $2$-cells, the same discussion can be applied to the $2$-cells. In other words, picking only one basis vector from Eq.~\eqref{eq:d1n}, we can discuss the action of  $d_{1}^{2,0}$ on the picked basis, as is the case of $E_{1}^{1,0}$ (see Fig.~\ref{fig:diagnosis} for an intuitive illustration).}
If there exist the basis vectors such that $d_{1}^{2,0}(\frakb_{D^{2}_{i},\alpha}^{(2)})\neq 0$ in Eq.~\eqref{eq:d1n}, the gapless point on the 1-cell is part of a surface node. Otherwise, it is part of a line node.

Next, we discuss the case (B) where $d_{1}^{1,0}(\frakn^{(1)})=\bm{0}$. Since the relation indicates the absence of any domain walls discussed above, one might expect that the gapless point on the 1-cell is a genuine point node. Indeed, this is not always true. 
\add{The gapless point on the $1$-cell is a genuine point node only if $\frakn^{(1)}$ is a member of gapless point classifications, i.e., $\frakn^{(1)}$ can be expanded by the obtained band labels from gapless point classifications in Sec.~\ref{sec4:cc-formulas}. If not, the gapless point on the line is part of line nodes extended from the 1-cell to 3-cells, generic momenta.}

Using the above scheme, we classify nodes on all 1-cells for any MSG $\calM$, taking into account all the possible one-dimensional irreducible representations of the superconducting gaps, \add{the conditions $\calC^2=\pm 1$}, and the spinful/spinless nature of the systems. The results are tabulated in the Supplementary Materials. In Appendix~\ref{app:cell_3D}, we explain the cell decomposition for 3D BZ which we used in the classifications.

\subsection{Examples}
\label{sec5:example}
In the following, we will apply the above scheme to concrete symmetry settings. As mentioned in Secs.~\ref{sec1} and \ref{sec2}, our scheme is applicable to complex symmetry settings, e.g., noncentrosymmetric systems and rotation axes in the glide planes, which are out of the scope of previous studies. 
After we reproduce the results of previous works for spinful superconductors in MSG $P2/m1'$ with $B_g$ pairing by our method, we show that our method can detect nodal structures for those in MSG $P21'$, $P4$, and $Pmc2_11'$, which are noncentrosymmetric, TR breaking, or nonsymmorphic MSGs. The results are summarized in Table~\ref{tab:overview}.

\begin{table*}[t]
	\begin{center}
		\caption{\label{tab:overview}Summary of classification results for examples discussed in this work. 
			Space groups and pairing symmetries are shown in the first and second columns.
			The third and fourth ones represent the boundary points of the line where a gapless point exists. The fifth column is the label of an irreducible representation (irrep), which follows the notation in Ref.~\onlinecite{Bilbao}.
			The sixth column shows the classification $\mZ$ or $\mZ_2$, and the seventh column means the type of nodes. Here P, L, and S denote point, line, and surface nodes, respectively. In addition, while (A) means that the shape of the node is determined only by compatibility relations, (B) indicates that point-node classifications are necessary.
		}
		\begin{tabular}{c|c|c|c|c|c|c}
			\hline
			SG & pairing & HSP1 & HSP2 & irrep & classification & type of node \\
			\hline\hline 
			\multirow{2}{*}{$P2/m$ with TRS} & \multirow{2}{*}{$B_g$} & $(0,0,0)$ & $(0,\frac{1}{2},0)$ &$\bar{\Lambda}_3$ & $\mZ_2$ & L(B) \\
			& & $(0,0,0)$ & $(\frac{1}{2},0,0)$ & $\bar{\text{F}}_3$ & $\mZ_2$ & L(A) \\
			\hline
			$P2$ with TRS & $B$ & $(0,0,0)$ & $(0,\frac{1}{2},0)$ & $\bar{\Lambda}_3$ & $\mZ$ & L(B) \\
			\hline
			\multirow{4}{*}{$P4$ without TRS} & \multirow{4}{*}{$^1E$} & \multirow{4}{*}{$(0,0,0)$} & \multirow{4}{*}{$(0,0,\frac{1}{2})$} & $\bar{\Lambda}_5$ & $\mZ$ & S(A) \\
			& & & & $\bar{\Lambda}_6$ & $\mZ$ & S(A) \\
			& & & &$\bar{\Lambda}_7$ & $\mZ$ & S(A) \\
			& & & & $\bar{\Lambda}_8$ & $\mZ$ & S(A) \\
			\hline
			$Pmc2_1$ with TRS & $A_2$ & $(0,0,\frac{1}{2})$ & $(\frac{1}{2},0,\frac{1}{2})$ & $\bar{\text{A}}_3$ & $\mZ_2$ & L(B) \\
			\hline
		\end{tabular}
	\end{center}
\end{table*}

\subsubsection{$P2/m1'$ with $B_g$ pairing}
\label{sec5:p2mBg}
Let us begin with the 1-cell $D^1$ in Fig.~\ref{fig:cell_p2m} (a), which is the rotation axis in BZ for $P2/m1'$ with $B_g$ pairing. On the 1-cell, there are two irreducible representations listed in Table~\ref{tab:irreps_P2}. Ref.~\onlinecite{Sumita-Nomoto-Shiozaki-Yanase} has shown that line nodes pinned to the rotation axes can exist in this symmetry setting, although the derivation has not been shown. Here, we show that the line nodes pinned to the rotation axes can be stable by $d_{1}^{1,0}$ and our point-node classifications. 

Let us suppose that we have $\frakn^{(1)} = \frakb_{D^1}^{(1)}$ in which $\frakp_{D^1}^{1}, \frakp_{D^1}^{2}$, $\frakp_{\calT D^1}^{1}$, and $\frakp_{\calT D^1}^{2}$ equal $1$. We first define adjacent 2-cells to the 1-cell $D^1$ by Fig.~\ref{fig:cell_p2m} (a). The EAZ classes at the 2-cells are class DIII due to the existence of $I\calT$ and $I\calC$ with $(I\calT)^2=-1$ and $(I\calC)^2=+1$, and then compatibility relations among them do not exist. This results in $d_{1}^{1,0}(\frakn^{(1)}) = 0$, which indicates the gapless point on the 1-cell is not extended to the 2-cells. 

Next, we classify stable point nodes on the 1-cell. 
As discussed in Sec.~\ref{sec4:p2m}, we find there are no stable point nodes on the 1-cell. Therefore, we conclude that the gapless point is part of a line node extended from the 1-cell to 3-cells. This line node is protected by one-dimensional winding number $W$ defined by the chiral symmetry at the 3-cells. 
This is precisely what Ref.~\onlinecite{Sumita-Nomoto-Shiozaki-Yanase} has proposed. 

We then discuss the mirror plane in the $k_y = 0$. Let us focus on the 1-cell $b$ in Fig.~\ref{fig:p4mm} and suppose that we have $\frakn^{(1)} = \frakb_{b}^{(1)}$ which has $\frakp_{b}^{\pm}=1$ for irreducible representations $U_{b}^{\pm}(M_y) = \pm i$. The 2-cells $\alpha$ and $\alpha_7$ are adjacent to the 1-cell $b$ and the same symmetry class. Consequently, compatibility relations among them exist, and $d_{1}^{1,0}( \frakb_{b}^{(1)}) = \frakb_{\alpha}^{(2)}+\frakb_{\alpha_7}^{(2)}$. Here, $\frakb_{\alpha}^{(2)}$ $(\frakb_{\alpha_7}^{(2)})$ is a basis of $E_{1}^{2,0}$ in which  $\frakp_{\alpha}^{\pm}$ $(\frakp_{\alpha_7}^{\pm})$ and associated band labels equal $1$. As discussed in Sec.~\ref{sec5:general}, $d_{1}^{1,0}(\frakn^{(1)}) \neq 0$ indicates that the gapless point on the 1-cell $b$ should be extended to the adjacent 2-cells. Since EAZ classes of all 3-cells are class DIII, there are no compatibility relations, i.e., $d_{1}^{2,0} = 0$. 
Therefore, we conclude that the gapless point on the 1-cell $b$ is classified into $\mZ_2$ and is part of the line node in the mirror plane. Our result is consistent with the result of group theoretical analysis in Ref.~\onlinecite{Sumita-Yanase}

\subsubsection{$P21'$ with $B$ pairing}
\label{sec5:p2}
Next, we consider the same 1-cell as that in Sec.~\ref{sec5:p2mBg}, but without the inversion symmetry. In this case, the system can have line nodes pinned to the rotation axes. Irreducible representations $U_{D^1}^{\beta=1,2}$ and their EAZ classes are listed in Table~\ref{tab:irreps_P2}. 

We again assume that we have $\frakn^{(1)} = \frakb_{D^1}^{(1)}$ in which $\frakcn_{D^1}^{1}=-\frakcn_{D^1}^{2}=+1$ and associated band labels equal $1$ or $-1$.
%We again assume that we have $\frakn^{(1)} = \frakb_{D^1}^{(1)}$ generated by an irreducible representation in Table~\ref{tab:irreps_P2}. 
Unlike the above case, the 2-cells are invariant only under $\Gamma$, and then their EAZ classes are class AIII. As with the case of Sec.~\ref{sec5:p2mBg}, this implies that there are no compatibility relations among them, i.e., $d_{1}^{1,0}(\frakn^{(1)}) = 0$, and the gapless point on the 1-cell is not extended to the 2-cells. As shown in Sec.~\ref{sec4:p2}, 
%Since $(\frakcn_{D^1}^{1}, \frakcn_{D^1}^{2}) = (1, -1)$ is not the generator of point nodes in
since $(\frakcn_{D^1}^{1}, \frakcn_{D^1}^{2}) = (1, -1)$ is not a member of the gapless point classification in Eq.~\eqref{eq:gene_P2}, we conclude that $\frakb_{D^1}$ indicates the existence of line nodes pinned to the rotation axes. This is consistent with the fact that the winding number $W$ does not change after breaking the inversion symmetry of the system in Sec.~\ref{sec5:p2mBg}.

\begin{figure}[t]
	\begin{center}
		\includegraphics[width=0.65\columnwidth]{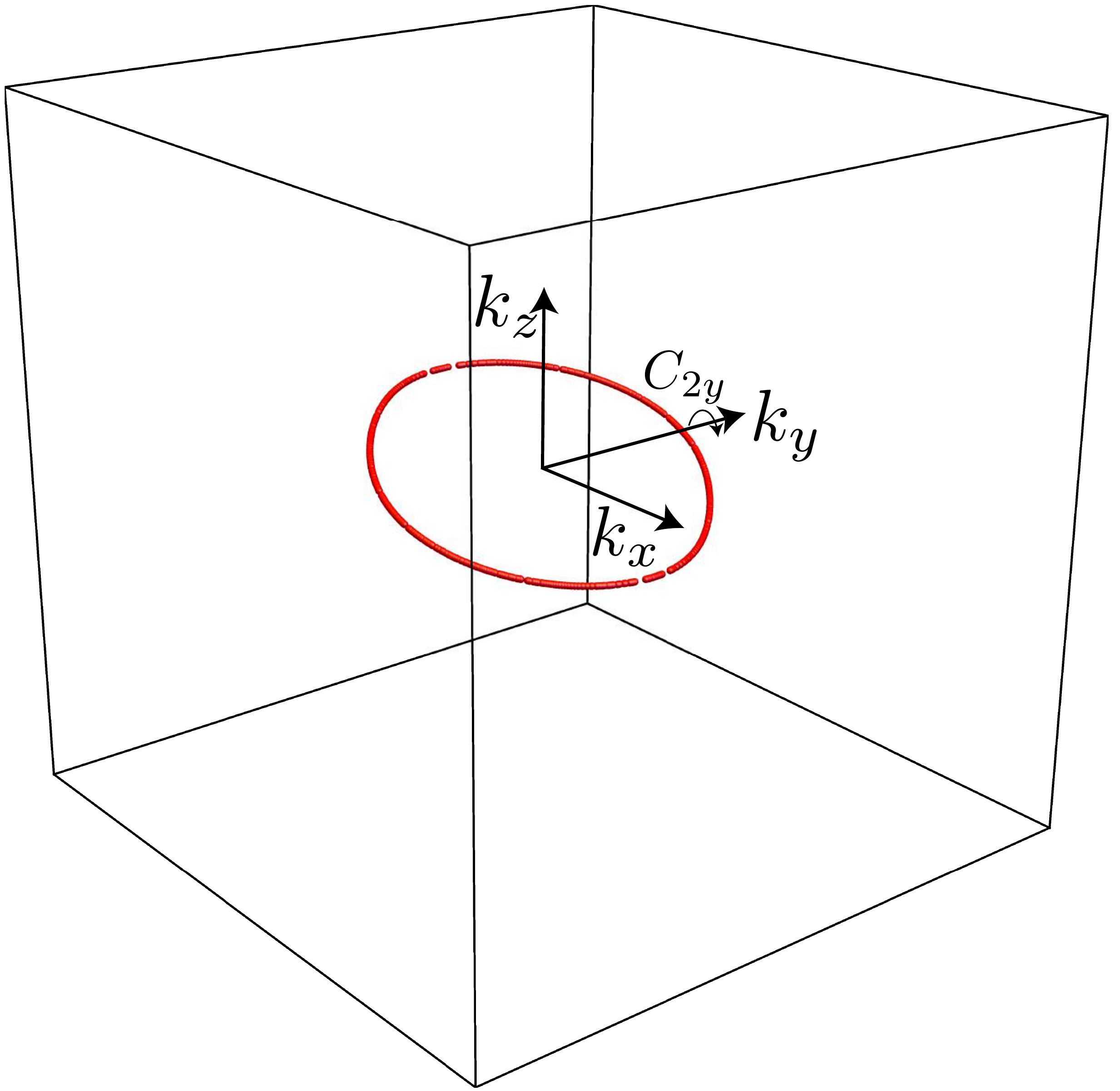}
		\caption{\label{fig:p2_line}The nodal line of the tight-binding model in Eq.~\eqref{eq:p2_model} for $\mu = +1$.} 
	\end{center}
\end{figure}

To verify the existence of such line nodes, let us consider the following model
\begin{align}
	\label{eq:p2_model}
	H_{\bk}&=(3-\cos k_x - \cos k_y - \cos k_z-\mu)\tau_z \nonumber\\
	&\quad\quad\quad\quad\quad\quad+ (\sin k_x+2\sin k_z)\tau_y,\\
	\rho(C_{2}^{y}) &= -i\tau_z\sigma_y,\\
	\rho(\calT) &= i\sigma_y,\\
	\rho(\calC) &= \tau_x,
\end{align}
where $\sigma_{i=x,y,z}$ and $\tau_{j=x,y,z}$ are Pauli matrices which represent different degree of freedom. After computing the region where the spectrum is gapless, we find a line node in Fig.~\ref{fig:p2_line}. This is the line node that we have discussed above.
%This is nothing but the line node discussed above.

\begin{figure}[t]
	\begin{center}
		\includegraphics[width=0.99\columnwidth]{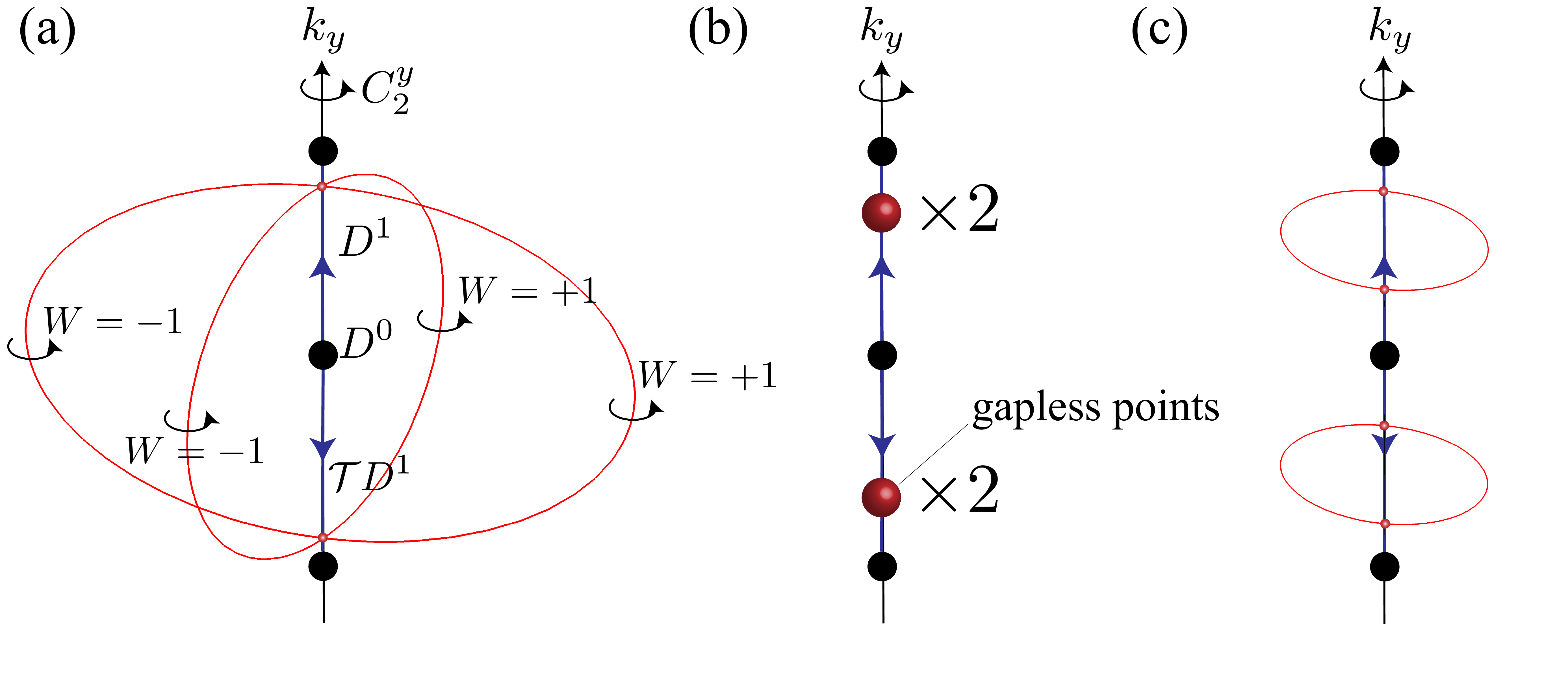}
		\caption{\label{fig:p2}Deformation of nodal structures in MSG $P21'$. Two line nodes pinned to the rotation axis are protected by 1D winding numbers (a). These line nodes can be deformed to point nodes without closing gap at 0-cells (b).  Since there are no reasons why two point nodes are at the same position, each of two split gapless points is again part of a line node.}
	\end{center}
\end{figure}

The question is whether $\frakn^{(1)} = 2\frakb_{D^1}^{(1)}$ is the point node or not. In the following, we show that the above line node can exist even in the case. To explain this, we start with the case where there are two the above line nodes generated by $\frakn^{(1)} = 2\frakb_{D^1}^{(1)}$ illustrated in Fig.~\ref{fig:p2} (a). By rotating one of the lines, the winding numbers can be cancelled. Then, we get two pair of point nodes in Fig.~\ref{fig:p2} (b). However, in the absence of other symmetries than MSG $P21'$ with PHS, there are no reasons why two gapless points on the 1-cell exist at the same point. Finally, each pair again forms a line node illustrated in Fig.~\ref{fig:p2} (c). As a result, $\frakn^{(1)} = 2\frakb_{D^1}^{(1)}$ indicates the existence of line nodes in Fig.~\ref{fig:p2} (c), and therefore nodes on the 1-cell are classified into $\mZ$, whose elements are line nodes of \textit{case (B)}. 

\subsubsection{$P4$ with $^1E$ pairing}
\label{sec5:p4}
\begin{figure}[t]
	\begin{center}
		\includegraphics[width=0.99\columnwidth]{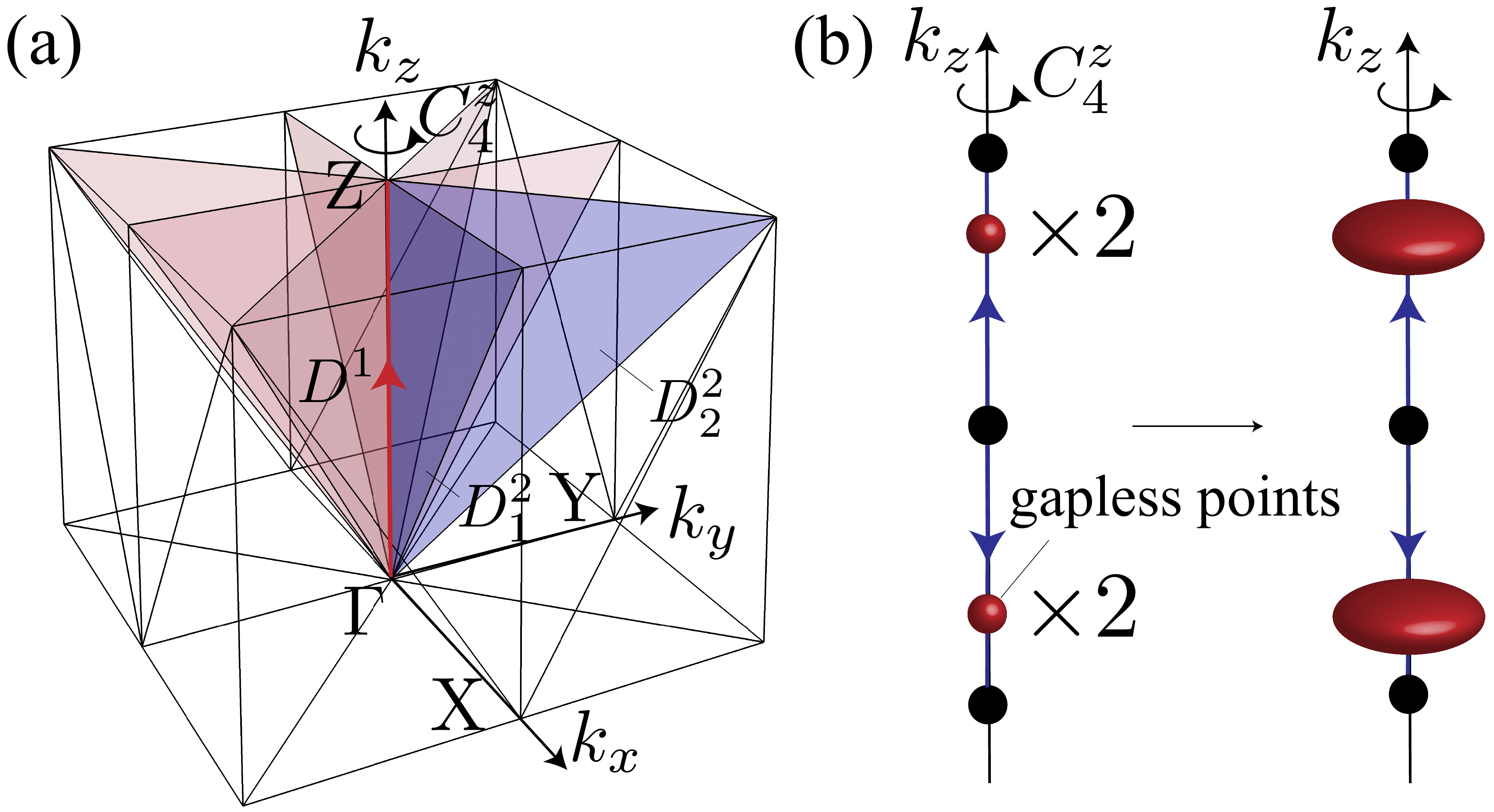}
		\caption{\label{fig:p4}(a) A half BZ in MSG $P4$. Here, the blue planes $D^{2}_{1}$ and $D^{2}_{2}$ are adjacent 2-cells to $D^1$ and red ones are symmetry-related to $D^{2}_{1}$ and $D^{2}_{2}$. (b) Deformation of nodal structures in MSG $P4$.}
	\end{center}
\end{figure}

Next, we discuss MSG $P4$ with $^1E$ pairing, which is generated by the four-fold rotation symmetry $C_{4}^{z}$. We consider the 1-cell $D^1$ in Fig.~\ref{fig:p4} (a). In the following, we show that a gapless point on the 1-cell is part of surface nodes. Irreducible representations $U_{D^1}^{\beta}\ (\beta=1,2,3,4)$ and their EAZ classes are tabulated in Table~\ref{tab:P4}. 

Suppose that we have $\frakn^{(1)} = \frakb_{D^1, \beta=1}^{(1)}$ which has $\frakcn^{1}_{D^1} = -\frakcn^{3}_{\calC D^1}  = +1$. Although there exist eight adjacent 2-cells to $D^1$ [colored in Fig.~\ref{fig:p4} (a)], only two of them are independent due to the presence of $C_{4}^{z}$. Here, we choose blue planes $D^{2}_{1}$ and $D^{2}_{2}$ in Fig.~\ref{fig:p4} (a) as independent adjacent 2-cells. Since the EAZ classes at $D^1$, the adjacent 2-cells, and 3-cells are the same, compatibility relations exist. Accordingly, $d_{1}^{1,0}(\frakn^{(1)}) = \frakb_{D^{2}_1}^{(2)} - \frakb_{D^{2}_2}^{(2)}$, in which  $\frakcn_{D_{i=1,2}^2}=+1$ and associated band labels equal $1$ or $-1$. We further find $d_{1}^{2,0}(\frakb_{D^{2}_1} )\neq 0$ and $d_{1}^{2,0}(\frakb_{D^{2}_2} ) \neq 0$, which implies that a gapless point on the 1-cell is part of surface nodes. Note that the disucssions and results for other values of $\beta$ do not change. 

As shown in Sec.~\ref{sec4:p4}, when $\frakn^{(1)}$ is a linear combination of $\{\frakb_{D^1, \beta}^{(1)}\}_{\beta=1}^{4}$, point nodes on the 1-cell can exist. However, the same logic in Sec.~\ref{sec5:p2} is valid, and therefore the point nodes can be inflated, which results in sphere nodes (Bogoliubov Fermi surfaces) pinned at the 1-cell like the right panel in Fig.~\ref{fig:p4} (b). Ref.~\onlinecite{PhysRevLett.125.237004} has discussed such Bogoliubov Fermi surfaces in multi-components superconductors without the inversion symmetry, although Ref.~\onlinecite{PhysRevLett.125.237004} has not discussed the symmetry-protection of them.

\subsubsection{$Pmc2_11'$ with $A_{2}$ pairing}
\label{sec5:pmc21}
Finally, we discuss nonsymmorphic and noncentrosymmetric MSG $Pmc2_11'$ with $A_{2}$ pairing. We focus on the 1-cell $D^1$ in the boundary of BZ [see Fig.~\ref{fig:cell_p2m} (b)], which is invariant under the glide symmetry $G_{y}$. There are two irreducible representations $U_{D^1}^{\pm}(G_y) = \pm 1$ of $\calG_{D^1}$, and their EAZ classes are class D. Let us consider that we have $\frakn^{(1)}=\frakb^{(1)}_{D^1}$ in which $\frakp_{D^1}^{\pm} = 1$ and associated band labels are nontrivial. As shown in Fig.~\ref{fig:cell_p2m} (b), three adjacent 2-cells to $D^1$ exist. The EAZ classes of the 2-cells in the $k_y=0$ plane and the $k_z=\pi$ are class A and class DIII, respectively. Consequently, there are no compatibility relations, i.e., $d_{1}^{1,0}(\frakn^{(1)}) = \bm{0}$. In addition, as shown in Sec.~\ref{sec4:pmc2}, there are no locally stable point nodes. As a result, we arrive at the line node pinned to the 1-cell $D^1$, which is extended from the 1-cell $D^1$ to 3-cells. 
Interestingly, such line nodes on the 1-cell do not exist in symmorphic MSG $Pmm21'$ with $A_{2}$ pairing, whose point group is the same as $Pmc2_11'$. In $Pmm21'$ with $A_{2}$ pairing, the line node pinned to the 1-cell $D^1$ is understood by the compatibility relations.
This is an example where nonsymmorphic symmetries change the classifications of nodes. As shown in this example, our method can capture the shape of nodes even in the presence of nonsymmorphic symmetries and in the absence of the inversion symmetry.

\section{Applications to materials}
\label{sec6}
In this section, we provide an efficient algorithm to diagnose the shape of nodes, which needs only the zero-dimensional topological invariants at 0-cells as input data. Since the energy scale of the superconducting gaps in most superconductors is believed to be much smaller than that of normal phases~\cite{PhysRevB.81.134508,PhysRevB.81.220504,PhysRevLett.105.097001,Ono-Yanase-Watanabe2019,Ono-Po-Watanabe2020,Ono-Po-Shiozaki2020}, assuming the pairing symmetry, we can obtain the input data from DFT calculations by the following formulas:
\begin{align}
	\frakp_{\bk}^{\alpha} &= n_{\bk}^{\alpha}\vert_{\text{occ}},\\
	\frakcn_{\bk}^{\alpha} &= n_{\bk}^{\alpha}\vert_{\text{occ}} - n_{-\bk}^{\tilde{\alpha}}\vert_{\text{occ}},
\end{align}
where $n_{\bk}^{\alpha}\vert_{\text{occ}}$ is the number of irreducible representations labeled by $\alpha$ in the normal phase, and $\tilde{\alpha}$ is a label of the particle-hole conjugate irreducible representation of $\alpha$.
We also demonstrate our scheme through a simple tight-binding model and a recently discovered superconductor CaPtAs.

\subsection{Efficient algorithm for detection of nodal structures}
In Sec.~\ref{sec5}, we have classified nodes on the 1-cells, and we have shown that the basis of $E_{1}^{1,0}$ can largely determine the shape of nodes. Here we recall that $d_{1}^{0,0}$ is a map from $E_{1}^{0,0}$ to $E_{1}^{1,0}$. This enable us to know nodal structures on the 1-cells from information at the 0-cells. First, let us assume that we have the set of band labels at the 0-cells $\frakn^{(0)}$ and $d_{1}^{0,0}(\frakn^{(0)})\neq 0$. By expanding $d_{1}^{0,0}(\frakn^{(0)})$ by the basis of $E_{1}^{1,0}$, we find which coefficients are nontrivial. Referring to the results of classifications in Sec.~\ref{sec5}, we diagnose the shape of nodal structures, i.e., gapless points on the 1-cells are point nodes or part of line/surface nodes. 

\begin{figure}[t]
	\begin{center}
		\includegraphics[width=0.7\columnwidth]{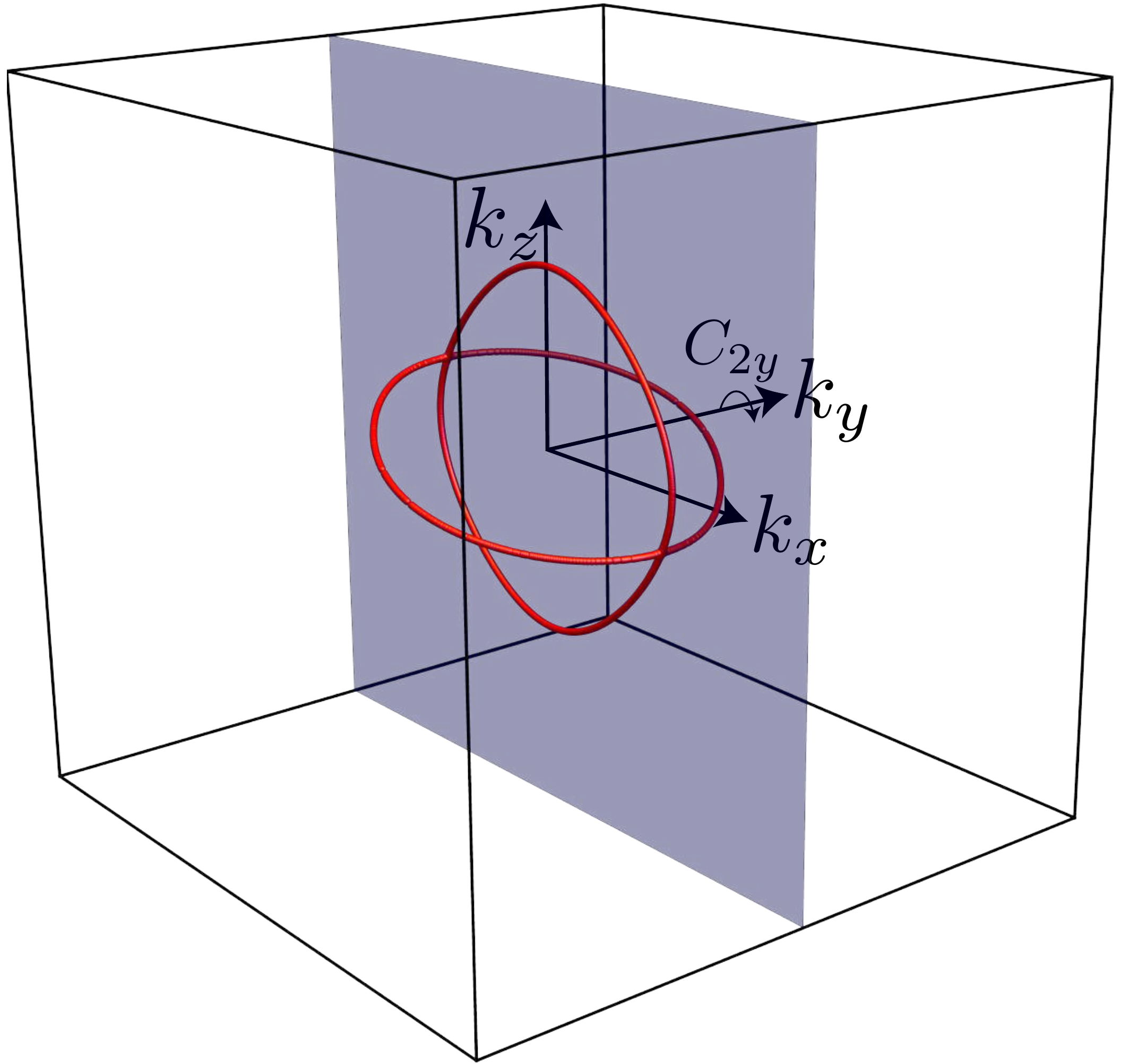}
		\caption{\label{fig:p2m}The nodal lines of the tight-binding model in Eq.~\eqref{eq:p2m_model} for $\mu = +1$. The blue plane is the mirror symmetric plane.}
	\end{center}
\end{figure}

To demonstrate the scheme, we consider a simple tight-binding model of MSG $P2/m1'$ with $B_g$ pairing:
\begin{align}
	\label{eq:p2m_model}
	H_{\bk}&=(3-\cos k_x - \cos k_y - \cos k_z-\mu)\tau_z \nonumber\\
	&\quad\quad\quad\quad+ (\sin k_x+2\sin k_z)\sin k_y \tau_y\sigma_y,\\
	\rho(I) &= \mathds{1},\\
	\rho(C_{2}^{y}) &= -i\tau_z\sigma_y,\\
	\rho(\calT) &= i\sigma_y,\\
	\rho(\calC) &= \tau_x,
\end{align}
where $\sigma_{i=x,y,z}$ and $\tau_{j=x,y,z}$ are Pauli matrices which represent different degree of freedom. Using this model, we show that the above algorighm can detect nodal structures discussed in Sec.~\ref{sec5:p2mBg}. After computing Pfaffian invariants in Eq.~\eqref{eq:Pf} for all 0-cells, we find $\frakp_{\Gamma}^{1} = \frakp_{\Gamma}^{2} = 1$ and others equal zero, where $\frakp_{\Gamma}^{1}$ and $\frakp_{\Gamma}^{2}$ are band labels for irreducible representations $(U_{\Gamma}^{1}(I), U_{\Gamma}^{1}(C_{2}^{y}) )=(1,+i)$ and $(U_{\Gamma}^{2}(I),U_{\Gamma}^{2}(C_{2}^{y}) )=(1,-i)$.
This set of band labels correspond to a basis of $E_{1}^{0,0}$ denoted by $\frakb_{\Gamma, 1}^{(0)}$, and we get $d_{1}^{0,0}(\frakb_{\Gamma, 1}^{(0)}) = \frakb_{a}^{(1)}+\frakb_{b}^{(1)}+\frakb_{a_1}^{(1)}+\frakb_{b_1}^{(1)}+\frakb_{D^1}^{(1)}$, where we use the same labels of 1-cells in Figs.~\ref{fig:p4mm} and \ref{fig:cell_p2m}(a). This indicates that gapless points exist on the 1-cells $a,b,a_1,b_1$, and $D^1$. As discussed in Sec.~\ref{sec5:p2mBg}, the gapless point on the 1-cell $b$ is part of line nodes in the mirror plane. Similar to the case, gapless points on the 1-cells $a, a_1,$ and $b_1$ are also extended to their adjacent 2-cells in the plane. Taking into account symmetry relations among 2-cells, we find that a line node in the mirror plane encircles $\Gamma$ point. On the other hand, we have shown that a gapless point in the rotation axis is also part of a line node pinned to the axis. We verify that our method correctly captures the nodes of the tight-binding model shown in Fig.~\ref{fig:p2m}. 

\subsection{Material example}
In this subsection, we apply the above algorithm to realistic superconductors CaPtAs, whose MSG is $I4_1md1'$.
A recent experiment~\cite{PhysRevLett.124.207001} has reported the time-reversal breaking and the signature of point nodes. Breaking TRS indicates that the order parameter belongs to $^1E$ or $^2E$ representations of the point group $C_4$. Then, MSG $I4_1md1'$ is reduced to $I4_1$. Here, we assume that the superconducting gap belongs to $^1E$ representation. Ref.~\onlinecite{Ono-Po-Shiozaki2020} has computed irreducible representations by QUANTUM-ESPRESSO~\cite{qe1,qe2} and \textit{qeirreps}~\cite{qeirreps} and found that $\frakp_{\Gamma}^{4} = 1$ and $\frakcn_{\Gamma}^{1} = -\frakcn_{\Gamma}^{3}=-1$, where the labels of irreducible representations follow Table~\ref{tab:P4}. Then, the set of band labels $\frakn^{(0)}$ corresponds to $-\frakb_{\Gamma,1}^{(0)} + \frakb_{\Gamma,4}^{(0)}$. In the following, we show that this superconducting material is expected to have small Bogoliubov Fermi surfaces. 

We check if this material satisfies compatibility relations, i.e., $d_{1}^{0,0}(\frakn^{(0)}) = \bm{0}$. After computing $d_{1}^{0,0}(\frakn^{(0)})$, we find $d_{1}^{0,0}(\frakn^{(0)}) = -\frakb_{D^1,1}^{(1)}+\frakb_{D^1,3}^{(1)}$, where $D^1$ denotes the rotation symmetric line between $\Gamma=(0,0,0)$ and $\text{Z}=(0,0,2\pi)$. In fact, the symmetry setting in this line is the completely same as that in Sec.~\ref{sec5:p4}, and then the nodal structures are also the same. 
Since $d_{1}^{0,0}(\frakn^{(0)})$ correspond to a set of band labels listed in Eq.~\eqref{eq:p4-point}, we expect that this material has small Bogoliubov Fermi surfaces as discussed in Sec.~\ref{sec5:p4} (see Fig.~\ref{fig:p4} (b)).

Our result might not contradict the experimental observation.
Since the superconducting gaps in most superconductors are considered to be much small, it is natural to think the Bogoliubov Fermi surfaces are also small. Further experiments to distinguish between this case and \textit{exact} point nodes are awaited.

\section{Further extension to nodes at generic points}
\label{sec7}
Thus far, we have focused on nodes pinned to 1-cells. However, in general, nodes can exist at generic points. In this section, we discuss how to extend our symmetry-based approach to nodes at generic points through the mirror plane in MSG $P2/m1'$ with $B_u$ pairing.

Here, we decompose the mirror plane into the cell decomposition in Fig.~\ref{fig:p4mm} and discuss the 1-cell denoted by $b$ in Fig.~\ref{fig:p4mm}. After applying the method in Sec.~\ref{sec4} to the 1-cell, we find that the classification of gapless points is $\mZ$. The generating Hamiltonian is 
\begin{align}
	H_{(k_1,k_2)} &= k_1 \tau_y + k_2 \tau_{x}\sigma_z+\delta k_3\tau_z,\\
	\sigma(I\calC) &= i \tau_y K \\
	\sigma(I\calT) &= i \tau_z \sigma_y K\\
	\sigma(M_y) &= i \tau_z \sigma_x,
\end{align}
where $k_1$ is perpendicular to both the mirror plane, the 1-cell, $k_2$ is perpendicular to the 1-cell but parallel to the mirror plane, and $\delta k_3$ is a displacement from the gapless point in the direction of the 1-cell. The gapless point is protected by the mirror winding number~\cite{PhysRevLett.113.046401}. On the other hand, since the EAZ class at the 1-cell is class AIII, there are no topological invariants, which implies that gapless points pinned to the 1-cell do not exist. In fact, we can add the symmetric perturbation terms which shift the gapless point to the $k_2$-direction. Therefore, gapless points can locally exist everywhere in the mirror plane. 

\begin{figure}[t]
	\begin{center}
		\includegraphics[width=0.9\columnwidth]{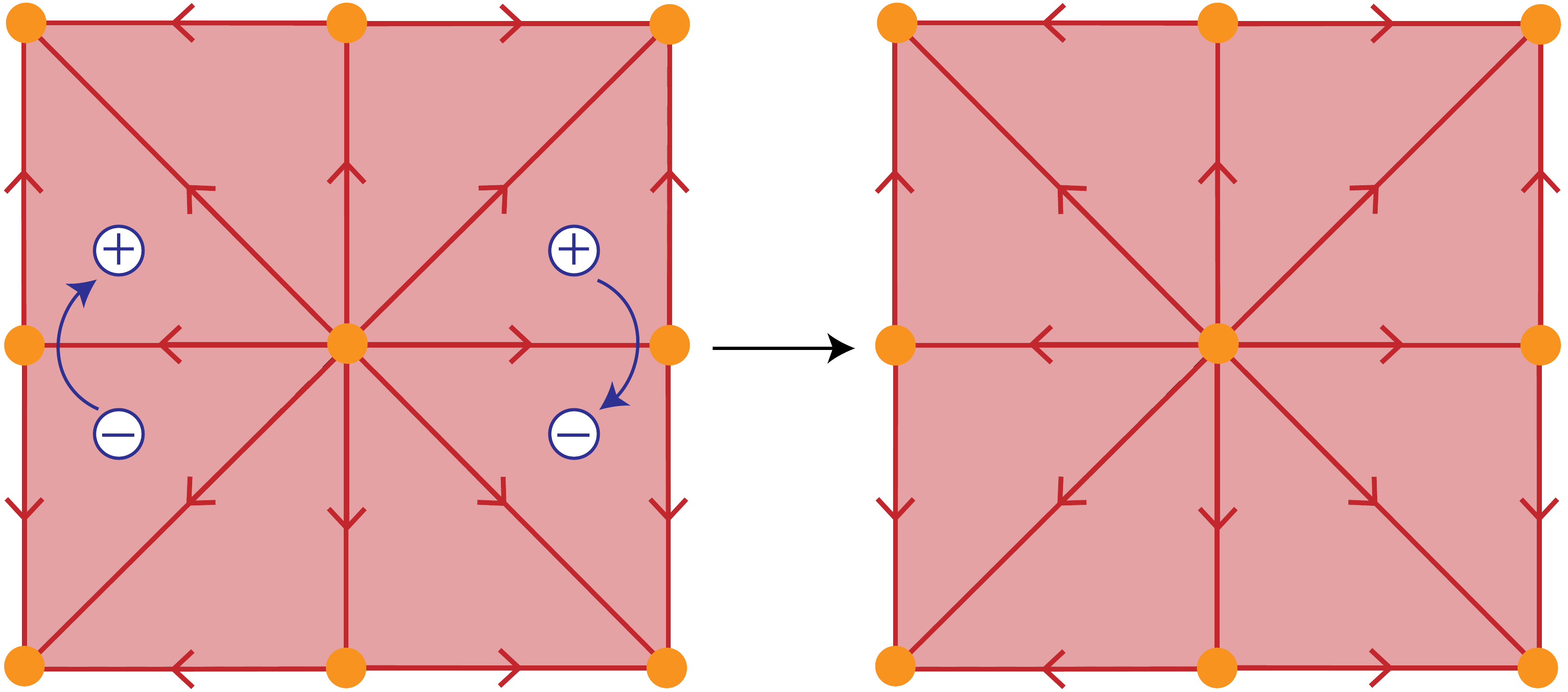}
		\caption{\label{fig:mirror}Illustration of annihilation process of gapless points. Here white solid circles denote gapless points and $\pm$ represent the sign of the winding numbers.}
	\end{center}
\end{figure}

The question is whether these gapless points are globally stable. In the following, we show that there can globally exist only two gapless points in the plane. To explain this, let us suppose that there are four gapless points in the plane as shown in Fig.~\ref{fig:mirror}. Since $C_{2}^{y}$ anticommutes with PHS, $C_{2}^{y}$ changes the sign of the winding number (see Appendix~\ref{app:winding}). As discussed above, the gapless points can freely move in the plane, and therefore two winding numbers with opposite signs can be canceled. This indicates that only one pair of gapless points can globally exist. 

Symmetry indicators in this symmetry class can detect the globally stable gapless points. The symmetry indicator group is $(\mZ_2)^2 \times \mZ_4$, whose $\mZ_2$-parts originate from lower dimensions.
The $\mZ_4$ index is defined by
\begin{align}
	z_4 &= \frac{1}{4}\sum_{K \in \text{TRIMs}}\left(\frakcn_{K}^{+}-\frakcn_{K}^{-}\right)\mod 4,
\end{align}
where $\frakcn_{K}^{\pm}$ is the band label for irreducible representations $U_{K}^{\pm}(I) = \pm 1$ at the time-reversal invariant momenta (TRIMs). If the system is fully gapped, $z_4 = 1,3$ indicate the mirror Chern number modulo $2$ equals $1$. However, the nontrivial mirror Chern numbers are forbidden in this symmetry setting~\cite{PhysRevLett.111.056403}. Therefore, we conclude that $z_4 = 1,3$ indicate the existence of gapless points. 

Actually, the above annihilation procedure can be understood as ``second differential'' $d_{2}^{p,0}$ in the theory of Atiyah-Hirzebruch Spectral Sequence~\cite{Shiozaki2018}. Although establishing full classifications of nodes at generic points and relationship between symmetry indicators and the nodes are interesting issues, they are out of scope of this paper. 

%[The relation between symmetry indicators and nodal structures (e.g. $P2/m1'$ with $B_u$ pairing)]

\section{Conclusion and Outlook}
\label{sec8}
In this work, we have established a systematic framework to classify superconducting nodes pinned to any line in momentum space. After decomposing BZ of all MSGs into points (0-cells), lines (1-cells), planes (2-cells), and polyhedrons (3-cells), we have applied our method to the lines and obtained comprehensive classifications of nodes pinned to the lines. Moreover, our theory has resulted in a highly efficient way to diagnose the superconducting nodes in superconducting materials. As a demonstration, we have analyzed the nodes in CaPtAs assuming the time-reversal broken pairing and pointed out that this material can have small Bogoliubov Fermi surfaces. 

Our work opens up various possibilities for future studies. Although our results cover a wide range of nodes, nodes at generic points are missing as discussed in Sec.~\ref{sec7}. The symmetry-based approach can be more refined to detect such nodes, and we leave deriving full relationships between symmetry indicators and the nodes as future works. This type of study will give us more information of nodes pinned to lines as follows. Suppose that a system violates compatibility relations, which indicates the existence of nodes pinned to 1-cells as discussed in Sec.~\ref{sec6}. Since we can always forget about symmetries that impose the violated compatibility relations on the system, we can apply symmetry indicators for lower symmetry classes to the system as discussed in Ref.~\onlinecite{PhysRevResearch.2.022066}. Then, the symmetry indicators will clarify topological nature behind the nodes.

The integration of our algorithm with DFT calculations enables a comprehensive investigation of nodes in the materials listing in the database. Such studies help to find the possible pairings of unconventional superconductivity compatible with experimental observations. We hope that our study will lead to a deep understanding of superconductivity in discovered superconductors.

\begin{acknowledgments}
	We thank Hoi Chun Po, Shuntaro Sumita, Takuya Nomoto, and Haruki Watanabe for fruitful discussions. In particular, KS thanks Takuya Nomoto for sharing ideas on how the first differential detects the nodal structure in the early stages of the project.
	SO is also grateful to Yohei Fuji for valuable comments on the manuscript. 
	The work of SO is supported by The ANRI Fellowship and  KAKENHI Grant No. JP20J21692 from he Japan Society for the Promotion of Science.
	The work of KS is supported by PRESTO, JST (Grant No. JPMJPR18L4) and CREST, JST (Grant No. JPMJCR19T2).
	
	\textit{Note added}.---After posting the preprint of this work (arXiv:2102.07676), Ref.~\onlinecite{wu2021symmetryenforced} appeared, which is based on a similar idea and discusses only gapless states in the normal phases. However, this work is different from Ref.~\onlinecite{wu2021symmetryenforced} in terms of the formulation and the mathematical approach. Note that, as stressed in this paper, compatibility relations do not determine superconducting nodes pinned to lines in the momentum space completely. Therefore, our unification of compatibility relations and point-nodes classifications plays a vital role in the classifications of the superconducting nodes.
\end{acknowledgments}

\appendix
\section{Cell decomposition for representative space groups}
\label{app:cell_3D}
\begin{figure}[t]
	\begin{center}
		\includegraphics[width=1.01\columnwidth]{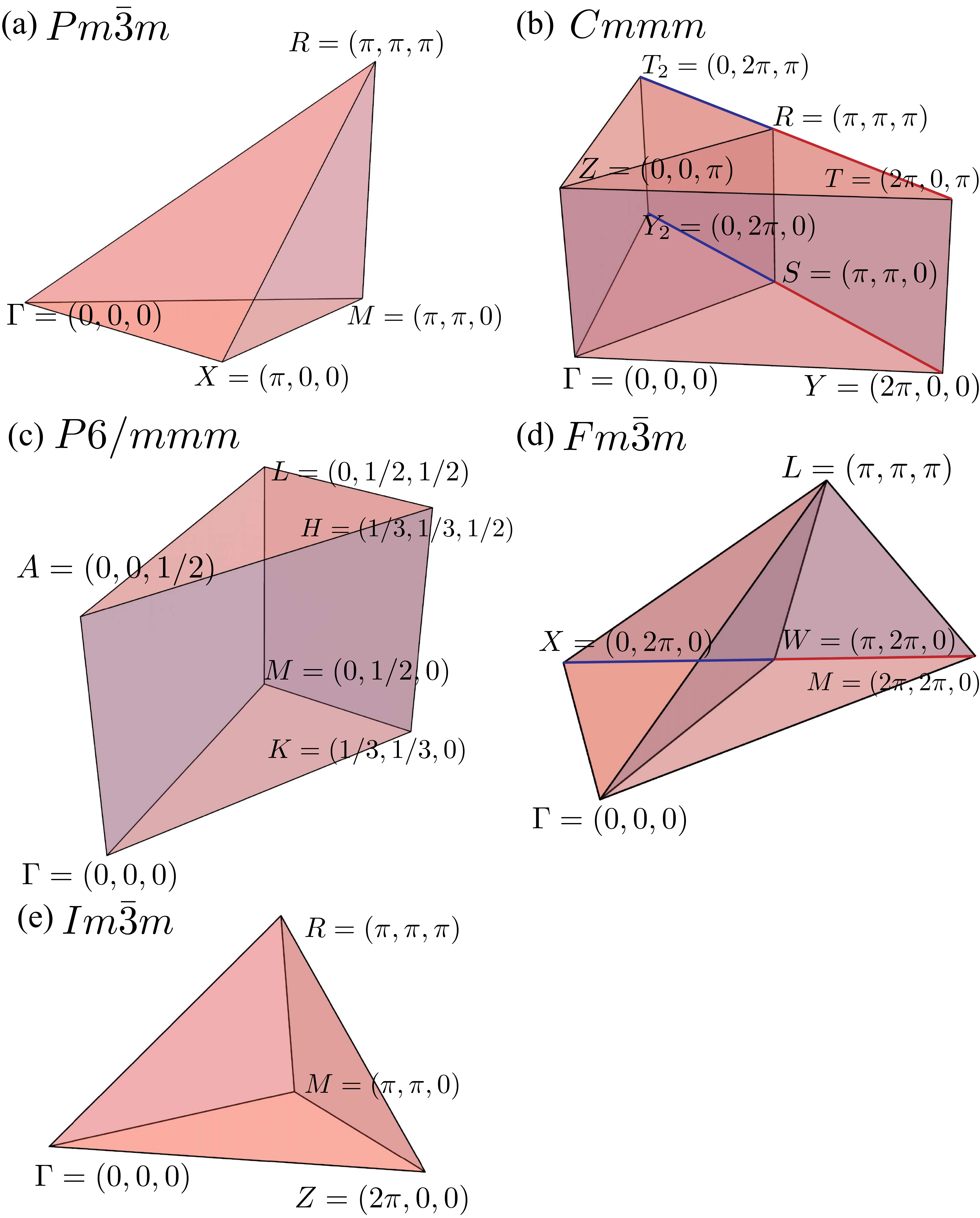}
		\caption{\label{fig:3D-cell}Units of BZ for $Pm\bar{3}m$ (a), $Cmmm$ (b), $P6/mmm$ (c), $Fm\bar{3}m$ (d), and $Im\bar{3}m$ (e). Note that coordinates in $P6/mmm$ are denoted by coefficients of primitive reciprocal lattice vectors. Orientations except for blue and red lines in $Cmmm$ and $Fm\bar{3}m$ can be arbitrarily chosen. Orientations of these colored lines are choosen by symmetric manners.}
	\end{center}
\end{figure}
In this appendix, we present units of BZ for each type of lattices, which can fill the entire BZ by symmetry operations. In fact, it is enough to define the units for SG $Pm\bar{3}m, Cmmm, P6/mmm, Fm\bar{3}m,$ and $Im\bar{3}m$. Note that the cell decomposition for $Amm2$ is the same as that for $Cmmm$ with axes exchanged and the cell decomposition for $R\bar{3}m1$ is constructed by \{$k_{j,x}\bb_1+k_{j,y}\bb_2+k_{j,z}\bb_3\}_{j},$ where $(k_{j,x},k_{j,y}, k_{j,z})$ is a cell for $Pm\bar{3}m$ and $b_{i}$ is a primitive reciprocal lattice vector.  
When we discuss a lower symmetry setting than them, we use the cell decomposition of one whose lattice is the same as the system. 

\section{Derivation other formulas}
\label{app:formulas}
In this appendix, we derive formulas to obtain elements of $E_{1}^{p,0}$ corresponding to generating Dirac Hamiltonians. To achieve this, we find the generating Hamiltonian and symmetries like Eqs.~\eqref{eq:gene_ham_AC}-\eqref{eq:gene_rep_AC}. We tabulate Gamma matrices $\gamma_{0,1,2}$ and symmetry representations $\widetilde{\sigma}(g)$ in Table~\ref{tab:gene}. By substituting them into Eqs.~\eqref{eq:formula_Z} and \eqref{eq:formula_Z2}, one can obtain formulas in Table~\ref{tab:formulas}. 

\begin{table*}[t]
	\begin{center}
		\caption{\label{tab:gene}Gamma matrices in Eq.~\eqref{eq:cc-ham} and onsite unitary symmetries. Here, $\widetilde{u}^{\widetilde{\alpha}}$ is an irreducible representation of the onsite unitary symmetry group $\widetilde{\calG}$. In addition, $\widetilde{\calT}\widetilde{\alpha}$, $\widetilde{\calC}\widetilde{\alpha}$, and $\widetilde{\Gamma}\widetilde{\alpha}$ are labels of the time-reversal, the particle-hole, and the chiral symmetry related irreducible representations.}
		\begin{tabular}{c|c|c|c|c}
			\hline
			EAZ & $\gamma_0$ & $\gamma_1$ & $\gamma_2$ & $\widetilde{\sigma}(g)$\\
			\hline\hline
			A & $\sigma_z$& $\sigma_x$ & $\sigma_y$ & $\mathds{1}\otimes\widetilde{u}^{\widetilde{\alpha}}(g)$\\
			\hline 
			A$_\calT$  & $\tau_z$ & $\tau_x$ & $\tau_y\sigma_z$ & $\tau_0\text{diag}(\widetilde{u}^{\widetilde{\alpha}}(g), \widetilde{u}^{\widetilde{\calT}\widetilde{\alpha}}(g))$\\
			\hline
			A$_\calC$ & $\tau_z$ & $\tau_x$ & $\tau_y$ & $\tau_0\text{diag}(\widetilde{u}^{\widetilde{\alpha}}(g), \widetilde{u}^{\widetilde{\calC}\widetilde{\alpha}}(g))$\\
			\hline
			A$_\Gamma$ &$\tau_z$ & $\tau_x$ & $\tau_y\sigma_z$ & $\tau_0\text{diag}(\widetilde{u}^{\widetilde{\alpha}}(g), \widetilde{u}^{\widetilde{\Gamma}\widetilde{\alpha}}(g))$\\
			\hline
			A$_{\calT,\calC}$ & $\tau_z$ & $\tau_x$ & $\tau_y(\mathds{1}\otimes\sigma_z)$& $\tau_0\text{diag}(\widetilde{u}^{\widetilde{\alpha}}(g), \widetilde{u}^{\widetilde{\calT}\widetilde{\alpha}}(g), \widetilde{u}^{\widetilde{\calC}\widetilde{\alpha}}(g), \widetilde{u}^{\widetilde{\Gamma\alpha}}(g))$\\
			\hline
			C & $\sigma_z$& $\sigma_x$ & $\sigma_y$ & $\mathds{1}\otimes\widetilde{u}^{\widetilde{\alpha}}(g)$\\
			\hline 
			C$_\calT$ & $\tau_z$& $\tau_y$ & $\tau_x\sigma_z$ & $\tau_0\text{diag}(\widetilde{u}^{\widetilde{\alpha}}(g), \widetilde{u}^{\widetilde{\calT}\widetilde{\alpha}}(g))$\\
			\hline
			D & $\tau_z$& $\tau_x\sigma_y$ & $\tau_y\sigma_y$ & $\tau_0\text{diag}(\widetilde{u}^{\widetilde{\alpha}}(g), \widetilde{u}^{\widetilde{\alpha}}(g))$\\
			\hline 
			D$_\calT$ & $s_z$& $s_y\tau_y$ & $s_x\tau_y\sigma_z$ & $s_0\tau_0\text{diag}(\widetilde{u}^{\widetilde{\alpha}}(g), \widetilde{u}^{\widetilde{\calT}\widetilde{\alpha}}(g))$\\
			\hline
			AI & $\sigma_z$& $\sigma_x$ & $\tau_y\sigma_y$ & $\tau_0\text{diag}(\widetilde{u}^{\widetilde{\alpha}}(g), \widetilde{u}^{\widetilde{\alpha}}(g))$\\
			\hline
			AI$_\calC$ & $s_z$ & $s_x$ & $s_y\tau_z$ & $s_0\tau_0\text{diag}(\widetilde{u}^{\widetilde{\alpha}}(g), \widetilde{u}^{\widetilde{\calC}\widetilde{\alpha}}(g))$\\
			\hline
			CI & $\tau_z$ & $\tau_x$ & $\tau_y\sigma_z$ & $\tau_0\text{diag}(\widetilde{u}^{\widetilde{\alpha}}(g), \widetilde{u}^{\widetilde{\alpha}}(g))$\\
			\hline
		\end{tabular}
	\end{center}
\end{table*}

\add{
\section{Remark on the results of classifications}
\label{app:remark}
In this appendix, we provide points to be noted in our classification results. 
\subsection{S(A) and L(A)}
In this work, we have classified nodes into only four categories: S(A) and L(A) represent surface and line nodes diagnosed by compatibility relations; L(B) and P(B) denote line and point nodes not explained by compatibility relations. However, two types of nodes are included in S(A) and L(A).
One is that each of some $2$-cells has only one gapless line. The other is that at least one $2$-cell contains multiple gapless lines. Indeed, these can be distinguished by Eq.~\eqref{eq:d1n}. 
For the case where the expansion in Eq.~\eqref{eq:d1n} contains two different generators of $E_{1}^{2,0}$ for the same $2$-cells or where an expansion coefficient is not one, multiple gapless lines on a $2$-cell are extended from a gapless point on the $1$-cell. In the following, we discuss such a case in S(A) and L(A) through two examples.

Let us begin by discussing case S(A). 
As with the case where a loop or surface node is shrunk to a point, nodes classified into S(A) are sometimes shrinkable to lines. 
To see this, let us discuss MSG $Pmm2$ with $A_2$ representation. Suppose that a gapless point exists on the line $(1/2,0,0)-(1/2,1/2,0)$, which corresponds to a generator of $E_{1}^{1,0}$. Then, we construct an effective model near the gapless point
\begin{align}
	\label{eq:effective_model1}
	H_{\delta \bk} &= \delta k_y \sigma_3 + \delta k_z \sigma_0 + \delta k_x \sigma_1 , \\
	\rho_{\bk}(M_x) &= i\sigma_3,\\ 
	\rho_{\bk}(C_{2}^{z}\calC) &= \sigma_2 K,
\end{align}
where $\delta \bk = (\delta k_x, \delta k_y, \delta k_z)$ is the displacement vector from the gapless point. Then, we find the energy dispersion $E_{\delta \bk} = \pm \sqrt{\delta k_x^2 + \delta k_y^2} + \delta k_z$. By solving $E_{\delta \bk} = 0$, we see that the system exhibits surface nodes shown in Fig.~\ref{fig:SA-LA} (a). One can see that there are two gapless lines in the mirror plane, which are part of surface nodes. When we deform the gapless lines such that the two gapless lines lie in the same positions, the surface nodes can shrink to lines nodes. However, since there are no reasons why the gapless lines are forced to be in such a way, it is natural to think that the gapless point on the $1$-cell is part of surface nodes.

\begin{figure}[t]
	\begin{center}
		\includegraphics[width=0.99\columnwidth]{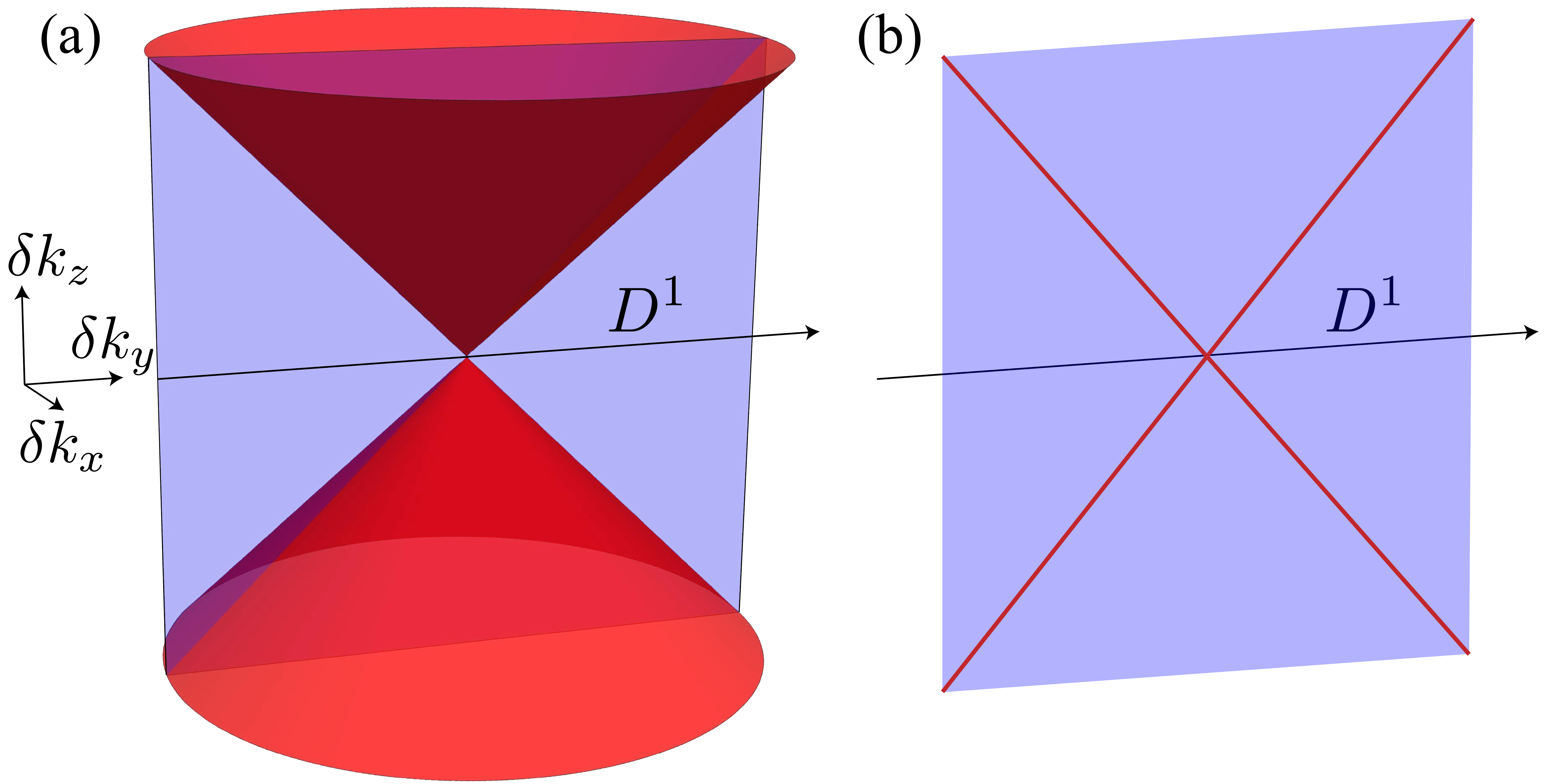}
		\caption{\label{fig:SA-LA}Nodal structures for effective low-energy models in Eqs.~\eqref{eq:effective_model1} and \eqref{eq:effective_model2}. The blue planes represent the mirror invariant planes, and red cones in (a) and the red lines in (b) are surface nodes for Eq.~\eqref{eq:effective_model1} and line nodes for Eq.~\eqref{eq:effective_model2}.
		}
	\end{center}
\end{figure}

As for case L(A), a gapless point on a $1$-cell is sometimes part of several nondegenerate line nodes. In other words, several line nodes on a $2$-cell can be extended from a gapless point on a $1$-cell. 
To show this, we discuss the line $(1/2,0,1/2)-(1/2,1/2,1/2)$ in MSG $P_Cmm2$ with $B_1$ representation. This MSG is generated by space group $Pmm2$ and $\{\calT\vert \frac{\bm{e}_z}{2}=(0,0,1/2)^T\}$, where Seitz symbol $\{p_g \vert \bm{t}_g\}$, a point-group operation $p_g$ and a translation $\bm{t}_g$, is adopted. 
We again construct an effective model near the gapless point $\bm{K} = (K_x, K_y, K_z)$:
\begin{align}
	\label{eq:effective_model2}
	H_{\delta \bk} &= \delta k_y \sigma_3 + \delta k_z \sigma_0 + \delta k_x \sigma_1 , \\
	\rho_{\bk}(M_x) &= i\tau_3,\\ 
	\rho_{\bk}(C_{2}^{z}\calC) &= i \tau_1\sigma_1 K,\\
	\rho_{\bk}(\{C_{2}^{z}\calT\vert \bm{e}_z/2\}) &= \left(
	\begin{array}{cccc}
		0 & e^{i (K_z + \delta k_z)} & 0 & 0 \\
		1 & 0 & 0 & 0 \\
		0 & 0 & 0 & e^{i  (K_z + \delta k_z)} \\
		0 & 0 & 1 & 0 \\
	\end{array}
	\right)K.
\end{align}
The energy dispersion is $E_{\delta \bk} = \pm \sqrt{(\delta k_y \pm \delta k_z)^2 + \delta k_x^2}$. We find the solutions $\delta k_z = \pm \delta k_y$ with $\delta k_x=0$ for $E_{\delta \bk} = 0$, which indicates the existence of two nondegenerate line nodes in the mirror plane (see Fig.~\ref{fig:SA-LA} (b)).

\subsection{P(B)}
To distinguish a genuine point node from line nodes extended from 1-cells to 3-cells, we classified two-dimensional massive Dirac Hamiltonians~\eqref{eq:cc-ham}. However, one might sometimes notice that the degeneracy of the point node is smaller than the dimension of the corresponding Dirac Hamiltonians. 

Let us discuss the four-fold rotation axis in spinless MSG $P4$ as an example. After performing the procedures discussed in Sec.~\ref{sec4}, we obtain the following generating Hamiltonians
\begin{align}
	H &= k_1 \sigma_x + k_2 \sigma_y + \delta k_3 \sigma_3,\\
	\sigma(C_4) &= \text{diag}(1,\pm i)\text{  or  }\text{diag}(\pm i,-1),
\end{align}
where corresponding band labels are $(\frakcn^{1},\frakcn^{-1}, \frakcn^{i},\frakcn^{-i}) = (1,0,-1,0), (1,0,0,-1), (0,-1,0,1), $ and $(0,-1,1,0)$.
One can see that $(\frakcn^{1},\frakcn^{-1}, \frakcn^{i},\frakcn^{-i}) = (1,-1,0,0)$, which is a generator of $E_{1}^{1,0}$, does not correspond to any generating Hamiltonian. 
Although one might think that the generator does not correspond to a point node, this is untrue. To show this, let us consider stacking two of the above generating Hamiltonians with a coupling term. For example, we here discuss $H' = H\oplus H$ and $\sigma(C_4) = \text{diag}(1,i)\oplus\text{diag}(i,-1)$. Since the eigenvalue $i$ appears in both occupied and unoccupied bands, one of the gapless points can be gapped out by coupling these two states. As a result, we obtain the band labels $(\frakcn^{1},\frakcn^{-1}, \frakcn^{i},\frakcn^{-i}) = (1,-1,0,0)$.
However, while the degeneracy of the gapless point is two, the dimension of the stacking Hamiltonian is four.
Indeed, this mismatch originates from the restriction of linear dependence of $k_1$ and $k_2$. When we consider quadratic terms instead, we get another generating Dirac Hamiltonian
\begin{align}
	H &= k_1 k_2 \sigma_x + (k_1^2 - k_2^2) \sigma_y + \delta k_3 \sigma_3,\\
	\sigma(C_4) &= \text{diag}(1,-1),
\end{align}
where the dimension of Dirac Hamiltonian equals the degeneracy of the gapless point.

On the one hand, the existence of such large dimensional linear Dirac Hamiltonians in Eq.~\eqref{eq:cc-ham} ensures that the point node is stable in the sense of K-theory, i.e., against adding trivial degrees of freedom. On the other hand, it does not rule out the possibility of nodes in the sense of \textit{fragile topological phases}~\cite{PhysRevLett.121.126402}. 
Actually, we find that the mismatches sometimes happen in two-dimensional point groups $4, 4mm, 6$, and $6mm$. To check if we can construct a minimal dimension Dirac Hamiltonians, we generalize the above discussion for MSG $P4$ to any symmetry setting. We redefine generating Dirac Hamiltonians by
\begin{align}
	\label{eq:Dirac_C4}
	H &= k_1 k_2  \gamma_1 + (k_1^2 - k_2^2)  \gamma_2 + \delta k_3 \gamma_0
\end{align}
for point groups $4$ and $4mm$;
%\begin{align}
%	\label{eq:Dirac_C6}
%	H = \begin{cases}
%		(k_1^3 -3 k_1 k_2^2)  \gamma_1 + (k_1^2 k_y -\tfrac{k_2^3}{3})  \gamma_2 + \delta k_3 \gamma_0\\
%		2 k_1 k_2  \gamma_1 + (-k_1^2 + k_2^2)  \gamma_2 + \delta k_3 \gamma_0
%	\end{cases}
%\end{align}

\begin{align}
	\label{eq:Dirac_C6-1}
	H = 
		(k_1^3 -3 k_1 k_2^2)  \gamma_1 + (k_1^2 k_y -\tfrac{k_2^3}{3})  \gamma_2 + \delta k_3 \gamma_0
\end{align}
or
\begin{align}
	\label{eq:Dirac_C6-2}
	H = 
	2 k_1 k_2  \gamma_1 + (-k_1^2 + k_2^2)  \gamma_2 + \delta k_3 \gamma_0
\end{align}
for $6$ and $6mm$.
 Indeed, the classification procedures can be performed just by redefining $\theta_g$ in Eq.~\eqref{eq:rep2}. The new definitions for rotation symmetry and mirror symmetry about the $yz$-plane are $\theta_{C_4} = \pi$ for point group $4$, $(\theta_{C_4}, \theta_{M_x}) = (\pi, 0)$ for point group $4mm$, $\theta_{C_6} = \pi$ [for Eq.~\eqref{eq:Dirac_C6-1}] or $\tfrac{2\pi}{3}$ [for Eq.~\eqref{eq:Dirac_C6-2}] for point group $6$, and $(\theta_{C_6}, \theta_{M_x}) = (\pi, 0)$ [for Eq.~\eqref{eq:Dirac_C6-1}] or $ (\tfrac{2\pi}{3}, 0)$ [for Eq.~\eqref{eq:Dirac_C6-2}] for point group $6mm$. 
 Since other elements are products of these two symmetries, $\theta_g$ is automatically determined. 
 After performing classifications of Dirac Hamiltonians in Eqs.~\eqref{eq:cc-ham} and~\eqref{eq:Dirac_C4}-\eqref{eq:Dirac_C6-2}, we find that all point nodes, classified into P(B), have corresponding Dirac Hamiltonians whose dimensions are equal to degeneracy of the point nodes. 
This implies that any node on $1$-cells is not fragile
}

\section{Stability of genuine point nodes against perturbations}
Suppose that we have a 2D massive Dirac Hamiltonian in Eq.~\eqref{eq:cc-ham} that is mapped to a generator $\frakb$ of $E_1^{1,0}$ by formulas in Table~\ref{tab:formulas}. In fact, this is the case (i) in Sec.~\ref{sec4:cc-formulas}.
The correspondence to a generator of $E_1^{1,0}$ ensures that the gapless point on the $1$-cell is not split. However, there still remains the possibility of part of shrinkable line or surface nodes, which has a single nodal point on the 1-cell [see Fig.~\ref{fig:SA-LA} (a) and (b)].
For instance, one should exclude a line node in the shape of the Arabic numeral ``$8$'', where the gapless point on the $1$-cell is the knot of eight.
We show that this is not the case: the generator is a genuine point node on the $1$-cell for a generic parameter region
To see this, let us denote the Dirac Hamiltonian near the $1$-cell in a slightly generic way than Eq.~\eqref{eq:cc-ham} as
\begin{align}
	\label{eq:cc-ham_v}
	H_{(k_1,k_2)} = v_1 k_1 \gamma_1 + v_2 k_2 \gamma_2 + v_3 \delta k_3\gamma_0 + O(k^2),
\end{align}
where $v_1, v_2$, and $v_3$ are constants, and $O(k^2)$ represents the order of $k_1^2, k_2^2$ and $\delta k_3^2$.
No constant terms compatible with symmetry can be added to Eq.~\eqref{eq:cc-ham_v}, since $\frakb$ is a generator of $E_1^{1,0}$.
The determinant of $H_{(k_1,k_2)}$ takes a form of
\begin{align}
	&\det H_{(k_1,k_2)} \nonumber \\
	&= \big\{(v_1 k_1)^2 + (v_2 k_2)^2 + (v_3 \delta k_3)^2 + O(k^3)\big\}^{N/2}
\end{align}
with the rank of Gamma matrices $N$.
It is clear that in a sufficiently small three dimensional ball near $(k_1,k_2,\delta k_3) = (0,0,0)$, $\det H_{(k_1,k_2)} > 0$ except for the point $(k_1,k_2,\delta k_3) = (0,0,0)$, unless either of $v_1, v_2$ or $v_3$ is zero.
Since there are no symmetry constraints enforcing $v_1, v_2$, or $v_3$ to be zero, we conclude that in a generic parameter region the gapless point $\frakb$ of $E_1^{1,0}$ represents a genuine point node.

\section{Symmetry property of the winding number}
\label{app:winding}
\begin{figure}[h]
	\begin{center}
		\includegraphics[width=0.4\columnwidth]{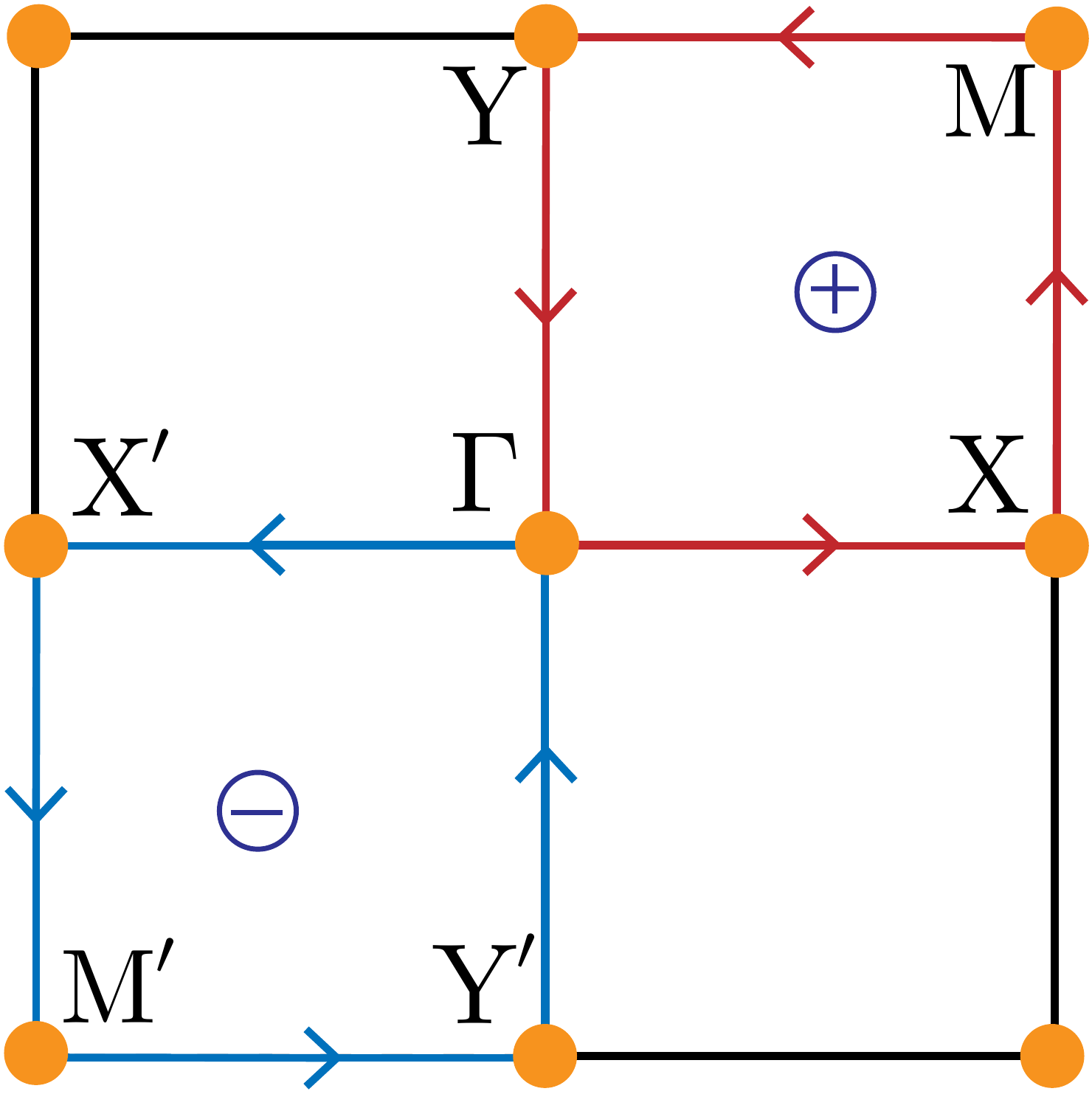}
		\caption{\label{fig:winding_BZ}Illustration of the interval of integral in the winding number. Note that the path colored by blue is symmetry-related to the red path. Here, white solids circles denote gapless points and $\pm$ represent the sign of the winding numbers.}
	\end{center}
\end{figure}

In this appendix, we show that $C_{2}^{y}$ which anticommutes with PHS changes the sign of the winding number. The winding number is defined by 
\begin{align}
	W[C] \equiv \oint_{C} \mathrm{tr}[U(\Gamma)\left(H_{\bk}^{-1}\partial_{k}H_{\bk}\right)] ds,
\end{align}
where we consider $\Gamma-\text{X}-\text{M}-\text{Y}-\Gamma$ in Fig.~\ref{fig:winding_BZ} as $C$. We first compute the integrand
\begin{align}
	&\mathrm{tr}[U(\Gamma)\left(H_{\bk}^{-1}\partial_{k_i}H_{\bk}\right)]  \nonumber\\
	&=\mathrm{tr}[U(C_{2}^{y})U(\Gamma)\left(H_{\bk}^{-1}\partial_{k_i}H_{\bk}\right)U^{-1}(C_{2}^{y})] \nonumber\\
	&=  -\mathrm{tr}[U(\Gamma)U(C_{2}^{y})\left(H_{\bk}^{-1}\partial_{k_i}H_{\bk}\right)U^{-1}(C_{2}^{y})]  \nonumber\\
	&=  -\mathrm{tr}[U(\Gamma)\left(H_{-\bk}^{-1}\frac{\partial}{\partial k_i}H_{-\bk}\right)].
\end{align}
Using the identity, we derive the following relation
\begin{align}
	&\int_{\Gamma}^{\text{X}} \mathrm{tr}[U(\Gamma)\left(H_{(k_x,0)}^{-1}\partial_{k_x}H_{(k_x,0)}\right)] dk_x \nonumber\\
	&\quad = -\int_{\Gamma}^{\text{X}'} \mathrm{tr}[U(\Gamma)\left(H_{(k_x,0)}^{-1}\partial_{k_x}H_{(k_x,0)}\right)] dk_x
\end{align}
For other integral intervals, one finds the same transformation. As a result, we obtain the relation $W[C] = -W[C_{2}^{y}C]$. Actually, the relation can be generalized to other point group symmetries as $W[C] = \chi_g\det p_gW[gC]\ (\chi_g =\pm 1)$, where $U(g)U(C) = \chi_g U(C)[U(g)]^{*}$.

\bibliography{ref}
\end{document}